\documentclass[a4paper,11pt]{article}

\usepackage{graphicx}
	
\usepackage{jheppub}
\usepackage{epsfig}
\usepackage{amsmath}
\usepackage{amsfonts}
\usepackage{amssymb}

\usepackage{mathrsfs}

\usepackage{graphicx}
\usepackage{dcolumn}
\usepackage{bm}
\usepackage{epsfig}

\usepackage[usenames]{color}

\usepackage{booktabs}
\usepackage{hyperref}

\newcommand{\dhd}{{\textstyle d}
	\lower.03ex\hbox{\kern-0.38em$^{\scriptstyle-}$}\kern-0.05em{}}
\newcommand{\dbar}{{\textstyle \delta}
	\lower.03ex\hbox{\kern-0.38em$^{\scriptstyle-}$}\kern-0.05em{}}
\newcommand{\half}{{1\over 2}}

\newcommand{\barq}{{\bar q}}
\newcommand{\barv}{{\bar v}}
\newcommand{\baru}{{\bar u}}

\newcommand{\barz}{{\bar z}}

\newcommand{\calf}{{\cal F}}
\newcommand{\calg}{{\cal G}}

\newcommand{\calk}{{\cal K}}

\newcommand{\calm}{{\cal M}}
 
\newcommand{\calo}{{\cal O}} 
 
\newcommand{\calq}{{\cal Q}}

\newcommand{\calt}{{\cal T}} 
\newcommand{\calu}{{\cal U}} 
 
\newcommand{\calw}{{\cal W}}

\newcommand{\barpsi}{{\bar \psi}}

\newcommand{\hatp}{{\hat p}}

\newcommand{\tildeD}{{\tilde D}}

\newcommand{\tildeQ}{{\tilde Q}}

\newcommand \ket [1] {|{#1}\rangle}
\newcommand \bra [1] {\langle {#1}|}

\newcommand{\ketx}{\ket{x}}
\newcommand{\kety}{\ket{y}}

\newcommand{\ketz}{\ket{z}}

\newcommand{\brax}{\bra{x}}

\newcommand{\braz}{\bra{z}}

\newcommand{\ketyp}{\ket{y_\perp}}
\newcommand{\ketzp}{\ket{z_\perp}}

\newcommand{\braxp}{\bra{x_\perp}}

\newcommand{\brazp}{\bra{z_\perp}}

\newcommand \sslash [1] {\slash\hspace{-0.2cm}{#1}}
\newcommand{\ssp}{\sslash{p}}
\newcommand{\ssn}{\sslash{n}}
\newcommand{\ssk}{\sslash{k}}
\newcommand{\ssq}{\sslash{q}}

\newcommand{\slashd}{\sslash{\partial}}

\newcommand \Slash [1] {\slash\hspace{-0.23cm}{#1}}
\newcommand{\Sa}{\Slash{A}}

\newcommand{\Sp}{\Slash{P}}

\newcommand{\Tr}{{\rm Tr}} 
\newcommand{\tr}{{\rm tr}}

\newcommand{\vecp}{\vec{p}}
\newcommand{\vecq}{\vec{q}}
\newcommand{\veck}{\vec{k}}

\begin{document}
	
\title{Sub-eikonal Structure of High-Energy Deep-Inelastic Scattering}

\author[]{Giovanni Antonio Chirilli}
\affiliation[]{Theoretical Physics Division, National Centre for Nuclear Research,\\ Pasteura 7, Warsaw 02-093, Poland}
\emailAdd{giovanni.chirilli@ncbj.gov.pl}

\abstract{	

I develop a mixed-space formulation of high-energy deep-inelastic scattering at sub-eikonal order,
deriving shock-wave Feynman rules through LSZ reduction, including instantaneous contributions. I calculate
the inclusive quark--antiquark, quark--gluon, and antiquark--gluon channels, which complement the
one-particle channels $\gamma^*\to q$ and $\gamma^*\to\bar q$ that give the leading sub-eikonal quark
contribution I derived in  ref.~\cite{Chirilli:2026pkv}. 
When the explicit Bjorken-$x$ dependence of the sub-eikonal operators is neglected, these
channels give vanishing corrections to the longitudinal and transverse structure functions. This follows
from charge conjugation: the $C$-even vector light-ray operator is odd in the momentum fraction of the
target quark and vanishes when that fraction is set to zero, so the cancellation is not the small-$x$
limit of the physical result. 
Keeping the longitudinal phase of the photon vertex, the quark–gluon and antiquark–gluon channels reproduce the 
Altarelli--Martinelli quark contribution in the massless, dilute leading-twist limit. The helicity-dependent structure function receives gluonic contributions and a
correction of relative order $\alpha_s$ to the leading quark term. The quark contribution exhibits
longitudinal endpoint and transverse short-distance logarithms. Its endpoint logarithm, including the
mixing with the gluonic operator, is one step of the rapidity evolution of the leading quark term.
}

\maketitle
\section{Introduction}
\label{sec:intro}

Deep-inelastic scattering (DIS) has played a central role in establishing Quantum Chromodynamics (QCD) as the theory of the strong interaction and in turning the notion of 
hadronic structure into a quantitative and experimentally testable framework. Inclusive DIS observables, encoded in structure functions such as $F_2(x_B,Q^2)$ and 
$g_1(x_B,Q^2)$, separate the short-distance dynamics probed by the photon virtuality $Q^2$ from the long-distance dynamics of quarks and gluons in the target. Over 
the years, this program has revealed the partonic structure of hadrons, the scaling violations driven by the DGLAP evolution equations, and the emergence of collective 
effects at high parton density.

A particularly important frontier is the high-energy, small-Bjorken-$x_B$ regime. In this limit, perturbation theory is enhanced by large logarithms of the energy, or 
equivalently of $1/x_B$, whose resummation is described by the Balitsky-Fadin-Kuraev-Lipatov (BFKL) evolution equation~\cite{Kuraev:1977fs,Balitsky:1978ic}. The resulting 
rapid growth of gluon densities eventually drives the system toward the saturation regime, where multiple scattering and gluon recombination can no longer be 
neglected~\cite{Gribov:1983ivg,Mueller:1985wy,McLerran:1993ka}. In DIS, this regime is naturally described in the dipole 
picture~\cite{Nikolaev:1990ja,Nikolaev:1991et,Mueller:1993rr}, in which the virtual photon fluctuates into a quark-antiquark pair that subsequently scatters off the
target background through Wilson lines. The phenomenological relevance of this region was clearly demonstrated by the DIS measurements 
at HERA, whose combined inclusive data 
provide the standard benchmark for analyses of QCD dynamics at small $x_B$~\cite{H1:2015ubc}.

High-energy DIS is therefore most naturally formulated in the language of Wilson lines and high-energy operator expansion~\cite{Balitsky:1995ub,Balitsky:2001gj}, within the 
Color Glass Condensate (CGC) framework~\cite{Gelis:2010nm}. In this description, the interaction with the target is eikonal at leading power in the energy and is encoded in 
Wilson lines extending along nearly light-like trajectories. The energy dependence of the corresponding operators is governed by the BK/B-JIMWLK evolution 
equations~\cite{Balitsky:1995ub,Kovchegov:1999yj,Kovchegov:1999ua,Jalilian-Marian:1997qno,Jalilian-Marian:1997jhx,Jalilian-Marian:1997ubg,
Weigert:2000gi,Iancu:2000hn,Ferreiro:2001qy,Gelis:2010nm} (see ref.~\cite{Balitsky:2001gj, Balitsky:2025bup, Kovchegov:2012mbw} for reviews).

The phenomenological importance of this regime is one of the main motivations for the Electron-Ion Collider (EIC), which will provide high-luminosity electron-hadron and 
electron-ion collisions with polarized beams over a broad kinematic range. Besides precision studies of the onset of non-linear QCD dynamics, the EIC will offer a unique 
opportunity to investigate spin-dependent observables at small $x_B$, including helicity-dependent parton distributions and their manifestation in polarized structure 
functions. This makes it necessary to develop a formulation of high-energy DIS that is at the same time adapted to the Wilson-line description of dense targets and 
sufficiently precise to retain the leading spin-sensitive power corrections.

From the theoretical point of view, the eikonal approximation is intrinsically spin-blind. At leading power in the high-energy limit, an eikonal Wilson line resums the color 
phase accumulated by a fast parton propagating through the target field, but it does not resolve the subleading couplings responsible for genuine helicity sensitivity. As a 
consequence, polarized observables, most notably the structure function $g_1(x_B,Q^2)$, cannot be described within the eikonal approximation alone. In operator language, 
one has to enlarge the eikonal Wilson-line basis by including sub-eikonal corrections built from field strengths and quark fields. These gauge-invariant operator insertions 
describe the first subleading interactions of a fast parton with the target background and therefore provide the natural building blocks for the high-energy description of 
polarized scattering.

For observables such as $F_L$ and $F_T$, the eikonal dipole picture already captures the leading contribution. However, once one aims at a systematic treatment of power 
corrections, or at a unified description of polarized and unpolarized observables within the same high-energy formalism, the sub-eikonal sector becomes unavoidable. In 
particular, the same framework that is needed to access the helicity-dependent sector also determines the domain of validity of the eikonal approximation and organizes the 
first energy-suppressed corrections to the dipole structure functions.

A study of polarized scattering at small $x_B$ has been developed from several complementary perspectives. In particular, the small-$x_B$ helicity program of 
refs.~\cite{Kovchegov:2016zex, Kovchegov:2017lsr, Kovchegov:2018znm} showed that the polarized structure function $g_1$ and helicity-dependent parton 
distributions can be described in terms of polarization-dependent Wilson-line operators obeying dedicated small-$x_B$ evolution equations. Our goal here is different, though 
closely related. Rather than starting from helicity evolution as an independent framework, we develop the sub-eikonal high-energy expansion directly in the shock-wave 
formalism and construct its mixed-space realization for DIS amplitudes and structure functions. In this way, the operator content responsible for the first spin-sensitive 
corrections emerges directly from the same Wilson-line expansion that underlies the dipole picture.

In collinear factorization the polarized structure function $g_1$ is known to high perturbative order.
The polarized splitting functions were obtained at two loops by Mertig and van Neerven
\cite{Mertig:1995ny} and by Vogelsang \cite{Vogelsang:1996im}, and the two-loop coefficient functions of
$g_1$ by Zijlstra and van Neerven \cite{Zijlstra:1993sh}. At three loops the helicity-dependent splitting
functions were computed by Moch, Vermaseren and Vogt \cite{Moch:2014sna}. Bl\"umlein, Marquard, Schneider
and Sch\"onwald recomputed them with an independent method, based on off-shell operator matrix elements
\cite{Blumlein:2021ryt,Blumlein:2021lmf}, and also obtained the massless three-loop Wilson coefficient of
$g_1$ \cite{Blumlein:2022gpp}. At small $x_B$ the leading terms of both the splitting functions and the
coefficient functions are double logarithms $\alpha_s^{l+1}\ln^{2l}(1/x)$. The double-logarithmic
asymptotics of quark-exchange amplitudes were obtained by Kirschner and Lipatov \cite{Kirschner:1983di},
and the double logarithms were resummed for $g_1$ by Bartels, Ermolaev and Ryskin
\cite{Bartels:1995iu,Bartels:1996wc}. Bl\"umlein and Vogt resummed the leading small-$x_B$ terms in the
evolution kernels of the non-singlet and singlet parts of $g_1$ and predicted the leading small-$x_B$ terms
of the three-loop splitting functions \cite{Blumlein:1995jp,Blumlein:1996hb}. In the non-singlet case they
found the effect of the resummation to be at the per-cent level or below in the kinematic range then
accessible \cite{Blumlein:1995jp}. In the singlet case the effect is large, but less singular terms of
comparable size and opposite sign can compensate it, so that these terms must be calculated before firm
conclusions can be drawn \cite{Blumlein:1996hb}. The three-loop results agree with the predicted leading
small-$x_B$ logarithms and exhibit a large double-logarithmic enhancement at small $x_B$
\cite{Moch:2014sna}. These limits also provide a benchmark for the Wilson-line approach. The revised
small-$x_B$ helicity evolution \cite{Cougoulic:2022gbk} reproduces the small-$x_B$ limit of the fixed-order
polarized anomalous dimensions through three loops, up to a scheme transformation at three loops in the
quark sector \cite{Borden:2023ugd,Borden:2024bxa}, while it appears
to differ from the result of ref.~\cite{Bartels:1996wc} starting at four loops
\cite{Borden:2023ugd,Borden:2024bxa}. In the present framework, the longitudinal-endpoint and transverse
logarithms identified in Sec.~\ref{sec:divergence} have the structure of the first step of this double-logarithmic series.
A quantitative comparison with the fixed-order results requires the finite-$x_B$ kinematics and a common
factorization scheme.

The framework used throughout this work is the background-field method, which provides a natural formulation of high-energy factorization in QCD~\cite{Balitsky:1998ya, Balitsky:2001re, Balitsky:2001gj, Balitsky:2025bup}. 
The fast projectile fields are treated perturbatively in the presence of the slowly varying target background, whose interaction with energetic partons is encoded in light-like Wilson lines where multiple interactions with the background are resummed, while the first power-suppressed terms appear as sub-eikonal insertions of gluon field strengths and quark fields along the Wilson lines. The shock-wave formalism employed below is the high-energy realization of this background-field expansion and provides the starting point for the systematic derivation of the sub-eikonal DIS amplitudes.

The starting point for such a program is the propagation of quarks through a background shock-wave field beyond the eikonal approximation. In ref.~\cite{Chirilli:2018kkw}, 
we derived the quark and gluon propagators including sub-eikonal corrections in the shock-wave background. 
In ref.~\cite{Chirilli:2021lif} we extended to sub-eikonal accuracy in coordinate space the high-energy operator product expansion for 
the time-ordered product of two electromagnetic currents, the enlarged operator basis beyond the eikonal limit 
was identified, and the corresponding rapidity evolution equations were derived. These results provide the coordinate-space foundation for a systematic treatment of DIS 
beyond the eikonal approximation. The purpose of the present paper is to develop the corresponding mixed-space formulation and to apply it directly to the dipole structure 
functions.

We develop and use a mixed-space shock-wave formalism at sub-eikonal accuracy in order to compute the first sub-eikonal corrections to the dipole structure functions. The use of LSZ reduction to extract scattering amplitudes in the presence of a strong classical background field has appeared previously in the high-energy literature~\cite{Dumitru:2001jn,Dumitru:2002qt}. Here we formulate the LSZ reduction directly in the shock-wave background and extend it beyond the eikonal approximation. An important feature of this construction is that the background-field propagator has restricted support in light-cone time. Consequently, the derivatives entering the LSZ reduction act not only on the propagating fields but also on the step functions specifying the position of the external points relative to the shock wave, generating instantaneous contributions localized at the interaction. These terms are required for a consistent derivation of the mixed-space Feynman rules at sub-eikonal accuracy.

An important outcome of the calculation is the different behavior of the
unpolarized and helicity-dependent projections. We compute the $q\barq$,
$qg$, and $\bar q g$ final states. The minimal configuration in which quark
operators enter DIS beyond the dipole approximation is instead the
one-particle final state, $\gamma^*\to q$ or $\gamma^*\to\bar q$, with one
end-point of the quark propagator inside the shock wave; it was analyzed in
ref.~\cite{Chirilli:2026pkv}. In that channel the target quark field enters the amplitude
without any power of $g$, whereas in the channels considered here the
quark-background propagator is of order $g^2$ and the $qg$ and $\barq g$
amplitudes are of order $g$. The quark-operator contributions computed here
are therefore of relative order $\alpha_s$ with respect to the one-particle
channel. Within the high-energy approximation used for the background
operators, in which their $x_B$ dependence is neglected, the first sub-eikonal
corrections of the $q\barq$, $qg$, and $\bar q g$ final states to the
inclusive longitudinal and transverse structure functions cancel. This
includes the gluonic and quark-background contributions to the $q\bar q$
channel and the sum of the charge-conjugate $qg$ and $\bar q g$ channels. In
the same approximation the one-particle channel does not contribute to $F_L$
and $F_T$ either: its longitudinal contribution lies beyond sub-eikonal
accuracy, and its quark and antiquark contributions to $F_T$ cancel when the momentum fraction in the quark operators is set to zero. 
At finite $x_B$ it yields the parton-model result
$F_T=x_B\sum_f e_f^2\,(q_f+\bar q_f)$~\cite{Chirilli:2026pkv}. 
These cancellations have a common origin. Charge conjugation maps the vector light-ray operator at
momentum fraction $x$ onto minus the one at $-x$, so the $C$-even combination that enters $F_L$ and $F_T$ is
odd in $x$ and vanishes at $x=0$. Its matrix element, however, tends to $q(0)+\bar q(0)$ as $x\to0^+$, and
the vanishing at $x_B=0$ is therefore not the $x_B\to 0^+$ limit of the physical result. We
show this explicitly for $F_L$: when the longitudinal phase of the photon vertex is kept, the $qg$ and
$\bar qg$ channels give the Altarelli--Martinelli quark contribution~\cite{Altarelli:1978tq}
(Sec.~\ref{sec:xBreal}). The $C$-even axial combination is instead even in $x$ and survives at $x=0$,
which is why $g_1$ receives a sub-eikonal quark correction in this limit.

The quark contribution to the helicity-dependent result requires combining the different final states before
interpreting its operator content. The explicit local terms in the $q\bar q$ contribution cancel those generated by the
$qg$ and $\bar qg$ channels. The remaining color traces, however, contain a local contribution when expressed in
terms of subtracted bilocal operators. Thus, the vanishing of the bilocal operators at zero dipole size does not imply
that the complete helicity-dependent expression vanishes in this limit.

The main results of this work are the following. First, we construct the mixed-space LSZ reduction
and the corresponding Feynman rules at sub-eikonal accuracy. Second, we find that the sub-eikonal
corrections of the $q\bar q$, $qg$, and $\bar qg$ final states to the inclusive unpolarized projections
cancel when $x_B$ is set to zero in the operators, and we show that this cancellation follows from charge
conjugation. When the $x_B$ dependence is kept, the $qg$ and $\bar qg$ channels give instead the
Altarelli--Martinelli quark contribution to $F_L$. Third, we obtain the explicit coordinate-space
expression for the helicity-dependent contribution of these final states, which adds to the leading
one-particle term of ref.~\cite{Chirilli:2026pkv}. This expression identifies the gluonic and axial quark
operators relevant to $g_1$, together with logarithmic longitudinal and transverse integration regions.
The coefficient of the longitudinal endpoint logarithm has the operator structure of one step of the
rapidity evolution of $Q_5$, including its mixing with $\calf$. This singularity is therefore absorbed
into the evolution of the one-particle term, except for the $\calf_i$ contribution, which has no
counterpart in that equation. In the flavor non-singlet sector the corresponding evolution is closed.

The paper is organized as follows. In section~~\ref{sec:notation} we present
the kinematics and conventions. In section~~\ref{sec:quark-subeikcor} we rewrite the
background-field quark propagator at sub-eikonal accuracy,
and in  section~\ref{sec:Feynmanrule} we develop the corresponding LSZ reduction formula.
In section~\ref{sec:dipolecrossection} we reproduce the eikonal dipole cross sections.
In section~\ref{sec:dipolewithsbeikonal} we calculate the sub-eikonal contributions with
a $q\bar q$ final state, while in section~\ref{sec:crossection-qg} we consider the $qg$ and
$\bar qg$ channels. In section~\ref{sec:summaryresults} 
we collect the resulting structure functions and discuss their singular regions
and relation to rapidity factorization. In section~\ref{sec:conclusions} we present
conclusions and outlook.


\section{Preliminaries and notation}
\label{sec:notation}

Before we proceed to the calculation of the polarized and unpolarized structure functions, 
in this section we summarize the kinematics and conventions we will use. 

We work in Minkowski space with metric
$g^{\mu\nu}=\mathrm{diag}(1,-1,-1,-1)$ and introduce two light-like vectors along the beam directions. We will use
the light-cone vectors $n_1^\mu$ and $n_2^\mu$ with $n_1^2=n_2^2=0$
and $n_1\!\cdot n_2=1$.

Given an arbitrary four-vector $k^\mu$, the Sudakov decomposition with respect to $n_1$ and $n_2$ is
\begin{eqnarray}
k^\mu = k^+n_1^\mu + k^- n_2^\mu + k_\perp^\mu\,,
\qquad 
k_\perp\!\cdot n_1 = k_\perp\!\cdot n_2 = 0\,,
\label{sudakov-k}
\end{eqnarray}
with $k_\perp^\mu = (0,k^1,k^2,0)$ and 
\begin{eqnarray}
p_\perp^\mu k^\perp_\mu = p^ik_i = - (p,k)_\perp = -(p^1k^1+p^2k^2)
\end{eqnarray}
with $i=1,2$. In particular, we will use $k^2_\perp = -k^\mu_\perp k_\mu = k^ik^i$.

The virtuality of the photon is $q^2=-Q^2$, and for $q_\perp=0$ one has $q^2 = 2q^+q^-$, it follows that
\begin{eqnarray}
q^\mu = q^+ n_1^\mu - {Q^2\over 2q^+}n_2^\mu\,,
\label{q-parameterization}
\end{eqnarray}

In the high-energy kinematics relevant for this paper, the virtual photon carries a large
``plus'' component and a small ``minus'' component fixed by the virtuality $Q^2$.
Therefore, once $q_\perp=0$ is chosen, the whole dependence on the photon momentum is
encoded in the two longitudinal variables $q^+$ and $q^-$, related by the 
condition $q^2=-Q^2$. 

We choose the longitudinal polarization vector as
\begin{eqnarray}
\varepsilon_L^\mu = \alpha^+ n_1^\mu + \beta^- n_2^\mu\,,
\label{epsL-ab}
\end{eqnarray}
and impose the conditions
\begin{eqnarray}
\varepsilon_L^\mu \varepsilon_{L\,\mu} = 1\,,
\qquad\qquad
\varepsilon_L^\mu q_\mu = 0\,.
\label{epsL-conds}
\end{eqnarray}
Therefore,
\begin{eqnarray}
\varepsilon^\mu_L = {q^+\over Q}n_1^\mu + {Q\over 2q^+}n_2^\mu\,.
\label{epsL-final}
\end{eqnarray}

With this choice, the longitudinal polarization vector is normalized to unity and orthogonal
to the photon momentum. We will use this form repeatedly when separating longitudinal and
transverse contributions to the DIS cross section. Notice also that, in the high-energy limit,
$\varepsilon_L^\mu$ contains a large component along $n_1^\mu$ and a compensating small
component along $n_2^\mu$, as required by the condition $\varepsilon_L\cdot q=0$.

For transverse polarization, we choose the transverse polarization vectors
\begin{eqnarray}
	\varepsilon_\lambda^k = -{1\over \sqrt{2}}(\lambda,i)\,,
	\qquad \lambda = \pm 1\,,
\end{eqnarray}
where $i$ denotes the imaginary unit.

We will also use the $\hbar$-inspired notation
\begin{eqnarray}
\dhd^n k \equiv {d^n k\over (2\pi)^n}\,,
\qquad
\dbar^{(n)}(k)\equiv (2\pi)^n \delta^{(n)}(k)\,,
\end{eqnarray}
so that
\begin{eqnarray}
\int \dhd^n k\, \dbar^{(n)}(k)=1\,.
\end{eqnarray}

Since the proton moves predominantly along the $n_2^\mu$ direction, we parameterize its momentum as
\begin{eqnarray}
P^\mu = \sqrt{s\over 2}n_2^\mu + {M^2\over \sqrt{2s}}n_1^\mu\,,
\end{eqnarray}
where $M$ is the hadron mass and $s$ is the Mandelstam variable
so that we are in a frame in which the hadronic target has a large $P^-=\sqrt{s/2}$ component and the virtual photon has a large $q^+=\sqrt{s/2}$ 
component.

We will use $\tr$ for trace over spinor indexes and $\Tr$ for trace over color indexes in the fundamental representation.

\section{Quark propagator up to sub-eikonal corrections}
\label{sec:quark-subeikcor}

In this section we summarize the quark propagator in the external ``shock-wave'' gluon
background, keeping terms up to sub-eikonal accuracy. The expressions we need were
derived in ref.~\cite{Chirilli:2018kkw}. Our purpose here is not to rederive them, but to
rewrite each contribution in a form convenient for the subsequent application of the LSZ
reduction formula to DIS amplitudes.

The expressions we need are conveniently written in the Schwinger (operator) notation,
which makes the separation between free propagation and interaction with the shock wave
fully transparent. In this way, the propagator is naturally organized as free propagation
from $y$ to the shock-wave plane, followed by an interaction localized on the shock wave,
and then free propagation from the shock-wave plane to $x$. At eikonal level this
interaction is encoded in Wilson lines, while at sub-eikonal level it is supplemented by
local operator insertions.

We denote the time-ordered propagator in the background quark and gluon fields by
\begin{eqnarray}
S(x,y)\equiv \langle {\rm T}\{\psi(x)\bar\psi(y)\}\rangle_{A,\psi,\bar\psi}\,.
\label{eq:Sxy-def}
\end{eqnarray}
Since the target field is localized near a light-cone hypersurface, the interaction region is confined to an infinitesimal interval in one light-cone coordinate. The propagator can therefore be represented as free propagation from $y$ to the shock-wave plane, followed by an instantaneous interaction with the shock wave encoded in Wilson lines and, at sub-eikonal order, local operator insertions, and then free propagation from the shock-wave plane to $x$~\cite{Balitsky:1995ub}.
To make this structure explicit, we employ the Schwinger notation for transverse coordinates and momenta,
\begin{eqnarray}
\langle x_\perp|\,\hat{\mathcal O}\,|y_\perp\rangle\,,
\qquad
\hat p_\perp^2 \equiv -\partial_\perp^2\,,
\end{eqnarray}
and rewrite the propagator in a form where the free transverse evolution operators appear on the left and on the right of an operator insertion localized on the shock wave. In the kinematics relevant for high-energy scattering, this yields representations of the schematic form
\begin{eqnarray}
S(x,y) =
\theta(x^{+})\,\theta(-y^{+})\int d^4 z\;\delta(z^{+})\;
\langle x|\,S_0\,|z\rangle\;\mathcal{W}(z_\perp)\;\langle z|\,S_0\,|y\rangle
\;+\;\ldots\,,
\label{eq:S-factorized}
\end{eqnarray}
where $S_0$ is the free quark propagator operator, while $\mathcal{W}(z_\perp)$ represents the Wilson-line structures which can be at eikonal or sub-eikonal level.
The ellipsis stands for other possible time orderings (e.g.\ both points on the same side of the shock wave) and for terms that are beyond the accuracy 
we are considering.

The step functions in \eqref{eq:S-factorized} are characteristic of the shock-wave formalism: the background field effectively ``cuts'' spacetime into two 
half-spaces along a light-cone direction (here $x^{+}$). As a consequence, the LSZ reduction formula, in addition to the usual amputation by the free 
inverse propagator acting on the external legs, has to take into account the fact that derivatives with respect to $x^{+}$ 
acting on time-ordered expressions generate contact terms due to derivatives of $\theta(x^{+})$, producing $\delta(x^{+})$ contributions. 
The structure of the propagator \eqref{eq:S-factorized} is designed to make 
the free propagation before and after the shock wave manifest, so that LSZ amputation can be implemented.

The quark propagator we will use for the calculation of the polarized and unpolarized structure functions is
made of the eikonal part plus the sub-eikonal corrections. The quark propagator with sub-eikonal corrections we are going to use was derived in
ref.~\cite{Chirilli:2018kkw, Chirilli:2021euj}. 

The quark propagator can be written as a sum of different contributions.
Our goal is to put the propagator in the following form
\begin{eqnarray}
\langle {\rm T}\{\psi(x)\barpsi(y)\}\rangle_{A,\psi,\barpsi} \sim
\int d^4z \delta(z^+)\brax{i\hat{\ssp}\over p^2+i\epsilon}\ketz\widehat{\calw}(z_\perp)\braz{i\hat{\ssp}\over p^2+i\epsilon}\kety
\label{sampleprop}
\end{eqnarray}

The advantage of the representation \eqref{eq:S-factorized} is that it isolates the part of
the propagator which is genuinely affected by the background field. In particular, all the
dependence on the shock wave is contained in the operator insertion $\widehat{\calw}(z_\perp)$,
while the factors to its left and to its right are ordinary free propagators. This is precisely
the form that will be needed in the next section when applying the LSZ reduction formula
to external quark legs.

In eq.~\eqref{sampleprop}, the operator $\hat W$ collects the effects of the quark
propagating in the external field, and therefore it may represent either the eikonal or the
sub-eikonal interaction with the shock wave.

\subsection{Eikonal contribution}

The quark propagator at eikonal level can be written in the following form
\begin{eqnarray}
\langle{\rm T}\{\psi(x)\barpsi(y)\}\rangle^{\rm eik} \equiv\!\!\!&&
\left[\int_0^{+\infty}\!\!{\dhd p^+\over 4(p^+)^2}\theta(x^+-y^+) - 
\int_{-\infty}^0\!\!{\dhd p^+\over 4(p^+)^2}\theta(y^+-x^+) \right]e^{-ip^+(x^- - y^-)}
\nonumber\\
&&\hspace{-0.2cm}
\times\braxp\, e^{-i{\hatp^2_\perp\over 2p^+}x^+}\hat{\ssp}\ssn_2[x^+,y^+]\hat{\ssp}e^{i{\hatp^2_\perp\over 2p^+}y^+}\ketyp
\nonumber\\
&&\!\!\!=\int d^4z\delta(z^+)\brax{i\,\hat{\ssp}\over p^2+i\epsilon}\ketz\, \ssn_2 
[x^+,y^+]_z
\nonumber\\
&&\times
\braz{i\,\hat{\ssp}\over p^2+i\epsilon}\kety\Big(\theta(x^+)\theta(-y^+) - \theta(-x^+)\theta(y^+)\Big)
\label{propeiko-1}
\end{eqnarray}
This representation makes explicit the case in which the quark crosses the shock wave, so
that the interaction with the target is localized at the plane $z^+=0$. In the physical
situation relevant for the dipole picture, the quark starts outside the shock wave, crosses
it, and ends again outside it. In that case, since the field outside the shock wave is a pure
gauge, the finite gauge link can be extended to an infinite Wilson line.
Thus the propagator (\ref{propeiko-1}) becomes
\begin{eqnarray}
\langle{\rm T}\{\psi(x)\barpsi(y)\}\rangle^{\rm eik-sw}
&&\!\!\!\equiv\int d^4z\delta(z^+)\brax{i\,\hat{\ssp}\over p^2+i\epsilon}\ketz\, \ssn_2 
\nonumber\\
&&\times\Big(
U_z\theta(x^+)\theta(-y^+) - U^\dagger_z\theta(-x^+)\theta(y^+)\Big)
\braz{i\,\hat{\ssp}\over p^2+i\epsilon}\kety
\label{propeiko-2}
\end{eqnarray}
Notice the different superscripts in (\ref{propeiko-2}) and (\ref{propeiko-1}).
In eq. (\ref{propeiko-2}), we are using the usual notation for infinite Wilson line in the fundamental representation
\begin{eqnarray}
U(x_\perp) = U_x =\!\!\!&& {\rm P}\exp\left\{ig\int dx^+A^-(x^+n_1 + x_\perp)\right\}
\nonumber\\
=\!\!\!&& [\infty n_1+x_\perp, -\infty n_1 + x_\perp] 
\nonumber\\
=\!\!\!&& [\infty n_1,-\infty n_1]_x
\end{eqnarray}
In the adjoint representation we will use the same notation but we will write explicitly the color indexes that run from 1 to $N^2_c-1$. 
So, in the adjoint representation we have
\begin{eqnarray}
U^{ab}(x_\perp) = U^{ab}_x =\!\!\!&& [\infty n_1+x_\perp, -\infty n_1 + x_\perp]^{ab}
\nonumber\\
=\!\!\!&& [\infty n_1,-\infty n_1]^{ab}_x\,.
\end{eqnarray}
For the finite light-cone gauge link we have 
\begin{eqnarray}
[x^+,y^+]_x = {\rm P}\exp\left\{ig\int_{y^+}^{x^+} dz^+A^-(z^+n_1 + x_\perp)\right\}
\end{eqnarray}
Further notations on the gauge link used throughout this work are presented in Appendix \ref{sec:notation-appendix}.

\subsection{Sub-eikonal corrections}

The sub-eikonal corrections have different sources. In the background of gluon fields, they
arise from operator insertions involving the field strength and transverse covariant
derivatives acting on the gauge links. In addition, there are contributions in which the
background contains quark fields. Since these different structures play different roles in the
DIS cross section, we discuss them separately.

The quark propagator we derived in ref.~\cite{Chirilli:2018kkw}, in the background of gluon field only is
\begin{eqnarray}
\brax{i\over \Sp+i\epsilon}\kety
=\!\!\!&& \left[\int_0^{+\infty}\!\!{\dhd p^+\over 4(p^+)^2}\theta(x^+ - y^+) - 
\int_{-\infty}^0\!\!{\dhd p^+\over 4(p^+)^2}\theta(y^+-x^+) \right] e^{-ip^+(x^- - y^-)}
\nonumber\\
&&\times \braxp\,e^{-i{\hatp^2_\perp\over 2p^+}x^+}\Bigg\{
\hat{\ssp}\ssn_2[x^+,y^+]\hat{\ssp}
+ \hat{\ssp}\ssn_2\,\hat{\mathcal{O}}_1(x^+,y^+;p_\perp)\,\hat{\ssp}
\nonumber\\
&&
+ \hat{\ssp}\ssn_2 \half\hat{\mathcal{O}}_2(x^+,y^+;p_\perp)
-  \half\hat{\mathcal{O}}_2(x^+,y^+;p_\perp)\ssn_2\hat{\ssp} \Bigg\}e^{i{\hatp^2_\perp\over 2p^+}y^+}\ketyp
\nonumber\\
&&+ O(\lambda^{-2})\,.
\label{quarksubnoedge3a}
\end{eqnarray}
where $\Sp = \ssp+g\Sa$, and the operators $\widehat{\calo}_1$, and $\widehat{\calo}_2$, appearing in eq. (\ref{quarksubnoedge3a}), and defined 
in eqs. \eqref{O1}, and \eqref{O2}, respectively, encode the sub-eikonal corrections. 
In Appendix~\ref{sec:O2}, we show that the contribution of operator $\widehat{\calo_2}$ does not contribute to the quark propagator
with the end-points outside the shock-wave.
So, the quark propagator at sub-eikonal level with only gluons in the background we will use in this work is
\begin{eqnarray}
&&\langle {\rm T}\{\psi(x)\barpsi(y)\}\rangle_A
\nonumber\\
=\!\!\!&& \left[\int_0^{+\infty}\!\!{\dhd p^+\over 4(p^+)^2}\theta(x^+-y^+) - 
\int_{-\infty}^0\!\!{\dhd p^+\over 4(p^+)^2}\theta(y^+-x^+) \right] e^{-ip^+(x^- - y^-)}
\nonumber\\
&&\times \braxp\,e^{-i{\hatp^2_\perp\over 2p^+}x^+}\Bigg\{
\hat{\ssp}\,\ssn_2[x^+,y^+]\hat{\ssp}
+ {ig\over 2p^+}\int^{x^+}_{y^+}\!\!\!d\omega^+\,\hat{\ssp}\,\ssn_2\bigg(
[x^+,\omega^+]\half \sigma^{ij}F_{ij}[\omega^+,y^+]
\nonumber\\
&&~~+ \big\{\hat{p}^i,[x^+,\omega^+]\,\omega^+\, {F_i}^{\;-}(\omega^+)\,[\omega^+,y^+]\big\}
\nonumber\\
&&~~+ g\!\!\int^{x^+}_{\omega^+}\!\!\!d\omega'^+\big(\omega^+ - \omega'^+\big)
[x^+,\omega'^+]F^{i-}[\omega'^+,\omega^+]\,{F_i}^{\;-}\,[\omega^+,y^+]\bigg)\,\hat{\ssp}\Bigg\}e^{i{\hatp^2_\perp\over 2p^+}y^+}\ketyp
\label{O1prop}
\end{eqnarray}

For related analyses of next-to-eikonal quark propagators and scattering amplitudes in the
CGC framework, see refs.~\cite{Altinoluk:2020oyd,Altinoluk:2021lvu}.

In the next section we will provide a representation of the quark propagator, with the $\widehat{\calo}_1$ operator, in a form suitable for the application of the LSZ reduction formula.

\subsubsection{Gluon field in the background}

The propagator with the gluon field in the background, \eqref{O1prop}, has three different types of operators that contribute. 
We have the $F_{ij}$ operator
\begin{eqnarray}
&&\langle{\rm T}\{\psi(x)\bar{\psi}(y)\}\rangle^{F_{ij}}
\nonumber\\
&&\equiv\Big[\int_0^{+\infty}\!{\dhd p^+\over 8(p^+)^3}\theta(x^+-y^+) 
- \int^0_{-\infty}\!{\dhd p^+\over 8(p^+)^3}\theta(y^+-x^+)\Big]\,e^{-ip^+(x^- - y^-)}
\nonumber\\
&&\times\!\!\int d^2z\,\braxp \hat{\ssp} \,e^{-i{\hatp^2_\perp\over 2p^+}x^+}\ketzp
\nonumber\\
&&\times ig\!\!\int_{y^+}^{x^+}\!dz^+\,\ssn_2[x^+,z^+]_z
\half \sigma^{ij}F_{ij}(z^+,z_\perp)[z^+,y^+]_z\brazp\hat{\ssp}\,e^{i{\hatp^2_\perp\over 2p^+}y^+}\ketyp
\nonumber\\
&& =  {1\over 4}\!\!\int d^4z\delta(z^+)\brax  {i\,\hat{\ssp}\over p^+( p^2+i\epsilon)}\ketz\, \ssn_2 
\gamma^5 \Big(ig\int_{y^+}^{x^+}\!\!dz^+[x^+,z^+]_z
\,\epsilon^{ij}F_{ij}\big(z^+,z_{\perp}\big)[z^+,y^+]_z\Big)
\nonumber\\
&&\times
\braz{i\,\hat{\ssp}\over p^2+i\epsilon}\kety\Big(\theta(x^+)\theta(-y^+) - \theta(-x^+)\theta(y^+)\Big)
\label{qpropFijterm_a}
\end{eqnarray}
where we used $\ssn_2\sigma^{ij}F_{ij} = \ssn_2\gamma^5\epsilon^{ij}F_{ij}$, with $\gamma^5 = i\gamma^0\gamma^1\gamma^2\gamma^3$,
and $\sigma^{ij} = {i\over 2}(\gamma^i\gamma^j - \gamma^j\gamma^i)$.

As done in the previous case, the gauge field is a pure gauge outside the shock-wave~\cite{Chirilli:2018kkw}, 
so, in the gauge link we may extend the limit of integration to $+\infty$ and $-\infty$, thus, from (\ref{qpropFijterm_a}) we obtain
\begin{eqnarray}
&&\hspace{-1.7cm}\langle{\rm T}\{\psi(x)\bar{\psi}(y)\}\rangle^{F_{ij}\rm sw}
 ={1\over 2s}\!\int d^4z\delta(z^+)\brax {i\,\hat{\ssp}\over p^+(p^2+i\epsilon)}\ketz\, \ssn_2 \gamma^5 
\nonumber\\
&&\hspace{2cm}\times\Big(\calf(z_\perp)\theta(x^+)\theta(-y^+) - \calf^\dagger(z_\perp) \theta(-x^+)\theta(y^+)\Big)
\braz{i\,\hat{\ssp}\over p^2+i\epsilon}\kety
\label{qpropFijterm_b}
\end{eqnarray}
with $\epsilon^{ij}F_{ij} = 2F_{12}$, and where we have defined~\cite{Chirilli:2021lif}
\begin{eqnarray}
&&\calf_z = \calf(z_\perp) \equiv ig{s\over 2}\int_{-\infty}^{+\infty}\!\!dz^+[\infty n_1,z^+]_z
\,\epsilon^{ij}F_{ij}\big(z^+,z_\perp\big)[z^+,-\infty n_1]_z\,.
\label{calfoperator_a}
\end{eqnarray}
We observe that the factor of ${1\over p^+}$  in the eq. (\ref{qpropFijterm_b})
 can be included either in the free propagator to the left or to the right on the shock-wave (Wilson line)
because the classical fields do not depend on the $x^-$ component. This will be convenient when we apply the LSZ reduction formula because its
application on the free propagator is easier.

The second contribution to sub-eikonal corrections with gluon field in the background is
(recall that $\{p_i, [x^+,y^+]_x\} = p_i[x^+,y^+]_x + [x^+,y^+]_xp_i$)
\begin{eqnarray}
&&\hspace{-1cm}\langle{\rm T}\{\psi(x)\bar{\psi}(y)\}\rangle^{F_i}
\nonumber\\
&&\hspace{-1cm}\equiv  ig\!\!\int_{y^+}^{x^+}d\omega^+
\left[\int_0^{+\infty}\!{\dhd p^+\over 8(p^+)^3}\theta(x^+-y^+) - \int^0_{-\infty}\!{\dhd p^+\over 8(p^+)^3}\theta(y^+-x^+)\right]
e^{-ip^+(x^- - y^-)}
\nonumber\\
&&\hspace{-0.5cm}\times
\braxp \ssp \,e^{-i{\hatp^2_\perp\over 2p^+}x^+}\,\ssn_2\Big\{\hatp^i\,,[x^+, \omega^+]
\omega^+ {F_i}^{\;-}(\omega^+)[\omega^+,y^+]\Big\}
\ssp\,e^{i{\hatp^2\over 2p^+}y^+}\ketyp
\label{Fi}
\end{eqnarray}
Extending the gauge link to infinity, as done above, from (\ref{Fi}) we have
\begin{eqnarray}
&&\hspace{-2cm}\langle{\rm T}\{\psi(x)\bar{\psi}(y)\}\rangle^{F_i\,{\rm sw}}
\nonumber\\
&&\hspace{-2cm}= {1\over s}\!\!\int d^4z\delta(z^+)\brax {i\,\hat{\ssp}\,\hatp^i\over p^+(p^2+i\epsilon)}\ketz\, 
\ssn_2\Big(\calf_{iz}\theta(x^+)\theta(-y^+) - \calf^\dagger_{iz}\theta(-x^+)\theta(y^+)\Big)
\braz{i\,\hat{\ssp}\over p^2+i\epsilon}\kety
\nonumber\\
&&\hspace{-2cm}
+ {1\over s}\!\!\int d^4z\delta(z^+)\brax {i\,\hat{\ssp}\over p^+(p^2+i\epsilon)}\ketz\, 
\ssn_2\Big(\calf_{iz}\theta(x^+)\theta(-y^+) - \calf^\dagger_{iz}\theta(-x^+)\theta(y^+)\Big)
\braz{i\,\ssp\,\hatp^i\over p^2+i\epsilon}\kety
\label{Fisw}
\end{eqnarray}
where we define
\begin{eqnarray}
\calf_{iz} = \calf_i(z_\perp) 
\equiv ig {s\over 2}\!\!\int_{-\infty}^{+\infty}d\omega^+[\infty n_1, \omega^+]_z\omega^+ {F_i}^{\;-}(\omega^+,z_\perp)[\omega^+, -\infty n_1]_z
\label{calFi}
\end{eqnarray}
with $\calf^\dagger_i(z_\perp)$ its adjoint conjugate.

The third contribution to the sub-eikonal corrections with gluon field in the background is
\begin{eqnarray}
&&\hspace{-0.5cm}\langle{\rm T}\{\psi(x)\bar{\psi}(y)\}\rangle^{F^2}
\nonumber\\
\equiv&&  ig^2\!\!\int_{y^+}^{x^+}\!\!d\omega^+\!\!\int_{\omega^+}^{x^+}\!\! d\omega'^+(\omega^+ - \omega'^+)
\left[\int_0^{+\infty}\!{\dhd p^+\over 8(p^+)^3}\theta(x^+-y^+) \right.
\nonumber\\
&&\hspace{-0.5cm}\left. - \int^0_{-\infty}\!{\dhd p^+\over 8(p^+)^3}\theta(y^+ - x^+)\right]
e^{-ip^+(x^- - y^-)}\int d^2z
\braxp \ssp \,\ssn_2\,e^{-i{\hatp^2_\perp\over 2p^+}x^+}\ketzp 
\nonumber\\
&&\hspace{-0.5cm}\times [x^+,\omega'^+]_zF^{i-}(\omega'^+,z_\perp)
[\omega'^+,\omega^+]_z{F_i}^{\;-}(\omega^+,z_\perp)[\omega^+,y^+]_z
\brazp\ssp\,e^{i{\hatp^2\over 2p^+}y^+}\ketyp
\label{Fsquared}
\end{eqnarray} 
Let us perform the shock-wave limit thus extending the gauge links to infinity and define
\begin{eqnarray}
\calg_2(z_\perp) 
\equiv\!\!\!&& ig^2{s\over 2}\!\!\int_{-\infty}^{+\infty}\!\!d\omega^+\!\!\int_{\omega^+}^{+\infty}\!\! d\omega'^+(\omega^+-\omega'^+)
\nonumber\\
&&\times[\infty n_1,\omega'^+]_zF^{i-}(\omega'^+,z_\perp)
[\omega'^+,\omega^+]_z{F_i}^{\;-}(\omega^+,z_\perp)[\omega^+,-\infty n_1]_z
\label{F2power2}
\end{eqnarray}
where we will often use the shorthand notation $\calg_{2z} =\calg_2(z_\perp)$.
We can also define the gluon distribution
\begin{eqnarray}
&&G(z_\perp) \equiv \Tr\{\calg_{2z}U_z ^\dagger\}
\nonumber\\ 
=\!\!\!&& i\,g^2\,{s\over 2}\!\!\int_{-\infty}^{+\infty}\!\!d\omega^+\!\!\int_{\omega^+}^{+\infty}\!\! d\omega'^+(\omega^+-\omega'^+)
\nonumber\\
&&\times\Tr\big\{[\infty n_1,\omega'^+]_zF^{i-}(\omega'^+,z_\perp)
[\omega'^+,\omega^+]_z{F_i}^{\;-}(\omega^+,z_\perp)[\omega^+,-\infty n_1]_zU^\dagger_z\big\}
\nonumber\\ 
=\!\!\!&& i\,g^2\,{s\over 4}\!\!\int_{-\infty}^{+\infty}\!\!d\omega^+\!\!\int_{\omega^+}^{+\infty}\!\! d\omega'^+(\omega'^+-\omega^+)
{F_i^a}^{,-}\,(\omega'^+,z_\perp)[\omega'^+,\omega^+]_z^{ab}{F_i^b}^{,-}(\omega^+,z_\perp)
\label{G}
\end{eqnarray}
where $a,b$ are color indexes in the adjoint representation.

With definition (\ref{F2power2}), the sub-eikonal correction (\ref{Fsquared}) becomes
\begin{eqnarray}
\hspace{-2cm}\langle{\rm T}\{\psi(x)\bar{\psi}(y)\}\rangle^{F^2sw}
=\!\!\!&& {1\over s}\!\!\int d^4z\delta(z^+)\brax {i\,\hat{\ssp}\over p^+( p^2+i\epsilon)}\ketz
\nonumber\\
&&
\times\ssn_2\Big(\calg_2(z_\perp)\theta(x^+)\theta(-y^+) - \calg^\dagger_2(z_\perp)\theta(-x^+)\theta(y^+)\Big)
\braz {i\,\hat{\ssp}\over p^2+i\epsilon}\kety
\label{Fsquared-sw}
\end{eqnarray}

For later use, it is convenient to combine the second and third sub-eikonal corrections,
eqs.~\eqref{Fsquared-sw} and \eqref{Fisw}, in such a way that the operator $\hat p_i$
appears only on one side of the shock wave, either to the left or to the right. Indeed,
both corrections contain one term in which $\hat p_i$ multiplies the free propagator to the
left of the shock wave and another term in which $\hat p_i$ multiplies the free propagator
to the right of the shock wave, the shock wave being located at $z^+=0$. It is therefore
convenient to rewrite their sum in a form in which $\hat p_i$ appears only on the left or
only on the right.

To this end, we recall that when $\hat p_i$ is evaluated at the edge of the shock wave,
\textit{i.e.} outside the support of the background field, so, it can be promoted to the covariant
momentum operator $\hat P_i$, since we are working in a gauge in which the transverse
component of the gauge field vanishes at that point~\cite{Chirilli:2018kkw,Chirilli:2021lif}.
In this way, the sum of the two sub-eikonal corrections can be rewritten as (see Appendix~\ref{app:pushingP})
\begin{eqnarray}
&&\langle{\rm T}\{\psi(x)\bar{\psi}(y)\}\rangle^{F_i\,{\rm sw}}
+
\langle{\rm T}\{\psi(x)\bar{\psi}(y)\}\rangle^{F^2\,{\rm sw}}
\nonumber\\
&&=
\langle{\rm T}\{\psi(x)\bar{\psi}(y)\}\rangle^{P_{\rm right}}
=
\langle{\rm T}\{\psi(x)\bar{\psi}(y)\}\rangle^{P_{\rm left}}\,.
\end{eqnarray}
 where
 \begin{eqnarray}
 &&\langle{\rm T}\{\psi(x)\bar{\psi}(y)\}\rangle^{P_{\rm right}}
\nonumber\\
&& = {1\over s}\!\!\int d^4z\delta(z^+)\Bigg\{\brax {i\,\hat{\ssp}\over p^2+i\epsilon}\ketz \ssn_2\Bigg[
 \calf_i(z_\perp)\braz{2i\,\hat{\ssp}\,\hatp^i\over p^+(p^2+i\epsilon)}\kety 
\nonumber\\
&&\hspace{4cm}+ \Big(\calf'(z_\perp) - \calf_{2'}(z_\perp)\Big)\braz{i\,\hat{\ssp}\,\over p^+(p^2+i\epsilon)}\kety
 \Bigg]\theta(x^+)\theta(-y^+)
 \nonumber\\
 &&- \brax {i\,\hat{\ssp}\over p^2+i\epsilon}\ketz \ssn_2\Bigg[
 \calf^\dagger_i(z_\perp)\braz{2i\,\hat{\ssp}\,\hatp^i\over p^+(p^2+i\epsilon)}\kety 
 \nonumber\\
 &&\hspace{3cm} + \Big(\calf^\dagger_2(z_\perp) - \calf'^\dagger(z_\perp)
\Big)\braz{i\,\hat{\ssp}\over p^+(p^2+i\epsilon)}\kety
 \Bigg]\theta(-x^+)\theta(y^+)
 \Bigg\}
 \label{PropP2right}
 \end{eqnarray}
and
\begin{eqnarray}
&&\langle{\rm T}\{\psi(x)\bar{\psi}(y)\}\rangle^{P_{\rm left}}
\nonumber\\
&& = {1\over s}\!\!\int d^4z\delta(z^+)\Bigg\{\Bigg[\brax{2i\,\hat{\ssp}\,\hat{p}^i\over p^+(p^2+i\epsilon)}\ketz\calf_i(z_\perp)
\nonumber\\
&&\hspace{3cm} + \brax {i\,\hat{\ssp}\over p^+(p^2+i\epsilon)}\ketz\Big(\calf_2(z_\perp) - \calf'(z_\perp)\Big)\Bigg]
\ssn_2\braz{i\,\hat{\ssp}\over p^2+i\epsilon}\kety\theta(x^+)\theta(-y^+)
\nonumber\\
&& - \Bigg[\brax{2i\,\hat{\ssp}\,\hat{p}^i\over p^+(p^2+i\epsilon)}\ketz\calf^\dagger_i(z_\perp) 
\nonumber\\
&&\hspace{1cm} + \brax {i\,\hat{\ssp}\over p^+(p^2+i\epsilon)}\ketz
\Big(\calf'^\dagger(z_\perp) - \calf^\dagger_{2'}(z_\perp)\Big)
\Bigg]\ssn_2\braz{i\,\hat{\ssp}\over p^2+i\epsilon}\kety\theta(-x^+)\theta(y^+)\Bigg\}
\label{PropP2left}
\end{eqnarray}
and where we defined
\begin{eqnarray}
&&\hspace{-1.3cm}\calf'_z = 
\calf'(z_\perp) \equiv ig{s\over 2}\!\!\int_{-\infty}^{+\infty}d\omega^+[\infty n_1, \omega^+]_z\omega^+ 
iD^i{F_i}^{\;-}(\omega^+,z_\perp)[\omega^+, -\infty n_1]_z
\label{Fprime}
\\
&&\hspace{-1.3cm}\calf_{2'z}
= \calf_{2'}(z_\perp) \equiv  isg^2\!\!\int_{-\infty}^{+\infty}\!\! d\omega^+\!\!\int_{-\infty}^{\omega^+}\! dz^+
[\infty n_1,\omega^+]_z\,\omega^+ {F_i}^{\;-} [\omega^+,z^+]_z F^{i-}[z^+,-\infty n_1]_z
\label{F2prime}
\\
&&\hspace{-1.3cm}\calf_{2z}
= \calf_2(z_\perp) \equiv isg^2\!\!\int_{-\infty}^{+\infty}\!\! d\omega^+\!\!\int_{-\infty}^{\omega^+}\! dz^+
[\infty n_1,\omega^+]_z{F_i}^{\;-} [\omega^+,z^+]_z \,z^+ F^{i-}[z^+,-\infty n_1]_z
\label{F2}
\end{eqnarray}
and the $\calf'^\dagger_z$, $\calf^\dagger_{2'z}$,  $\calf^\dagger_{2z}$,
are obtained by taking the adjoint conjugation of \eqref{Fprime},
\eqref{F2prime}, and \eqref{F2}, respectively.

The operator built from $F_{ij}$ will be particularly important for helicity-dependent
observables, while the operators involving ${F_i}^{\;-}$ and their composite combinations
contribute to the remaining sub-eikonal corrections to the dipole amplitude.

\subsection{Quark propagator with $\psi$ and $\barpsi$ in the background field}

\begin{figure}[t]
	\begin{center}
		\includegraphics[width=2.8in]{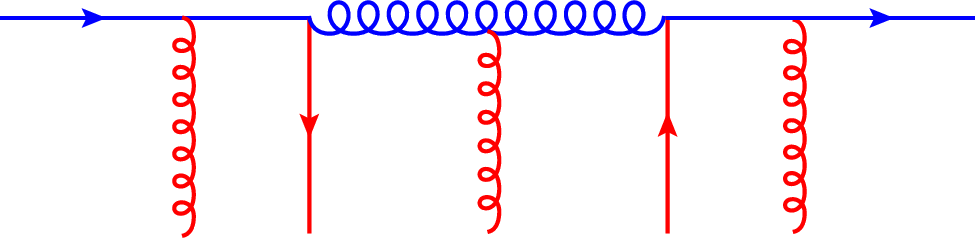}
		\caption{In the picture is shown a typical diagram contributing to the quark propagator in the background of quark fields. 
			As usual, we indicate in blue the quantum field while in red the background one.}
		\label{quarkprop-inq}
	\end{center}
\end{figure}
We now consider the contribution in which the background field contains quark fields, as illustrated in figure \ref{quarkprop-inq}.
This term provides the fermionic sub-eikonal correction to the quark propagator and will
contribute to the dipole cross section at sub-eikonal level.

The quark propagator with quark fields in the background is~\cite{Chirilli:2018kkw}
\begin{eqnarray}
&&\langle{\rm T}\{\psi(x)\bar{\psi}(y)\}\rangle_{\psi,\bar{\psi}}
\nonumber\\
=\!\!\!&& g^2\int_{y^+}^{x^+}\!\!\! dz^+\! \int_{y^+}^{z^+}\!\!\!dz'^+
\left[\int_0^{+\infty}\!{\dhd p^+\over 2p^+}\theta(x^+-y^+)
- \int^0_{-\infty}\!{\dhd p^+\over 2p^+}\theta(y^+-x^+)
\right]
e^{-ip^+(x^- - y^-)}
\nonumber\\
&&
\times { 1\over 16(p^+)^4}
\braxp e^{-i{\hatp^2_\perp\over 2p^+}x^+}
\ssp\ssn_2\ssp\,[x^+ ,z^+] \gamma^\mu t^a\psi(z^+)\left(\delta^\xi_\mu - {n_{2\mu}\over p^+}p^\xi\right)[z^+,z'^+]^{ab}
\nonumber\\
&&
\times\!\left(g_{\xi\nu} - p_\xi {n_{2\nu}\over p^+}\right)
\bar{\psi}(z'^+)t^b\gamma^\nu [z'^+,y^+]\,\ssp 
\ssn_2\ssp \,e^{i{\hatp^2_\perp\over 2p^+}y^+}\ketyp\,.
\label{qprpinquark-1}
\end{eqnarray}
In the propagator (\ref{qprpinquark-1}), we use the approximation (see also the Appendix, section~\ref{sec:notation-appendix})
\begin{eqnarray}
\ssn_1\psi = \gamma^-\psi \sim \lambda \,, ~~~ \gamma_\perp\psi \sim \lambda^{0}\,, 
~~~ \ssn_2\psi = \gamma^+\psi \sim\lambda^{-1}
\label{goodbadcomp}
\end{eqnarray}
to isolate the $O(\lambda^{-1})$ corrections and obtain
\begin{eqnarray}
&&\ssp\ssn_2\ssp\,[\infty n_1,z^+] \gamma^\mu t^a\psi(z^+)\left(\delta^\xi_\mu - {n_{2\mu}\over p^+}p^\xi\right)[z^+,z'^+]^{ab}
\nonumber\\
&&\times\!\left(g_{\xi\nu} - p_\xi {n_{2\nu}\over p^+}\right)
\bar{\psi}(z'^+)t^b\gamma^\nu [z'^+,-\infty n_1]\,\ssp \ssn_2\ssp
\nonumber\\
=\!\!\!&& 4(p^+)^2(p^+\ssn_1+\ssp_\perp)\,[\infty n_1,z^+] \gamma^\mu_\perp t^a\psi(z^+)[z^+,z'^+]^{ab}
\nonumber\\
&&\times\barpsi(z'^+)t^b\gamma^\perp_\mu[z'^+,-\infty n_1]
(p^+\ssn_1+\ssp_\perp) + O(\lambda^{-2})
\end{eqnarray}
We are interested in the case in which the beginning and the end of the propagation are outside the background field, and since
the field outside the shock-wave is a pure gauge, we can extend the gauge link to infinity ~\cite{Balitsky:1995ub, Chirilli:2018kkw}, thus obtaining
\begin{eqnarray}
\hspace{-1cm}\langle{\rm T}\{\psi(x)\bar{\psi}(y)\}\rangle_{\psi,\bar{\psi}} 
= && {1\over s}\int d^4z\delta(z^+)\brax {i\,\ssp\over p^+(p^2+i\epsilon)}\ketz\, 
\nonumber\\
&&\times\gamma^\mu_\perp\Big( Q(z_\perp)\theta(x^+)\theta(-y^+) -  \tilde{Q}(z_\perp)\theta(-x^+)\theta(y^+)\Big)\gamma_\mu^\perp 
\nonumber\\
&&\times\braz{i\,\ssp\over p^2+i\epsilon}\kety + O(\lambda^{-2})
\label{qprpinquark-0}
\end{eqnarray}
where we defined
\begin{eqnarray}
&&\hspace{-1.7cm}Q^{\alpha\beta}_{ij}(x_\perp)
\equiv g^2{s\over 2}\!\int_{-\infty}^{+\infty}\!\!\! dz^+\! \int_{-\infty}^{z^+}\!\!dz'^+
\nonumber\\
&&\hspace{0.2cm}\times\Big([\infty n_1,z^+]_xt^a  \psi^\alpha(z^+,x_\perp)[z^+,z'^+]^{ab}_x
\bar{\psi}^\beta(z'^+,x_\perp) t^b[z'^+,-\infty n_1]_x\Big)_{ij}
\label{Q}
\end{eqnarray}
and
\begin{eqnarray}
&&\hspace{-1.7cm}\tilde{Q}^{\alpha\beta}_{ij}(x_\perp) 
\equiv g^2{s\over 2}\!\int_{-\infty}^{+\infty}\!\!\! dz^+\! \int^{+\infty}_{z^+}\!\!dz'^+
\nonumber\\
&&\hspace{0.2cm} 
\times\Big([-\infty n_1,z^+]_xt^a  \psi^\alpha(z^+,x_\perp)[z^+,z'^+]^{ab}_x
\bar{\psi}^\beta(z'^+,x_\perp) t^b[z'^+,\infty n_1]_x\Big)_{ij}
\label{tildeQ}
\end{eqnarray}
In the definitions (\ref{Q}) and (\ref{tildeQ}) we have $\alpha,\beta$ spinor indexes, $i,j$ color indexes in the fundamental representation, and $a,b$ color 
indexes in the adjoint representation. We also introduce the gauge link in the adjoint representation $[x^+,y^+]^{ab}_x$. 
In subsequent equations we will omit these indexes except the color indexes in the adjoint representation.
We will use the propagator in the eq. (\ref{qprpinquark-0}) to calculate the dipole cross section at sub-eikonal level.

\subsection{Summary of contributions to the quark propagator}

Let us now collect all the terms contributing to the quark propagator. In the background of
quark and gluon fields, using eqs.~\eqref{propeiko-2}, \eqref{qpropFijterm_b}, \eqref{PropP2right}, and
\eqref{qprpinquark-0}, we obtain
\begin{eqnarray}
	\langle {\rm T}\{\psi(x)\bar\psi(y)\}\rangle_{A,\psi,\barpsi}
	&=&
	\langle x|\,{i \ssp\over p^2+i\epsilon}\,|y\rangle\,\theta(x^+ y^+)
	+\langle {\rm T}\{\psi(x)\bar\psi(y)\}\rangle^{{\rm eik-sw}}
	+\langle {\rm T}\{\psi(x)\bar\psi(y)\}\rangle^{F_{ij}\,{\rm sw}}
	\nonumber\\
	&&
	+\langle {\rm T}\{\psi(x)\bar\psi(y)\}\rangle^{P_{\rm right}}
	+\langle {\rm T}\{\psi(x)\bar\psi(y)\}\rangle_{\psi,\bar\psi}
	+ O(\lambda^{-2})\,.
	\label{sumpropa}
\end{eqnarray}
As explained above, the contribution
$\langle {\rm T}\{\psi(x)\bar\psi(y)\}\rangle^{P_{\rm right}}$
can be equivalently replaced by
$\langle {\rm T}\{\psi(x)\bar\psi(y)\}\rangle^{P_{\rm left}}$,
according to the situation at hand. Moreover, we have added the free propagator
proportional to $\theta(x^+ y^+)$, which takes into account the case in which the
propagation starts and ends on the same side of the shock wave.

The representation \eqref{sumpropa} is the form of the propagator that we will use
in the next section to derive the shock-wave Feynman rules through the LSZ reduction
formula.

\section{LSZ reduction formula in the shock-wave formalism}
\label{sec:Feynmanrule}

The Feynman rules in the presence of the shock-wave have been used several times in the
literature. However, they are usually introduced directly, rather than derived from the raw
application of the LSZ reduction formula to the Dirac matrix elements. This point is not
completely trivial in the shock-wave formalism, because the presence of the shock wave
effectively divides space-time into two half-spaces along one light-cone direction.
Therefore, when the LSZ reduction formula acts on the propagator, one does not obtain
immediately the usual four-dimensional delta function.

The essential difference with respect to the standard LSZ procedure is that, in the shock-wave background, the derivative acting on the step functions automatically generates the instantaneous contribution localized at the shock wave, 
and this contribution must be kept together with the usual derivative acting on the plane-wave factor.

\subsection{Propagation outside the shock-wave}

To make this point more explicit, let us first consider the simplest case, namely a quark
whose propagation starts and ends on the same side of the shock wave. In this case only
the free part of the propagator contributes, with both points lying outside the support of the
background field. The direct application of the LSZ reduction formula then gives
\begin{eqnarray}
\lim_{p^2\to 0}\int \! d^4x \, e^{ip\cdot x}\baru(p)i\slashd_x\Big(\theta(x^+)\theta(y^+)\left\langle{\rm T}\{\psi(x)\barpsi(y)\}\right\rangle\Big)
\nonumber\\
\end{eqnarray}
Notice the presence of the two theta-functions that signal that the space time has been halved due to the shock-wave.
In the absence of the theta-function in $x^+$, the result is straightforward. Instead, with the presence of the theta-function we have two terms
\begin{eqnarray}
&&\lim_{p^2\to 0}\int \! d^4x \, e^{ip\cdot x}\baru(p)i\slashd_x\Big(\theta(x^+)\theta(y^+)\left\langle{\rm T}\{\psi(x)\barpsi(y)\}\right\rangle\Big)
\nonumber\\
&&= -\lim_{p^2\to 0}\baru(p)\!\int d^4 x \dhd^4k\,e^{ip\cdot x}
\bigr(\ssn_2\delta(x^+) - i\theta(x^+)\ssk\bigl){\ssk \over k^2+i\epsilon}\,e^{-ik\cdot(x-y)}\theta(y^+)
\label{deriveFeyn1}
\end{eqnarray}
We observe that, while the integration over $x^-$ and $x_\perp$ gives a delta function which fixes the $k^+\to p^+$ and
$k_\perp\to p_\perp$, respectively, the integration over $x^+$, being restricted due to $\theta(x^+)$, forces us to calculate a residue in $k^-$. 
Thus, from eq. (\ref{deriveFeyn1}), we have
\begin{eqnarray}
&&\lim_{p^2\to 0}\int \! d^4x \, e^{ip\cdot x}\baru(p)i\slashd_x\Big(\theta(x^+)\theta(y^+)\left\langle{\rm T}\{\psi(x)\barpsi(y)\}\right\rangle\Big)
\nonumber\\
&&= - \lim_{p^2\to 0}\baru(p)\, e^{ip^+y^- - i (p,y)_\perp}\int \!\!\dhd k^- \,\theta(y^+)\,e^{ik^-y^+}
\nonumber\\
&&~~~~~~\times\left(\ssn_2 - {p^+\ssn_1+k^-\ssn_2+\ssp_\perp\over k^- - p^- - i\epsilon}\right)
{p^+\ssn_1+k^-\ssn_2+\ssp_\perp\over 2p^+k^- - p^2_\perp +i\epsilon}
\label{deriveFeyn2}
\end{eqnarray}
The final step is to take the residue over $k^-$, observing that the extra $k^-$ in the numerator cancels out. So,
distinguishing the different values of $p^+$, from (\ref{deriveFeyn2}), we arrive at
\begin{eqnarray}
&&\lim_{p^2\to 0}\int \! d^4x \, e^{ip\cdot x}\baru(p)i\slashd_x\Big(\theta(x^+)\theta(y^+)\left\langle{\rm T}\{\psi(x)\barpsi(y)\}\right\rangle\Big)
\nonumber\\
&&= - \lim_{p^2\to 0}\baru(p)\, e^{ip^+y^- - i (p,y)_\perp}\int \!\!\dhd k^- \,\theta(y^+)\,e^{ik^-y^+}
\nonumber\\
&&~~~~~~\times\left({-p^+ \ssn_1 - p^-\ssn_2-\ssp_\perp\over k^- - p^- - i\epsilon}\right)
{p^+\ssn_1 + k^-\ssn_2+\ssp_\perp\over 2p^+k^- - p^2_\perp +i\epsilon}
\nonumber\\
&&= i\, \baru(p)\theta(p^+)\theta(y^+)\, e^{ip^+y^- + i{p^2_\perp\over 2p^+}y^+ - i(p,y)_\perp}
\end{eqnarray}
So, at the end, the residue fixes the value of $k^-\to p^-={p^2_\perp\over2p^+}$, but multiplied by $\theta(p^+)$.
In the appendix we provide the Feynman rules for all the other cases which involve the quark in or out, antiquark in or out,
as well as the case with $\theta(-x^+)\theta(-y^+)$.

The important point is that, due to the presence of the step function, the LSZ reduction
formula no longer produces a full four-dimensional delta function. Instead, the integration
over $x^-$ and $x_\perp$ fixes the corresponding momentum components in the usual way,
while the integration over $x^+$ is restricted and therefore turns into a residue
calculation in the conjugate variable $k^-$. This is the basic mechanism behind the
shock-wave Feynman rules.

\subsection{Propagation crossing the shock-wave}

We now consider the case in which the quark starts its propagation before the shock wave,
interacts with it, and ends its propagation outside the shock wave. This is the case which
is directly relevant for the dipole picture of DIS, where the fast quark crosses the target
background and picks up a Wilson line.
We again have two terms coming from differentiating first the theta-function and then the exponential of the free quark propagator
\begin{eqnarray}
&&\hspace{-1.3cm}\lim_{p^2\to 0}\int\! d^4x\,e^{ip\cdot x}\baru(p)i\slashd_x
\langle{\rm T}\{\psi(x)\barpsi(y)\}\rangle^{\rm eik-sw}
\nonumber\\
&&\hspace{-1.3cm}
= \lim_{p^2\to 0}\int\! d^4x\,e^{ip\cdot x}\baru(p)i\slashd_x\int d^4z\delta(z^+)\brax{i\hat{\ssp}\over p^2+i\epsilon}
\ketz \ssn_2
\nonumber\\
&&\times\!\Big(U_z\theta(x^+)\theta(-y^+) - U^\dagger_z\theta(-x^+)\theta(y^+)\Big)\braz{i\hat{\ssp}\over p^2+i\epsilon}\kety
\nonumber\\
&&\hspace{-1.3cm}= - \lim_{p^2\to 0}\int d^4x \, e^{ip\cdot x}\baru(p)\int \dhd^4k d^4z\,\delta(z^+)\,e^{-ik\cdot(x-z)}\Bigg[
\Big(\ssn_2\delta(x^+) - i\ssk\theta(x^+)\Big)U_z\theta(-y^+) 
\nonumber\\
&&~~~~~+ \Big(\ssn_2\delta(x^+) + i\ssk\theta(-x^+)\Big)U^\dagger_z\theta(y^+)\Bigg]
{\ssk\over k^2+i\epsilon}\ssn_2\braz{i\ssp\over p^2+i\epsilon}\kety
\label{qSWderivation1}
\end{eqnarray}
As anticipated above, after differentiation we obtain two terms. One comes from
differentiating the exponential of the free propagator, while the other comes from
differentiating the step function and is proportional to $\ssn_2\,\delta(x^+)$.
The latter is precisely the contribution which, in the light-cone formalism, is usually
identified with the instantaneous interaction and treated as a separate diagram. Here,
instead, it appears automatically from the direct application of the LSZ reduction formula.
For this reason, in the shock-wave formalism it is natural to keep the two contributions
together from the very beginning.
Thus, from eq. (\ref{qSWderivation1}) we arrive at
\begin{eqnarray}
&&\hspace{-1.3cm} - \!\lim_{p^2\to 0}\int d^4x \, e^{ip\cdot x}\baru(p)\int \dhd^4k d^4z\,\delta(z^+)\,e^{-ik\cdot(x-z)}\Bigg[
\Big(\ssn_2\delta(x^+) - i\ssk\theta(x^+)\Big)U_z\theta(-y^+) 
\nonumber\\
&&\hspace{-1.3cm}~~~~~ + \Big(\ssn_2\delta(x^+) + i\ssk\theta(-x^+)\Big)U^\dagger_z\theta(y^+)\Bigg]
{\ssk\ssn_2\over k^2+i\epsilon}\braz{i\ssp\over p^2+i\epsilon}\kety
\nonumber\\
&&\hspace{-1.3cm}
= \lim_{p^2\to 0} \int \!\! d^4z\delta(z^+)\,e^{ip^+z^- -i(p,z)}\!\int\!\!\dhd k^-\,\baru(p)
{\ssp(p^+\ssn_1+\ssp_\perp)\ssn_2\over 2p^+k^- - p^2_\perp+i\epsilon}
\nonumber\\
&&\hspace{-.3cm} \times\!
\Bigg({U_z\theta(-y^+)\over k^- - p^- - i\epsilon} + {U^\dagger_z\theta(y^+)\over k^- - p^- +i\epsilon}\Bigg)
\braz{i\ssp\over p^2+i\epsilon}\kety
\label{qSWderivation2}
\end{eqnarray}
Let us observe again that in eq. (\ref{qSWderivation2}), the effect of the LSZ reduction formula was not that 
of obtaining a full 4-dimensional delta-function,
as in the usual situation (no shock-wave), rather we obtained a delta-function of the light-cone component in $n_1$ direction, 
$\dbar(p^+ - k^+)$, and
a delta-function for the transverse component, $\dbar^{(2)}(p-k)$. In the $n_2$ direction, instead, we will have to calculate a residue.

Taking the residue integrating over $k^-$, from (\ref{qSWderivation2}) we arrive at
\begin{eqnarray}
&&\hspace{-1.3cm}\lim_{p^2\to 0}\int\! d^4x\,e^{ip\cdot x}\baru(p)i\slashd_x
\langle{\rm T}\{\psi(x)\barpsi(y)\}\rangle^{\rm eik-sw}
\nonumber\\
&&\hspace{-1.3cm} = i\lim_{p^2\to 0}\int d^4z\delta(z^+) e^{ip^+ z^- - i(p,z)_{\!\perp}}\,\baru(p)\ssn_2
\Big(\theta(p^+)\theta(-y^+) U_z - \theta(-p^+)\theta(y^+) U^\dagger_z\Big)
\nonumber\\
&&\times\braz{i\hat{\ssp}\over p^2+i\epsilon}\kety
\label{qSWderivation3}
\end{eqnarray}
We can further simplify result (\ref{qSWderivation3}) to finally obtain
\begin{eqnarray}
&&\hspace{-1.3cm}\lim_{p^2\to 0}\int\! d^4x\,e^{ip\cdot x}\baru(p)i\slashd_x
\langle{\rm T}\{\psi(x)\barpsi(y)\}\rangle^{\rm eik-sw}
\nonumber\\
&&\hspace{-1.3cm} = i\lim_{p^2\to 0}\int d^2z \dhd^2 k\, e^{ip^+ y^- - i(p-k,z)_\perp - i(k,y)_\perp + i{k^2_\perp\over 2p^+}y^+}
\nonumber\\
&&~~~~~~\times\baru(p)\Big(\theta(p^+)\theta(-y^+)U_z + \theta(-p^+)\theta(y^+)U^\dagger_z\Big)
{\ssn_2(p^+\ssn_1+\ssk_\perp)\over 2p^+}
\label{FeynRule1a}
\end{eqnarray}
Result (\ref{FeynRule1a}) is one of the Feynman rules we will utilize for the calculation of the dipole cross-section.
This example makes clear the general pattern: once the propagator is written in the form of
Sec.~\ref{sec:quark-subeikcor}, the LSZ reduction formula can be applied directly, and the resulting shock-wave
Feynman rules follow from the interplay between the restricted light-cone support of the
background field and the pole structure of the free propagators.

In the appendix \ref{sec:FeynRulesB}, we collect all the other Feynman rules for a quark (and an anti-quark) 
crossing the shock-wave with free propagation before and after the interaction.

\section{Dipole cross-section in the eikonal approximation}
\label{sec:dipolecrossection}

In this section we re-derive the well-known dipole cross-section for longitudinal and
transverse photon polarization in the eikonal approximation. Besides providing a useful
check of normalization and conventions, this calculation allows us to introduce the
mixed-space formalism that we will extend to sub-eikonal accuracy in the subsequent
sections.

\begin{figure}[t]
	\begin{center}
		\includegraphics[width=2.7 in]{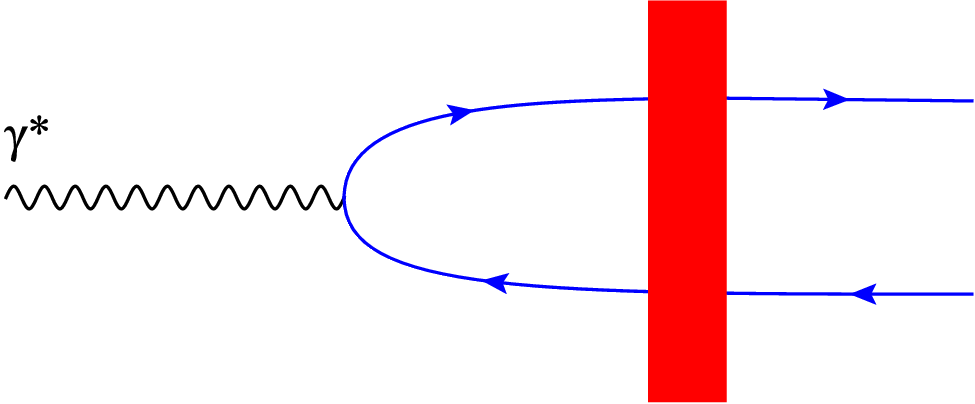}
		\caption{Diagram contributing to the transition amplitude $\gamma^*(q)\to q(p)\barq(k)$ in the eikonal approximation.}
		\label{Fig:dipole}
	\end{center}
	
\end{figure}

We start from the transition amplitude $\gamma^\ast(q)\to q(p)\bar q(k)$ in the shock-wave
background, shown in Fig.~\ref{Fig:dipole}. At eikonal level, the interaction with the target
is encoded in the Wilson lines due to the quark and antiquark crossing the
shock wave. To this end we consider the following matrix element
\begin{eqnarray}
&& \langle q(p)\barq(k)|\gamma^*(q)\rangle_{{\rm Fig.}\ref{Fig:dipole}}
\nonumber\\
=\!\!\!&& iee_f \int d^4z \,\varepsilon_\mu(q)e^{-iq\cdot z} d^4y d^4 x \,e^{ip\cdot x}
\nonumber\\
&&~~~\times\baru(p,\sigma)(i\sslash{\partial}_x)
\langle {\rm T}\psi(x)\barpsi(y)\barpsi(z)\gamma^\mu\psi(z)\rangle_A (-i\overleftarrow{\sslash{\partial}}_y)e^{ik\cdot y}v(k,\sigma')
\nonumber\\
\end{eqnarray}	
We have to apply the Feynman rules obtained in the previous section. In particular, using (\ref{FeynRule1}), and (\ref{FeynRule4}), we arrive at
\begin{eqnarray}
&&\langle q(p)\barq(k)|\gamma^*(q)\rangle_{{\rm Fig.}\ref{Fig:dipole}}	
\nonumber\\
=&& - ee_f{1\over 2}\int d^2z_1 d^2z_2 \dhd^2q_1 e^{-i(p-q_1-q,z_1)+i(k-q_1,z_2)}
	\dbar(p^+ + k^+-q^+){\theta(p^+)\theta(k^+)\over p^+ + k^+}
	\nonumber\\
	&&\times
	{\baru(p,\sigma)\Big(U_{z_1}U^\dagger_{z_2}-1\Big) \ssn_2[p^+\ssn_1+(\ssq_1+\ssq)_\perp]\sslash{\varepsilon}(k^+\ssn_1 - \ssq_{1\perp})\ssn_2v(k,\sigma')
		\over \left[\left(q_{1\perp} + {k^+\over p^++k^+}q_\perp\right)^2 + {k^+p^+\over (p^++k^+)^2}q^2_\perp
		- {2q^-p^+k^+\over p^+ + k^+} - i\epsilon\right]}
\label{LOdipoleA}
\end{eqnarray}
Setting $q_\perp = 0$, the LO Dirac-dipole matrix element is
\begin{eqnarray}
&&\baru(p,\sigma) \ssn_2[p^+\ssn_1+\ssq_{1\perp}]\gamma^\mu(k^+\ssn_1 - \ssq_{1\perp})\ssn_2v(k,\sigma')
\nonumber\\
=\!\!\!&&
\Big( 4n_1^\mu  p^+k^+ - 2 n_2^\mu q_{1\perp}^2 
- 2(p^+ - k^+)q_{1\perp}^\mu \Big) \baru(p,\sigma)\ssn_2 v(k,\sigma') 
\nonumber\\
&& - 2i\,q^+q_{1\nu}\epsilon_\perp^{\mu\nu}\baru(p,\sigma)\,
\gamma^5\ssn_2 v(k,\sigma')
\label{LOdipoleDiracmatrix}
\end{eqnarray}
with $\epsilon_\perp^{\mu\nu}$ the two dimensional antisymmetric tensor such that $\epsilon_\perp^{\mu\nu} = 0$ for $\mu,\nu\ne 1,2$ and
$\epsilon^{12} = -\epsilon^{21} = 1$ (for transverse indexes the symbol $\perp$ is redundant).
Using the result (\ref{LOdipoleDiracmatrix}) in (\ref{LOdipoleA}), we arrive at
\begin{eqnarray}
&&\hspace{-1.4cm}
\langle q(p)\barq(k)|\gamma^*(q)\rangle_{{\rm Fig.}\ref{Fig:dipole}}
\nonumber\\
&&\hspace{-0.3cm} = - {ee_f\over s}\int d^2z_1 d^2z_2 \dhd^2q_1 e^{-i(p-q_1,z_1)+i(k-q_1,z_2)}
\nonumber\\
&&\hspace{-0.3cm}\times 
{\theta(p^+)\theta(k^+)\over q^+}
{\dbar\left({p^+\over q^+} + {k^+\over q^+} - 1 \right)\over \left[q _{1\perp}^2 + {2\over s}Q^2 p^+k^+ - i\epsilon\right]}
 \baru(p,\sigma)\Big(U_{z_1}U^\dagger_{z_2}-1\Big)
\nonumber\\
&&\hspace{-0.3cm}\times\Bigg[
\varepsilon_\mu\Big(4n_1^\mu  p^+k^+ - 2 n_2^\mu q_{1\perp}^2 
- 2(p^+-k^+)q_{1\perp}^\mu \Big) q^+\ssn_2
- is\,\varepsilon_\mu q_{1\nu}\epsilon_\perp^{\mu\nu}\,\gamma^5\ssn_2\Bigg]\! v(k,\sigma')
\label{LOmatrixB}
\end{eqnarray}
To obtain the dipole cross section, we have to square the matrix element (\ref{LOmatrixB}), sum over the helicity $\sigma,\sigma'$ neglecting
quark masses, and sum over the flavor $f$, thus arriving at
\begin{eqnarray}
\calm_{\rm Eikonal} &&
= {1\over 2\pi\delta(0)}\int\!\dhd^4k\dhd^4p\,\dbar(k^2)\dbar(p^2) 
\theta(p^+)\theta(k^+)
|\langle q(p)\barq(k)|\gamma^*(q)\rangle_A|^2
\nonumber\\
&&= e^2\sum_fe_f^2 {2\over s^2}\int d^2z_1 d^2z_2 \dhd^2q_1 \dhd^2 q_2
\, e^{i(q_1-q_2,z_1-z_2)}
\nonumber\\
&&\hspace{0.5cm}
\times{1\over 2\pi}\int_0^1 dz{\varepsilon_\mu\varepsilon^*_\rho\over [q^2_{1\perp} + Q^2 z\barz ][q^2_{2\perp} + Q^2 z\barz ]}
\nonumber\\
&&\hspace{0.5cm}\times
2N_c\Big\langle1-{1\over N_c}\Tr\{U_{z_1}U^\dagger_{z_2}\}\Big\rangle\Bigg\{s^2\, q_{1\nu}\epsilon_\perp^{\mu\nu} q_{2\alpha}\epsilon_\perp^{\rho\alpha}
\nonumber\\
&& \hspace{0.5cm}
+ \Big(2sq^+n_1^\mu z\barz  - 2q^+ n_2^\mu q_{1\perp}^2  - s (z-\barz)q_{1\perp}^\mu \Big)
\nonumber\\
&&\hspace{0.5cm}\times
\Big(2sq^+n_1^\rho  z\barz - 2 q^+n_2^\rho q_{2\perp}^2 - s(z-\barz)q_{2\perp}^\rho \Big)  \Bigg\}
\label{eikonalsquare1}
\end{eqnarray}
To get eq. \eqref{eikonalsquare1} we have integrated over $\dhd^4k$ and $\dhd^4p$, and made the change of variable 
$z={p^+\over q^+}$ and $1-z\equiv\barz = {k^+\over q^+}$, and made use of the following Dirac matrices
\begin{eqnarray}
&&\tr\{\ssp\ssn_2\ssk\ssn_2\} =\tr\{\ssp\gamma^5\ssn_2\ssk\gamma^5\ssn_2\} = 8p^+k^+\nonumber\\
&&\tr\{\ssp\ssn_2\ssk\ssn_2\gamma^5\} = 0 
\end{eqnarray}
Moreover, the factor ${1\over 2\pi\delta(0)}$ is the infinite volume normalization factor 
which cancels out one of the $\dbar(p^+/q^+ +k^+/q^+ - 1)$
coming from squaring the scattering amplitude.

Equation \eqref{eikonalsquare1} already has the standard dipole structure: the dependence on the
target is entirely contained in the Wilson-line matrix element, while the remaining factors
are the photon wave function (impact factor) in momentum space. 

In the next two subsections we project
this expression onto longitudinal and transverse photon polarization.

\subsection{Eikonal dipole cross-section with Longitudinal polarization}

First, let us consider the Longitudinal polarization
$\varepsilon^\mu_L  = {q^+\over Q}n_1^\mu + {Q\over 2q^+}n_2^\mu$.
Using
\begin{eqnarray}
&&\varepsilon_{L\mu} \epsilon_\perp^{\mu\nu}q_{1\nu}  = 0
\label{LOlongypola1}
\\
&&\varepsilon_{L\mu} \Big(2sq^+n_1^\mu z\barz  - 2 q^+n_2^\mu q_{1\perp}^2  - s (z-\barz)q_{1\perp}^\mu \Big)
= Qs z\barz - {sq_{1\perp}^2 \over Q}
\label{LOlongypola2}
\end{eqnarray}
from (\ref{eikonalsquare1}) we have
\begin{eqnarray}
\calm^L_{\rm Eikonal} 
=\!\!\!&& {1\over 2\pi\delta(0)}\int\!\dhd^4k\dhd^4p\,\dbar(k^2)\dbar(p^2) 
\theta(p^+)\theta(k^+)
|\langle q(p)\barq(k)|\gamma^*_L(q)\rangle_A|^2
\nonumber\\
=\!\!\!&&{2N_c\over \pi}e^2\sum_fe_f^2 \int d^2z_1 d^2z_2 \dhd^2q_1 \dhd^2 q_2
\, e^{i(q_1-q_2,z_1-z_2)}\nonumber\\
&&\times\int_0^1 dz
{({q_{1\perp}^2\over Q} - Q z\barz)({q_{2\perp}^2\over Q} - Q z\barz)
	\over [q^2_{1\perp} + Q^2 z\barz ][q^2_{2\perp} + Q^2 z\barz ]}
\Big\langle1-{1\over N_c}\Tr\{U_{z_1}U^\dagger_{z_2}\}\Big\rangle
\label{EikonalLongy-a}
\end{eqnarray}
Making use of the unitarity constraint of the Wilson line matrix element $1-{1\over N_c}\Tr\{U_{z_1}U^\dagger_{z_2}\}$, we can re-write result 
(\ref{EikonalLongy-a}) as 
\begin{eqnarray}
&&\hspace{-2.5cm}\calm^L_{\rm Eikonal} =  {8\,e^2\over \pi} N_c\sum_fe_f^2\int d^2z_1 d^2z_2 \dhd^2q_1 \dhd^2 q_2
\, e^{i(q_1-q_2,z_1-z_2)}\nonumber\\
&&\times\int_0^1 dz\,
{Q^2 z^2\barz^2\over [q^2_{1\perp} + Q^2 z\barz][q^2_{2\perp} + Q^2 z\barz ]}\big\langle\calu(z_1,z_2)\big\rangle
\label{M_LOno4rward-1}
\end{eqnarray}
where as usual we defined  
\begin{eqnarray}
\calu(z_1,z_2) \equiv 1-{1\over N_c}\Tr\{U_{z_1}U^\dagger_{z_2}\}
\label{defcalu}
\end{eqnarray}
and we use the shorthand notation $\calu(z_1,z_2)=\calu_{z_1z_2}$.

From (\ref{M_LOno4rward-1}) we have 
\begin{eqnarray}
\calm^L_{\rm Eikonal}
  =\!\!\!&&   {8\,Q^2N_c\alpha_{\rm em}\over\pi^2}\sum_f e^2_f \int d^2z_1 d^2z_2 
\,\int_0^1 dz\, z^2\barz^2\left|K_0(\bar{Q} |z_{12}|)\right|^2
\big\langle\calu(z_1,z_2)\big\rangle
\label{LOdipoleLongy}
\end{eqnarray}
with $z_{12} = z_1-z_2$, $\bar{Q} = \sqrt{Q^2 z\barz}$, and
\begin{eqnarray}
K_0(\bar{Q}|x|) = \int {d^2q\over 2\pi}{e^{i(q, x)}\over \bar{Q}^2 +q^2}
\label{K0}
\end{eqnarray}
the Macdonald function.

Equation \eqref{LOdipoleLongy} is the standard dipole expression for the longitudinal photon
cross-section. As expected, the longitudinal photon wave function is proportional to
$Q\,z\bar z\,K_0(\bar Q |z_{12}|)$, while the interaction with the target is encoded in the
dipole operator $\calu(z_1,z_2)$. This is the form that will later serve as the eikonal part of
the sub-eikonal extension.

\subsection{Eikonal dipole cross-section with Transverse Polarization}

Let us consider transverse polarization $\varepsilon_\lambda^k = - {1\over \sqrt{2}}(\lambda,i)$ with $\lambda = \pm 1$. Using
\begin{eqnarray}
\varepsilon_{\perp\mu} q_{1\nu}\epsilon_\perp^{\mu\nu} = \varepsilon^1q^2-\varepsilon^2q^1 = \vec{\varepsilon}\times \vecq_1
\label{LOtranspola}
\end{eqnarray}
with $\varepsilon_{\perp\mu} = (0,\varepsilon_1,\varepsilon_2,0) = - (0,\varepsilon^1,\varepsilon^2,0)$, from (\ref{eikonalsquare1}), we have

\begin{eqnarray}
\calm^T_{\rm Eikonal}
=\!\!\!&& {1\over 2\pi\delta(0)}\int\!\dhd^4k\dhd^4p\,\dbar(k^2)\dbar(p^2) \theta(p^+)\theta(k^+)
\Big|\langle q(p)\barq(k)|\gamma^*_T(q)\rangle\Big|^2
\nonumber\\
=\!\!\!&& \int\!\dhd^4k\dhd^4p\,\dbar(k^2)\dbar(p^2) \theta(p^+)\theta(k^+)\dbar(p^+ + k^+-q^+)
\half\sum_{\lambda=\pm 1}\sum_{f,\sigma,\sigma'}
\nonumber\\
&&\times\Bigg|(- ee_f) {2\over s} \int d^2z_1 d^2z_2 {\dhd^2 q_1\over q^2_{1\perp}+ {2\over s}Q^2 p^+k^+ }
\,e^{ i(q_1-p,z_1)_\perp + i (k-q_1,z_2)}\,\baru(p,\sigma)\Big( U_{z_1}U^\dagger_{z_2}-1\Big)
\nonumber\\
&&\times\Big((p^+-k^+)(\varepsilon^\lambda, q_1)_\perp\ssn_2 v(k,\sigma') 
- iq^+(\vec{\varepsilon}^\lambda_\perp\times \vec{q}_1)\gamma^5\ssn_2 v(k,\sigma')\Big)
\Bigg|^2
\label{M_T-LO1}
\end{eqnarray}
Neglecting again  the quark masses, we have
\begin{eqnarray}
&&\sum_{\lambda=\pm 1}\sum_{\sigma,\sigma'}\Bigg|\int \dhd^2q_1\Big((p^+-k^+)(\varepsilon_\lambda, q_1)_\perp\baru(p,\sigma)\ssn_2 v(k,\sigma') 
- iq^+(\vec{\varepsilon}_\lambda\times \vec{q}_1)\baru(p,\sigma)\gamma^5\ssn_2 v(k,\sigma')\Big)\Bigg|^2
\nonumber\\
&& = \!\!\sum_{\lambda=\pm 1}2s^2z\barz\int \dhd^2q_1 \dhd^2q_2
\Big[(z-\barz)^2(\varepsilon_\lambda,q_1)_\perp(\varepsilon^{\,*}_\lambda,q_2)_\perp +(\vec{\varepsilon}_\lambda\times\vecq_1)(\vec{\varepsilon}^{\,*}_\lambda\times\vecq_2)\Big]
\nonumber\\
&& = 4s^2z\barz (z^2+\barz^2)\int \dhd^2q_1 \dhd^2q_2\,(q_1,q_2)_\perp
\label{sumTpoly}
\end{eqnarray}
So, using (\ref{sumTpoly}) in (\ref{M_T-LO1}) we arrive at
\begin{eqnarray}
&&\hspace{-1.5cm}\calm^T_{\rm Eikonal}
=  8N_c\alpha_{\rm em}\sum_fe^2_f\int_0^1\!\! dz (z^2+\barz^2) 
\nonumber\\
&&~~~\times\!\!\int \!\! d^2z_1 d^2z_2 \,
{\dhd^2 q_1\dhd^2 q_2e^{i(q_1-q_2,z_1-z_2)}\over [q^2_{1\perp} + Q^2 z\barz][q^2_{2\perp} + Q^2 z\barz]}
(q_1,q_2)_\perp\big\langle\calu(z_1,z_2)\big\rangle
\end{eqnarray}
Using the modified Bessel function
\begin{eqnarray}
&&\bar{Q}{i\,x^i\over |x_\perp|} K_1(\bar{Q}|x_\perp|) = \int {d^2 q\over 2\pi} \,{q^i e^{i(q,x)_\perp}\over q^2_\perp + \bar{Q}^2}
\label{K1}
\end{eqnarray}
we finally have
\begin{eqnarray}
&&\hspace{-1.5cm}\calm^T_{\rm Eikonal}
=  {2Q^2N_c\alpha_{\rm em}\over \pi^2}\sum_fe^2_f\int_0^1\!\! dz\, z\barz \,(z^2+\barz^2) 
\int \!\! d^2z_1 d^2z_2 \,
\left|K_1(\bar{Q} |z_{12}|)\right|^2\big\langle\calu(z_1,z_2)\big\rangle
\label{LOdipoleTransver}
\end{eqnarray}
Equation \eqref{LOdipoleTransver} is the standard dipole expression for transverse photon
polarization. In this case the photon wave function is proportional to
$\bar Q\,K_1(\bar Q |z_{12}|)$ and is weighted by the familiar factor $z\bar z (z^2+\bar z^2)$.
Again, all the target dependence is contained in the dipole operator $\calu(z_1,z_2)$.

Notice that the asymmetry contribution to the eikonal dipole scattering amplitude vanishes. Indeed, it is proportional to
 \begin{eqnarray}
 &&\Big[(z-\barz)^2(\varepsilon_+,q_1)_\perp(\varepsilon^{\,*}_+,q_2)_\perp 
 + (\vec{\varepsilon}_+\times\vecq_1)(\vec{\varepsilon}^{\,*}_+\times\vecq_2) \Big]
 \nonumber\\
&& - \Big[(z-\barz)^2(\varepsilon_-,q_1)_\perp(\varepsilon^{\,*}_-,q_2)_\perp 
+ (\vec{\varepsilon}_-\times\vecq_1)(\vec{\varepsilon}^{\,*}_-\times\vecq_2) \Big]
\nonumber\\
&&= 2i(z^2+\barz^2)\vecq_2\times\vecq_1
\label{asymmetrysum1}
\end{eqnarray}
which, after integration over the transverse momenta, gives zero contribution to the cross-section. 
This is one of the motivations for considering sub-eikonal corrections to the
dipole scattering amplitude. In the following sections we will study precisely these first
sub-eikonal contributions.

In strong and electromagnetic interactions, where parity is conserved, 
the hadronic tensor can be expanded in terms of the unpolarized structure functions $F_1$ and $F_2$ and 
the spin-dependent structure functions $g_1$ and $g_2$
\begin{eqnarray}
\hspace{-1cm}W_{\mu\nu}=\!\!&& \left(- g_{\mu\nu} + {q_\mu q_\nu\over q^2}\right)F_1(x_B, Q^2)
+ \left(P_\mu - q_\mu {q\cdot P\over q^2}\right) \left(P_\nu - q_\nu {q\cdot P\over q^2}\right){F_2(x_B, Q^2)\over P\cdot q}
\nonumber\\
\hspace{-1cm}&& + i\,\epsilon_{\mu\nu\lambda\sigma}\,q^\lambda S^\sigma {M\over P\cdot q}\,g_1(x_B,Q^2)
+  i\,\epsilon_{\mu\nu\lambda\sigma}q^\lambda\left(S^\sigma  - P^\sigma {q\cdot S\over q\cdot P}\right){M\over q\cdot P}\,g_2(x_B,Q^2)
\label{Htensor}
\end{eqnarray}
where $S^\mu$ is the spin of the target that satisfies  $S^2 = -1$ and $S\!\cdot\!P=0$.
Therefore, the longitudinal and transverse quantities computed above provide directly the
projections of the hadronic tensor relevant for extracting the corresponding structure
functions. In particular, the vanishing of the eikonal asymmetry is consistent with the fact
that the spin-dependent structure function $g_1$ requires sub-eikonal, and therefore
spin-sensitive, operator insertions.

We can relate the square of the dipole scattering amplitude to the hadronic tensor.
Thus,
\begin{eqnarray}
\varepsilon^\mu(q)\varepsilon^{*\nu}(q)W_{\mu\nu} = {Q^2\over 16\pi^2\alpha_{\rm em}x_B}\,\calm \, .
\label{extractFs}
\end{eqnarray}
where $\calm$ denotes either the longitudinal, or the transverse scattering amplitude square. Thus, we have
\begin{eqnarray}
&&\half\!\sum_{\lambda=\pm 1}\varepsilon^\mu_{\perp\lambda}\varepsilon^{*\nu}_{\perp\lambda} W_{\mu\nu} = F_1(x_B,Q^2)\,
\label{strF1}\\
&&\nonumber\\
&&\varepsilon^\mu_L\varepsilon^{*\nu}_L W_{\mu\nu} = - F_1(x_B,Q^2) + {F_2(x_B,Q^2)\over 2x_B}\,,
\label{strF1F2}
\end{eqnarray}
and in the small-x$_B$ limit we also have
\begin{eqnarray}
\half(\varepsilon_+^\mu\varepsilon^{*\nu}_+ - \varepsilon^\mu_-\varepsilon^{*\nu}_-)W_{\mu\nu} 
= g_1(x_B,Q^2) - g_2{4M^2x^2_B\over Q^2} \simeq  g_1(x_B,Q^2)
\label{extractq1}
\end{eqnarray}

It is customary to define the longitudinal and transverse structure functions as
\begin{eqnarray}
&&F_L(x_B,Q^2) = F_2(x_B,Q^2) - 2x_BF_1(x_B,Q^2)\,,
\label{strFl}
\nonumber\\
&&F_T(x_B,Q^2) = 2x_BF_1(x_B,Q^2)\label{strFt}\,.
\end{eqnarray}
Using the longitudinal and transverse dipole scattering amplitude square, eq. (\ref{LOdipoleLongy}) and eq. (\ref{LOdipoleTransver}), we have, respectively,
\begin{eqnarray}
\hspace{-0.5cm}F_L(x_B,Q^2) = {N_cQ^4\over \pi^4} \sum_fe^2_f \int d^2z_1 d^2z_2 \!\!
\int_0^1 dz\, z^2\barz^2\left|K_0(\bar{Q} |z_{12}|)\right|^2\big\langle\calu(z_1,z_2)\big\rangle
\label{LOstrFL}
\end{eqnarray}
and 
\begin{eqnarray}
\hspace{-1cm}F_T(x_B,Q^2) =  {N_cQ^4\over 4\pi^4}\sum_fe^2_f \int_0^1\!\! dz\, z\barz \,(z^2+\barz^2) \!\!
\int \!\! d^2z_1 d^2z_2 \,
\left|K_1(\bar{Q} |z_{12}|)\right|^2\big\langle \calu(z_1,z_2)\big\rangle
\label{LOstrFT}
\end{eqnarray}

One of the goals of this work is to determine the corrections to $F_L$ and $F_T$ due to sub-eikonal contributions.

The structure functions \eqref{LOstrFL} and \eqref{LOstrFT} are proportional to the matrix
element of the operator $\calu(z_1,z_2)$. Its small-$x$ evolution is governed by the
BK/B-JIMWLK equation~\cite{Balitsky:1995ub,Kovchegov:1999yj,Jalilian-Marian:1997ubg,Jalilian-Marian:1997qno,Iancu:2000hn,Ferreiro:2001qy}.
\begin{eqnarray}
{d\over d\eta}\calu_{xy} = {\alpha_s N_c\over 2\pi^2}
\int d^2z \, {(x-y)^2_\perp\over (x-z)^2_\perp(z-y)^2_\perp}
\Big[\calu_{xz} + \calu_{zy} -\calu_{xy} - \calu_{xz}\calu_{zy}\Big]
\label{BK}
\end{eqnarray}
where $\eta$ is the rapidity parameter. 
To solve the evolution equation (\ref{BK}), one must specify an initial condition for the dipole operator $\calu_{xy}$.
In practice, two standard choices are the McLerran-Venugopalan (MV) model, based on a semiclassical description of a dense target color field, 
and the Golec-Biernat-W\"usthoff (GBW) model, which parametrizes the dipole amplitude in a form that already incorporates saturation
effects~\cite{McLerran:1993ni,McLerran:1993ka,Golec-Biernat:1998js,Golec-Biernat:1999qd}. 

It is useful to recall here the factorization role played by the evolution equation
\eqref{BK}. The lowest-order dipole cross section identifies the Wilson-line operator
which couples the projectile to the target, while the dependence on the rapidity
factorization scale is carried by the matrix element of this operator. At one perturbative
order higher (at NLO), the piece of the coefficient function that is enhanced by the logarithm of energy
is generated by the rapidity-evolution of the operator obtained at lower (LO) order.
After this contribution is absorbed into the rapidity evolution of the Wilson
lines, the remaining higher-order impact factor is not enhanced by the logarithm of the
energy.

The same organization is expected at sub-eikonal accuracy. For the
axial quark sector this is shown explicitly in section \ref{sec:divergence}. The sub-eikonal dipole cross
section determines an enlarged operator basis, and the energy dependence of its matrix
elements is governed by the corresponding rapidity evolution equations. At the next
perturbative order, the part proportional to the high-energy logarithm must therefore be
consistent with the one-loop evolution of this enlarged operator basis. The remainder is
a genuine higher-order coefficient function which is not multiplied by the high-energy
logarithm. Notice, however, that being free of the high-energy logarithm does not by
itself imply that this remainder is free of ordinary UV or collinear singularities.

The sub-eikonal expansion introduces additional quark and
gluon operators into the scattering amplitudes. Their
contributions to an inclusive observable must be determined
after combining the relevant interference terms and final
states. As shown below, within the approximation in which the $x_B$ dependence of the sub-eikonal operators is neglected, 
the calculated first sub-eikonal corrections cancel in $F_L$ and $F_T$, whereas the helicity-dependent 
projection retains gluonic and axial quark contributions. The following sections identify
these surviving operators and their coefficient functions.

\section{Dipole cross-section at sub-eikonal level}
\label{sec:dipolewithsbeikonal}

\begin{figure}[t]
	\begin{center}
		\includegraphics[width=5in]{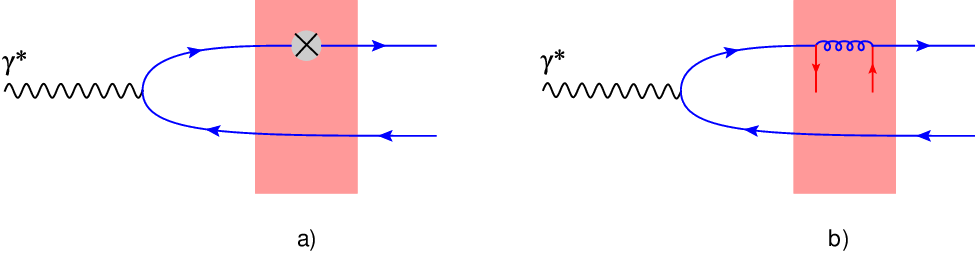}
		\caption{Diagrams for the dipole cross-section with sub-eikonal corrections. In the right panel we have
the diagram with quark field in the background. In the left panel we have 
the diagram with sub-eikonal contribution due to the gluon field. In particular we will consider
the $\calf(z_\perp)$, $\calf_2(z_\perp)$, $\calf_{2'}(z_\perp)$, and $\calf'(z_\perp)$ operators. }
		\label{Fig:DIS-Qsub}
	\end{center}
\end{figure}

In this section we derive the dipole cross-section including the first sub-eikonal
corrections with the quark propagator containing additional operator insertions localized on the shock wave.
As a consequence, besides the usual dipole operator, new non-eikonal operator structures
appear in the cross section.

We organize the calculation according to the different classes of operators appearing in the
sub-eikonal quark propagator. We first consider the contribution of the
$\calf_2(z_\perp)$, $\calf_{2'}(z_\perp)$, and $\calf_i(z_\perp)$ operators, then the one associated with
the $\calf(z_\perp)$ operator, and finally the contribution involving background quark fields.
After computing these terms separately, we collect them into the final expression for the
dipole cross-section at sub-eikonal level.

First, we need the sub-eikonal Feynman rule for a quark line:
\begin{eqnarray}
&&\hspace{-1cm}
\lim_{p^2\to 0}\int \!\! d^4 x\,e^{ip\cdot x}\Big(\baru(p)i\slashd_x \theta(x^+)\theta(-y^+)\langle {\rm T}\psi(x)\barpsi(y)\rangle_{\psi,\barpsi,A} \Big)
\nonumber\\
&&\hspace{-1cm}
= i\,\lim_{p^2\to 0}
\theta(p^+)\theta(-y^+)\!\!\int \!\!d^2 z_\perp \dhd^2 k\,
\baru(p)\bigg\{\ssn_2\, U_z
\nonumber\\
&&+ {1\over sp^+}\Big[\gamma^\mu_\perp Q(z_\perp)\gamma^\perp_\mu + \half\ssn_2\gamma^5 \calf_z
+ \ssn_2\Big(2 k^i \calf_{iz} + \calf'_z - \calf_{2'z}\Big)\Big]\bigg\} \nonumber\\
&&~~\times{p^+\ssn_1 + \ssk_\perp\over 2p^+}
\,e^{ ip^+ y^- + i{k^2_\perp\over 2p^+}y^+-i(p-k,z)_\perp - i(k,y)_\perp }
\label{FruleSub1a}
\end{eqnarray}
and for an anti-quark line:
\begin{eqnarray}
&&\hspace{-1cm}
\lim_{k^2\to 0}\int d^4 y\Big(\theta(-w^+)\theta(y^+)
\langle {\rm T}\psi^\delta_i(w)\barpsi_j^\sigma(y)\rangle_{\psi,\barpsi,A}\Big)\left(-i\overleftarrow{\slashd}_y \right)v^\sigma_j(k)\,e^{ik\cdot y}
\nonumber\\
&&\hspace{-1cm} 
= i\,\lim_{k^2\to 0}\theta(-w^+)\theta(k^+)\int d^2z \dhd^2 q {k^+\ssn_1  - \ssq_\perp\over 2k^+}\bigg\{\ssn_2\,U^\dagger_z
\nonumber\\
&&- {1\over sk^+}\Big[\gamma^\mu_\perp \tildeQ(z_\perp)\gamma^\perp_\mu + \half\ssn_2\gamma^5 \calf_z^\dagger
+ \ssn_2\Big(2 q^i \calf^\dagger_{iz} + \calf'^\dagger_z  - \calf^\dagger_{2'z}\Big)\Big]\bigg\}\nonumber\\
&&~~\times v(k)e^{ik^+ w^- + i{q^2_\perp\over 2k^+}w^+ -i(k+q,z)_{\perp} + i(q,w)}\,.
\label{FruleSub1b}
\end{eqnarray}

Notice that here we have to consider two different expressions of the quark propagator with
sub-eikonal corrections. For the quark line, we use the propagator with the
${P_{\rm right}}$ sub-eikonal contribution given in eq.~\eqref{PropP2right}, because in this
case the free propagator is on the left, where the LSZ reduction formula can be easily
applied, as explained in Sec.~\ref{sec:Feynmanrule}. Similarly, for the antiquark line, we
use the propagator with the ${P_{\rm left}}$ sub-eikonal contribution given in
eq.~\eqref{PropP2left}, where the free propagator is on the right. This is the reason why
the different operator structures appear in slightly different forms for the quark and
antiquark lines, even though they encode the same sub-eikonal content of the propagator.

Let us calculate diagrams in Fig. \ref{Fig:DIS-Qsub}.
The scattering amplitude we want to calculate is 
\begin{eqnarray}
&&\langle q(p)\barq(k)|\gamma^*(q)\rangle_{\small{\rm Fig.} \ref{Fig:DIS-Qsub}}
\nonumber\\
=\!\!\!&&iee_f\int d^4\omega\theta(-\omega^+)\,e^{-iq\cdot \omega}\varepsilon_\mu(q)\int d^4x \, e^{ip\cdot x}\baru(p)i\sslash{\partial}_x\theta(x^+)
\langle {\rm T}\psi(x)\barpsi(\omega)\rangle^{eik+sub}\gamma^\mu
\nonumber\\
&&\times\! \int d^4y
\langle {\rm T}\psi(\omega)\barpsi(y)\rangle^{eik+sub}\theta(y^+)\Big(-i\overleftarrow{\sslash{\partial}}_y\Big)v(k) \,e^{ik\cdot y}
\end{eqnarray}
Using the two Feynman rules (\ref{FruleSub1a}), and (\ref{FruleSub1b}), we have
\begin{eqnarray}
&&\langle q(p)\barq(k)|\gamma^*(q)\rangle_{\small{\rm Fig.} \ref{Fig:DIS-Qsub}}
\nonumber\\
=\!\!\!&& - {iee_f\over 4} \int^0_{-\infty}d\omega^+\int d^2\omega\, \dbar(p^++k^+-q^+){\theta(p^+)\theta(k^+)\over p^+k^+}
\int d^2z_1 d^2z_2 \dhd^2 q_1\dhd^2 q_2
\nonumber\\
\hspace{-2cm}&&\times
e^{i\omega^+\left({q^2_{1\perp}\over 2p^+} + {q^2_{2\perp}\over 2k^+}-q^-\right) + i(q_1-p,z_1)_\perp - i (q_2+k,z_2) 
- i(q_1 - q_2,\omega)_\perp}
\baru(p,\sigma)\Bigg[U_{z_1}\ssn_2
\nonumber\\
\hspace{-2cm}&&
 + {1\over sp^+ }\Big(2 q^i_1 \calf_{iz_1} + \calf'_{z_1} - \calf_{2'z_1}\Big)\ssn_2
+ {1\over sp^+ }\Big(\gamma^\mu_\perp Q(z_{1\perp})\gamma^\perp_\mu + \half\ssn_2\gamma^5 \calf_{z_1}\Big)\Bigg]
\nonumber\\
\hspace{-2cm}&&\times[p^+\ssn_1 + \ssq_{1\perp}]\sslash{\varepsilon}(q)
[k^+\ssn_1- \ssq_{2\perp}]\Bigg[U^\dagger_{z_2}\ssn_2 
- {1\over sk^+ }\Big(2 q_2^i \calf^\dagger_{iz_2} + \calf'^\dagger_{z_2} - \calf^\dagger_{2'z_2}\Big)\ssn_2
\nonumber\\
\hspace{-2cm}&&
- {1\over sk^+}\Big(\gamma^\mu_\perp \tildeQ(z_{2\perp})\gamma^\perp_\mu 
+ \half\ssn_2\gamma^5 \calf_{z_2}^\dagger\Big) \Bigg]v(k,\sigma')
\end{eqnarray}
In the above product we need only terms up to sub-eikonal terms, so, subtracting the no-interaction term, we have
\begin{eqnarray}
\hspace{-2cm}&&\langle q(p)\barq(k)|\gamma^*(q)\rangle_{\small {\rm Fig.} \ref{Fig:DIS-Qsub}}
\nonumber\\
\hspace{-2cm}&& 
= - {iee_f\over 4} \int^0_{-\infty}d\omega^+\int d^2\omega\, \dbar(p^++k^+-q^+)
{\theta(p^+)\theta(k^+)\over p^+k^+}
\int d^2z_1 d^2z_2 \dhd^2 q_1\dhd^2 q_2
\nonumber\\
\hspace{-2cm}&&\times
e^{i\omega^+\left({q^2_{1\perp}\over 2p^+} + {q^2_{2\perp}\over 2k^+} - q^- \right) 
+ i(q_1-p,z_1)_\perp - i (q_2+k,z_2) - i(q_1-q_2,\omega)_\perp}
\nonumber\\
\hspace{-2cm}&&\times
\Bigg\{
\baru(p,\sigma)\Big[\Big(U_{z_1}U^\dagger_{z_2}-1\Big) - {1\over sk^+}U_{z_1}\Big(2 q_2^i \calf^\dagger_{iz_2} 
+ \calf'^\dagger_{z_2}  - \calf_{2'z_2}^\dagger\Big)
+ {1\over sp^+}\Big(2 q^i_1 \calf_{iz_1} + \calf'_{z_1} - \calf_{2'z_1}\Big)U^\dagger_{z_2}\Big]
\nonumber\\
\hspace{-2cm}&&\times
\ssn_2[p^+\ssn_1 + \ssq_{1\perp}]\sslash{\varepsilon}(q)[k^+\ssn_1-\ssq_{2\perp}]\ssn_2v(k,\sigma')
\nonumber\\
\hspace{-2cm}&&
- {1\over sk^+}\baru(p,\sigma)\ssn_2[p^+\ssn_1 + \ssq_{1\perp}]\sslash{\varepsilon}(q)[k^+\ssn_1-\ssq_{2\perp}]
\gamma_\perp^\mu U_{z_1}\tilde{Q}(z_{2\perp})\gamma^\perp_\mu v(k,\sigma')
\nonumber\\
\hspace{-2cm}&&
+ {1\over sp^+}\baru(p,\sigma)\gamma_\perp^\mu Q(z_{1\perp})U^\dagger_{z_2}\gamma^\perp_\mu 
[p^+\ssn_1 + \ssq_{1\perp}]\sslash{\varepsilon}(q)[k^+\ssn_1-\ssq_{2\perp}]\ssn_2v(k,\sigma')
\nonumber\\
\hspace{-2cm}&&
+ \baru(p,\sigma)\ssn_2\gamma^5[p^+\ssn_1 + \ssq_{1\perp}]\sslash{\varepsilon}(q)[k^+\ssn_1-\ssq_{2\perp}]\ssn_2
\Big({1\over 2sp^+}\calf_{z_1}U^\dagger_{z_2} - {1\over 2sk^+}U_{z_1}\calf^\dagger_{z_2}\Big)v(k,\sigma')
\Bigg\} + O(\lambda^{-2})
\label{dipoleu2subeikona}
\end{eqnarray}
We now integrate over $d\omega^+$, and $d^2\omega d^2q_2$ we obtain
\begin{eqnarray}
&&\langle q(p)\barq(k)|\gamma^*(q)\rangle_{\small{\rm Fig.} \ref{Fig:DIS-Qsub}}
\nonumber\\
=\!\!\!&& - ee_f {1\over s}\, \theta(p^+)\theta(k^+)\dbar\!\left({p^+\over q^+} + {k^+\over q^+} -1\right)\!\!
\int d^2z_1 d^2z_2 {\dhd^2 q_1\over q^2_{1\perp}+ {2\over s}Q^2 p^+k^+ }
e^{i(q_1-p,z_1)_\perp - i (k + q_1,z_2)}
\nonumber\\
\hspace{-2cm}&&\times
\Bigg\{
\baru(p,\sigma)\Big[\Big(U_{z_1}U^\dagger_{z_2}-1\Big) 
- {1\over sk^+}U_{z_1}\Big(2 q_1^i \calf^\dagger_{iz_2} + \calf'^\dagger_{z_2}  - \calf_{2'z_2}^\dagger\Big)
+ {1\over sp^+}\Big(2 q^i_1 \calf_{iz_1} + \calf'_{z_1} - \calf_{2'z_1}\Big)U^\dagger_{z_2}\Big]
\nonumber\\
\hspace{-2cm}&&\times
\ssn_2[p^+\ssn_1 + \ssq_{1\perp}]\sslash{\varepsilon}(q)[k^+\ssn_1-\ssq_{1\perp}]\ssn_2v(k,\sigma')
\nonumber\\
\hspace{-2cm}&&
- {1\over sk^+}\baru(p,\sigma)\ssn_2[p^+\ssn_1 + \ssq_{1\perp}]\sslash{\varepsilon}(q)[k^+\ssn_1-\ssq_{1\perp}]
\gamma_\perp^\mu U_{z_1}\tilde{Q}(z_{2\perp})\gamma^\perp_\mu v(k,\sigma')
\nonumber\\
\hspace{-2cm}&&
+ {1\over sp^+}\baru(p,\sigma)\gamma_\perp^\mu Q(z_{1\perp})U^\dagger_{z_2}\gamma^\perp_\mu 
[p^+\ssn_1 + \ssq_{1\perp}]\sslash{\varepsilon}(q)[k^+\ssn_1-\ssq_{1\perp}]\ssn_2v(k,\sigma')
\nonumber\\
\hspace{-2cm}&&
+ \baru(p,\sigma)\ssn_2\gamma^5[p^+\ssn_1 + \ssq_{1\perp}]\sslash{\varepsilon}(q)[k^+\ssn_1-\ssq_{1\perp}]\ssn_2
\Big({1\over 2sp^+}\calf_{z_1}U^\dagger_{z_2} - {1\over 2sk^+}U_{z_1}\calf^\dagger_{z_2}\Big)v(k,\sigma')
\Bigg\} 
\nonumber\\
&&+ O(\lambda^{-2})
\label{eikoEsubeiko-a}
\end{eqnarray}
We can rewrite result (\ref{eikoEsubeiko-a}) as sum of four terms as 
\begin{eqnarray}
\langle q(p)\barq(k)|\gamma^*(q)\rangle_{\small{\rm Fig.} \ref{Fig:dipole}+\ref{Fig:DIS-Qsub}}
=\!\!\!&& \bigg( \langle q(p)\barq(k)|\gamma^*(q)\rangle_{\rm eik} + \langle q(p)\barq(k)|\gamma^*(q)\rangle_{\rm{F}_2}
\nonumber\\
&& + \langle q(p)\barq(k)|\gamma^*(q)\rangle_{\rm Gluon} + \langle q(p)\barq(k)|\gamma^*(q)\rangle_{\rm Quark}\bigg)
\end{eqnarray}
with (the following correspond to Fig. \ref{Fig:dipole})
\begin{eqnarray}
&&\langle q(p)\barq(k)|\gamma^*(q)\rangle_{\rm eik}
\nonumber\\
\equiv\!\!\!&& - {ee_f\over s} \,\theta(p^+)\theta(k^+)\dbar\!\left({p^+\over q^+} + {k^+\over q^+} -1\right)\!\!\!
\int d^2z_1 d^2z_2 {\dhd^2 q_1\over q^2_{1\perp}+ {2\over s}Q^2 p^+k^+ }
e^{i(q_1-p,z_1)_\perp  - i (k + q_1,z_2)}
\nonumber\\
&&\times
\baru(p,\sigma)\Big(U_{z_1}U^\dagger_{z_2} - 1\Big)
\ssn_2[p^+\ssn_1 + \ssq_{1\perp}]\sslash{\varepsilon}(q)[k^+\ssn_1-\ssq_{1\perp}]\ssn_2v(k,\sigma')
\label{dipoleikonal}
\end{eqnarray}
and
\begin{eqnarray}
&& \langle q(p)\barq(k)|\gamma^*(q)\rangle_{\rm{F}_2}
\nonumber\\
\equiv\!\!\!&& - {ee_f\over s}\,\theta(p^+)\theta(k^+)\dbar\!\left({p^+\over q^+} + {k^+\over q^+} -1\right)\!\!\!
\int d^2z_1 d^2z_2 {\dhd^2 q_1\over q^2_{1\perp} + {2\over s}Q^2 p^+k^+}
e^{i(q_1-p,z_1)_\perp  - i (k+ q_1,z_2)}
\nonumber\\
&&\times
\baru(p,\sigma)\Big[ - {1\over sk^+}U_{z_1}
\Big(2 q_1^i \calf^\dagger_{iz_2}  + \calf'^\dagger_{z_2}  - \calf_{2'z_2}^\dagger\Big)
\nonumber\\
&&+ {1\over sp^+}\Big(2 q^i_1 \calf_{iz_1} + \calf'_{z_1} - \calf_{2'z_1}\Big)U^\dagger_{z_2}\Big]
\ssn_2[p^+\ssn_1 + \ssq_{1\perp}]\sslash{\varepsilon}(q)[k^+\ssn_1-\ssq_{1\perp}]\ssn_2v(k,\sigma')
\label{diolesubeikoF2}
\end{eqnarray}
and
\begin{eqnarray}
&& \langle q(p)\barq(k)|\gamma^*(q)\rangle_{\rm Gluon}
\nonumber\\
\hspace{-2cm}
\equiv\!\!\!&&  - {ee_f \over s}\,\theta(p^+)\theta(k^+)\dbar\!\left({p^+\over q^+} + {k^+\over q^+} -1\right)\!\!\!
\int d^2z_1 d^2z_2 {\dhd^2 q_1\over q^2_{1\perp}+ {2\over s}Q^2 p^+k^+}
e^{i(q_1-p,z_1)_\perp  - i (k + q_1,z_2)}
\nonumber\\
&&\times\baru(p,\sigma)\ssn_2\gamma^5[p^+\ssn_1 + \ssq_{1\perp}]\sslash{\varepsilon}(q)[k^+\ssn_1-\ssq_{1\perp}]\ssn_2
\Big({1\over 2sp^+}\calf_{z_1}U^\dagger_{z_2} - {1\over 2sk^+}U_{z_1}\calf^\dagger_{z_2}\Big)v(k,\sigma')
\label{diolesubeikoG}
\end{eqnarray}
and finally,
\begin{eqnarray}
&& \langle q(p)\barq(k)|\gamma^*(q)\rangle_{\rm Quark}
\nonumber\\
\equiv\!\!\!&& - {ee_f\over s} \,\theta(p^+)\theta(k^+)\dbar\!\left({p^+\over q^+} + {k^+\over q^+} -1\right)\!\!\!
\int d^2z_1 d^2z_2 {\dhd^2 q_1\over q^2_{1\perp}+ {2\over s}Q^2 p^+k^+}
e^{i(q_1-p,z_1)_\perp  - i (k + q_1,z_2)}
\nonumber\\
&&\hspace{0.3cm}\times
\Bigg\{{1\over sp^+}\baru(p,\sigma)\gamma_\perp^\mu Q(z_{1\perp})U^\dagger_{z_2}\gamma^\perp_\mu 
[p^+\ssn_1 + \ssq_{1\perp}]\sslash{\varepsilon}(q)[k^+\ssn_1-\ssq_{1\perp}]\ssn_2v(k,\sigma')
\nonumber\\
&&\hspace{0.7cm}
- {1\over sk^+}\baru(p,\sigma)\ssn_2[p^+\ssn_1 + \ssq_{1\perp}]\sslash{\varepsilon}(q)[k^+\ssn_1-\ssq_{1\perp}]
\gamma_\perp^\mu U_{z_1}\tilde{Q}(z_{2\perp})\gamma^\perp_\mu v(k,\sigma')
\Bigg\}
\label{diolesubeikoQ}
\end{eqnarray}
So, eq. (\ref{eikoEsubeiko-a}) is the sum of the equations (\ref{dipoleikonal}), (\ref{diolesubeikoF2}), (\ref{diolesubeikoG}), and (\ref{diolesubeikoQ}).
In the next section we analyze each of them separately. To this end, we will need the longitudinal and transverse component of the 
eikonal term, eq. (\ref{dipoleikonal}). 

The eikonal Dirac matrix elements, already calculated in section \ref{sec:dipolecrossection},
are
\begin{eqnarray}
&&\baru(p,\sigma) \ssn_2[p^+\ssn_1+ \ssq_{1\perp}]\sslash{\varepsilon}^L(k^+\ssn_1 - \ssq_{1\perp})\ssn_2v(k,\sigma')
\nonumber\\
=\!\!\!&&  2q^+\Big(- {q_{1\perp}^2\over Q} + {2\over s}Q p^+k^+\Big) \baru(p,\sigma)\ssn_2 v(k,\sigma') 
\label{dipoleMElongy}
\end{eqnarray}
and
\begin{eqnarray}
&&\baru(p,\sigma) \ssn_2[p^+\ssn_1+ \ssq_{1\perp}]\sslash{\varepsilon}^T(k^+\ssn_1 - \ssq_{1\perp})\ssn_2v(k,\sigma')
\nonumber\\
=\!\!\!&& 2(p^+-k^+)(\varepsilon, q_1)_\perp\baru(p,\sigma)\ssn_2 v(k,\sigma') 
- 2i\,q^+\,(\vec{\varepsilon}_\perp\times \vec{q}_1)\baru(p,\sigma)\gamma^5\ssn_2 v(k,\sigma')
\label{dipoleMEtransver}
\end{eqnarray}
So, using (\ref{dipoleMElongy}), and (\ref{dipoleMEtransver}), we define
\begin{eqnarray}
&&\langle q(p)\barq(k)|\gamma^*_L(q)\rangle_{\rm eik}
\nonumber\\
=\!\!\!&& - {ee_f\over s} \,\theta(p^+)\theta(k^+)\dbar\!\left({p^+\over q^+} + {k^+\over q^+} -1\right)\!\!\!
\int d^2z_1 d^2z_2 {\dhd^2 q_1\over q^2_{1\perp}+{2\over s}Q^2 p^+k^+ }
e^{i(q_1-p,z_1)_\perp - i (k + q_1,z_2)}
\nonumber\\
&&\times
2q^+\baru(p,\sigma)\Big(U_{z_1}U^\dagger_{z_2} -1\Big)
\Big(- {q_{1\perp}^2\over Q} + {2\over s}Q p^+k^+\Big) \ssn_2 v(k,\sigma')
\label{dipoleikonallongy}
\end{eqnarray}
and
\begin{eqnarray}
&&\langle q(p)\barq(k)|\gamma^*_T(q)\rangle_{\rm eik}
\nonumber\\
=\!\!\!&& - {ee_f\over s} \,\theta(p^+)\theta(k^+)\dbar\!\left({p^+\over q^+} + {k^+\over q^+} -1\right)\!\!\!
\int d^2z_1 d^2z_2 {\dhd^2 q_1\over q^2_{1\perp} + {2\over s}Q^2 p^+k^+ }
e^{i(q_1-p,z_1)_\perp - i (k + q_1,z_2)}
\nonumber\\
&&\times
2\,\baru(p,\sigma)\Big(U_{z_1}U^\dagger_{z_2} -1\Big)
\Big[(p^+-k^+)(\varepsilon, q_1)_\perp\ssn_2 
- i\,q^+(\vec{\varepsilon}_\perp\times \vec{q}_1)\gamma^5\ssn_2 \Big]v(k,\sigma')
\label{dipoleikonaltransver}
\end{eqnarray}
In section \ref{sec:dipolecrossection}, we already calculated the longitudinal and transverse dipole cross-section using (\ref{dipoleikonallongy}), 
and (\ref{dipoleikonaltransver}), respectively.

\subsection{The gluon sub-eikonal correction: the $\calf'(z_\perp)$, $\calf_2(z_\perp)$, $\calf_{2'}(z_\perp)$, and $\calf_i(z_\perp)$ operators}

In this subsection, we consider the sub-eikonal contribution of the
$\calf'(z_\perp)$, $\calf_2(z_\perp)$, $\calf_{2'}(z_\perp)$, and $\calf_i(z_\perp)$ operators. This is the simplest
contribution, because the corresponding Dirac matrix element is the same as in the eikonal
case. Indeed, we have
\begin{eqnarray}
&& \Big(\langle q(p)\barq(k)|\gamma^*(q)\rangle_{\rm eik} + \langle q(p)\barq(k)|\gamma^*(q)\rangle_{\rm{F}_2}\Big)
\nonumber\\
= \!\!\!&& - {ee_f \over s}\,\theta(p^+)\theta(k^+)\dbar\!\left({p^+\over q^+} + {k^+\over q^+} -1\right)\!\!\!
\int d^2z_1 d^2z_2 {\dhd^2 q_1\over q^2_{1\perp}+ {2\over s}Q^2 p^+k^+ }
e^{i(q_1-p,z_1)_\perp - i (k + q_1,z_2)}
\nonumber\\
&&\times
\baru(p,\sigma)\Big[\Big(U_{z_1}U^\dagger_{z_2} -1\Big) 
- {1\over sk^+}U_{z_1}\Big(2 q_1^i \calf^\dagger_{iz_2} + \calf'^\dagger_{z_2} - \calf^\dagger_{2'z_2}\Big)
\nonumber\\
&&+ {1\over sp^+}\Big(2 q^i_1 \calf_{iz_1} + \calf'_{z_1} - \calf_{2'z_1}\Big)U^\dagger_{z_2}\Big]
\ssn_2[p^+\ssn_1 + \ssq_{1\perp}]\sslash{\varepsilon}(q)[k^+\ssn_1-\ssq_{1\perp}]\ssn_2v(k,\sigma')
\label{diole-Eik-subeikoF2}
\end{eqnarray}
 We now square the amplitude and consider separately the longitudinal and transverse
 photon polarizations. 
 
 \subsubsection{Longitudinal polarization with the $\calf'(z_\perp)$, $\calf_2(z_\perp)$, $\calf_{2'}(z_\perp)$, and $\calf_i(z_\perp)$ operators}
 \label{sec:LongyFprimeF2Fi}
 
From (\ref{diole-Eik-subeikoF2}), using (\ref{LOdipoleDiracmatrix}), (\ref{LOlongypola1}), and
(\ref{LOlongypola2}), we calculate the scattering amplitude square with longitudinal photon polarization.  We have
\begin{eqnarray}
&&{1\over 2\pi\delta(0)}\int\!\dhd^4k\dhd^4p\,\dbar(k^2)\dbar(p^2) \theta(p^+)\theta(k^+)
\Big|\langle q(p)\barq(k)|\gamma^*_L(q)\rangle_{\rm eik} + \langle q(p)\barq(k)|\gamma^*_L(q)\rangle_{\rm{F}_2}\Big|^2
\nonumber\\
= \!\!\!&& {2N_ce^2\over \pi}\sum_fe^2_f\int_0^1 dz\,
\int d^2z_1 d^2z_2 \dhd^2 q_1\dhd^2 q_2{Q z\barz - {q_{1\perp}^2\over Q}\over q^2_{1\perp}+Q^2 z\barz }
{Q z\barz - {q_{2\perp}^2\over Q}\over q^2_{2\perp}+Q^2 z\barz }
\,e^{ i(q_1 - q_2, z_1-z_2)_\perp }
\nonumber\\
&&\times \Bigg\{1 - {1\over N_c}\Tr\{U_{z_1}U^\dagger_{z_2}\} 
\nonumber\\
&&\hspace{0.2cm}+ {q^+\over N_cs^2}\Bigg[
{1\over \barz}\Big(2q_1^i\Tr\{U_{z_1}\calf^\dagger_{iz_2}\} + \Tr\{U_{z_1} \calf'^\dagger_{z_2}\} 
- \Tr\left\{[U_{z_1}-U_{z_2}]\calf^\dagger_{2'z_2}\right\}\Big)
\nonumber\\
&&\hspace{0.4cm} - {1\over z}\Big(2q_1^i\Tr\{\calf_{iz_1}U^\dagger_{z_2}\} + \Tr\{\calf'_{z_1}U^\dagger_{z_2}\} 
- \Tr\left\{\calf_{2'z_1}[U^\dagger_{z_2} - U^\dagger_{z_1}]\right\}\Big)\Bigg]
\nonumber\\
&&\hspace{0.4cm} + {q^+\over N_cs^2}\Bigg[{1\over \barz}
\Big(2q_2^i\Tr\{U^\dagger_{z_1}\calf_{iz_2}\} + \Tr\{U^\dagger_{z_1} \calf'_{z_2}\} 
- \Tr\left\{[U^\dagger_{z_1}-U^\dagger_{z_2}]\calf_{2'z_2}\right\}\Big)
\nonumber\\
&&\hspace{0.6cm} - {1\over z}\Big(2q_2^i\Tr\{\calf^\dagger_{iz_1}U_{z_2}\} + \Tr\{\calf'^\dagger_{z_1}U_{z_2}\} 
- \Tr\left\{\calf^\dagger_{2'z_1}[U_{z_2}-U_{z_1}]\right\}\Big)\Bigg]
\Bigg\} + O(\lambda^{-2})
\label{FiF2F2primeSquqre-1}
\end{eqnarray}
It is easy to show that, using the symmetry $q_1\leftrightarrow q_2$ and $z\leftrightarrow \barz$ and $z_1\leftrightarrow z_2$, the 
contribution of the operators $\calf_i(z_\perp)$ and $\calf'(z_\perp)$ and their adjoint conjugates cancel out
leaving only $\calf_{2'}(z_\perp)$, and its adjoint conjugate. So, from eq. (\ref{FiF2F2primeSquqre-1}), we obtain
\begin{eqnarray}
&&{1\over 2\pi\delta(0)}\int\!\dhd^4k\dhd^4p\,\dbar(k^2)\dbar(p^2) \theta(p^+)\theta(k^+)
\Big|\langle q(p)\barq(k)|\gamma^*_L(q)\rangle_{\rm eik} + \langle q(p)\barq(k)|\gamma^*_L(q)\rangle_{\rm{F}_2}\Big|^2
\nonumber\\
= \!\!\!&& {8Q^2N_c e^2 \over \pi}\sum_f e^2_f\int_0^1\! dz\,z^2\barz^2\!
\int d^2z_1 d^2z_2 \dhd^2 q_1\dhd^2 q_2{e^{i(q_1 - q_2, z_1-z_2)_\perp }\over [q^2_{1\perp}+Q^2 z\barz][q^2_{2\perp}+Q^2 z\barz] }
\Bigg\{\calu_{z_1z_2}
\nonumber\\
&&+ {q^+\over 2z\barz N_c s^2}\bigg[\Tr\{\big(U_{z_1}-U_{z_2}\big)\big(\calf^\dagger_{2'z_2} - \calf^\dagger_{2'z_2} \big)\} 
+ \Tr\{\big(\calf_{2'z_1} - \calf_{2'z_1}\big)\big(U^\dagger_{z_2}-U^\dagger_{z_1}\big)\}\bigg]\Bigg\} + O(\lambda^{-2})
\nonumber\\
=\!\!\!&&\calm^L_{\rm eikonal} + \calm^L_{G_2} + O(\lambda^{-2})
\end{eqnarray}
where $\calm^L_{\rm eikonal}$ is the eikonal contribution to the dipole cross-section we obtained in eq.  (\ref{LOdipoleLongy}), and 
\begin{eqnarray}
\calm^L_{G_2} \equiv \!\!\!&& {4\,Q^2e^2\over \pi}\sum_fe^2_f\int_0^1\! dz\,z\barz \!
\int d^2z_1 d^2z_2 \dhd^2 q_1\dhd^2 q_2{e^{ i(q_1 - q_2, z_1-z_2)_\perp }\over [q^2_{1\perp}+Q^2 z\barz][q^2_{2\perp}+Q^2 z\barz] }
\nonumber\\
&&\times {q^+\over s^2}\bigg[\Tr\{\big(U_{z_1}-U_{z_2}\big)\big(\calf^\dagger_{2'z_2} - \calf^\dagger_{2'z_2} \big)\} 
+ \Tr\{\big(\calf_{2'z_1} - \calf_{2'z_1}\big)\big(U^\dagger_{z_2}-U^\dagger_{z_1}\big)\}\bigg]
\nonumber\\
=\!\!\!&& 0
\label{dipoleG2subeik}
\end{eqnarray}

 \subsubsection{Transverse polarization with the $\calf'(z_\perp)$, $\calf_2(z_\perp)$, $\calf_{2'}(z_\perp)$, and $\calf_i(z_\perp)$ operators}
 
 For the transverse polarization contribution, using (\ref{LOmatrixB}), and \eqref{sumTpoly}, we have
 \begin{eqnarray}
 &&{1\over 2\pi\delta(0)}\Big|\langle q(p)\barq(k)|\gamma^*_T(q)\rangle_{\rm eik} + \langle q(p)\barq(k)|\gamma^*_T(q)\rangle_{\rm{F}_2}\Big|^2
 \nonumber\\
 = \!\!\!&& {2e^2\over \pi s^2}\sum_fe^2_f\int_0^1 dz \int d^2z_1 d^2z_2 \,
 {\dhd^2 q_1\dhd^2 q_2\,e^{i(q_1-q_2,z_1-z_2)}\over [q^2_{1\perp} + Q^2 z\barz][q^2_{2\perp} + Q^2 z\barz]}
\half\sum_{\lambda=\pm 1}
 \nonumber\\
 && \times 
 \Big[(z-\barz)^2(\varepsilon_\lambda,q_1)(\varepsilon^*_\lambda,q_2) +(\vec{\varepsilon}_\lambda\times\vecq_1)(\vec{\varepsilon^*}_\lambda\times\vecq_2)\Big]
\Bigg[s^2N_c\Big(1 - {1\over N_c}\Tr\{U_{z_1}U^\dagger_{z_2}\}\Big)
\nonumber\\
&& + {q^+\over \barz}\bigg(2q_1^i\Tr\{U_{z_1}\calf^\dagger_{iz_2}\} + \Tr\{U_{z_1} \calf'^\dagger_{z_2}\} 
- \Tr\{[U_{z_1}-U_{z_2}]\calf^\dagger_{2'z_2}\}\bigg)
 \nonumber\\
 && - {q^+\over z}\bigg(2q_1^i\Tr\{\calf_{iz_1}U^\dagger_{z_2}\} + \Tr\{\calf'_{z_1}U^\dagger_{z_2}\} 
 - \Tr\{\calf_{2'z_1}[U^\dagger_{z_2}-U^\dagger_{z_1}]\}\bigg)
 \nonumber\\
 &&+ {q^+\over \barz}\bigg(2q_2^i\Tr\{U^\dagger_{z_1}\calf_{iz_2}\} + \Tr\{U^\dagger_{z_1} \calf'_{z_2}\} 
 - \Tr\{[U^\dagger_{z_1}-U^\dagger_{z_2}]\calf_{2'z_2}\}\bigg)
 \nonumber\\
 && - {q^+\over z}\Big(2q_2^i\Tr\{\calf^\dagger_{iz_1}U_{z_2}\} + \Tr\{\calf'^\dagger_{z_1}U_{z_2}\} 
 - \Tr\{\calf^\dagger_{2'z_1}[U_{z_2}-U_{z_1}]\}\bigg)\Bigg]
 \label{FiF2F2primeSquqre-1tran}
 \end{eqnarray}
Using again the quark-anti-quark symmetry, and the $q_{1\perp}\leftrightarrow q_{2\perp}$ symmetry (see eq. \eqref{sumTpoly}),
we observe that the only operator contributing is $\calf_{2'}(z_\perp)$ 
and its adjoint conjugate. So, from (\ref{FiF2F2primeSquqre-1tran}) we obtain
\begin{eqnarray}
&&{1\over 2\pi\delta(0)}\int\!\dhd^4k\dhd^4p\,\dbar(k^2)\dbar(p^2) \theta(p^+)\theta(k^+)
\Big|\langle q(p)\barq(k)|\gamma^*_T(q)\rangle_{\rm eik} + \langle q(p)\barq(k)|\gamma^*_T(q)\rangle_{\rm{F}_2}\Big|^2
\nonumber\\
=\!\!\!&& 8\,N_c\alpha_{\rm em}\sum_fe^2_f\int_0^1 dz\, (z^2+\barz^2)\int d^2z_1 d^2z_2 \,
{\dhd^2 q_1\dhd^2 q_2\,e^{i(q_1-q_2,z_1-z_2)}\over [q^2_{1\perp} + Q^2 z\barz][q^2_{2\perp} + Q^2 z\barz]}
(q_1,q_2) 
\nonumber\\
&&\times\Bigg[\calu_{z_1z_2}
 + {q^+\over 2z\barz N_cs^2}\Bigg(\Tr\big\{\big(U_{z_2}-U_{z_1}\big)\big(\calf^\dagger_{2'z_1} - \calf^\dagger_{2'z_1}\big)\big\}
\nonumber\\
&&\hspace{2cm}+ \Tr\big\{\big(\calf_{2'z_2} - \calf_{2'z_2}\big)\big(U^\dagger_{z_1} - U^\dagger_{z_2}\big)\big\}\Bigg)
\Bigg]
 \nonumber\\
=\!\!\!&&\calm^T_{\rm Eikonal} + \calm^T_{G_2}
\label{FiF2F2primeSquqre-1tran-a}
\end{eqnarray}
where $\calm^T_{\rm Eikonal}$ is eq. (\ref{LOdipoleTransver}) and we define $\calm^T_{G_2}$ as
\begin{eqnarray}
\hspace{-1cm}\calm^T_{G_2}
\equiv\!\!\!&& {2Q^2\alpha_{\rm em}\over \pi^2}\sum_fe^2_f\int_0^1 dz\, (z^2+\barz^2)\int d^2z_1 d^2z_2 \,
\left|K_1(\bar{Q}|z_{12}|)\right|^2
\nonumber\\
&&\hspace{-0.2cm}\times {\sqrt{s/2}\over s^2}\bigg(\Tr\big\{\big(U_{z_2}-U_{z_1}\big)\big(\calf^\dagger_{2'z_1} - \calf^\dagger_{2'z_1}\big)\big\}
+ \Tr\big\{\big(\calf_{2'z_2} - \calf_{2'z_2}\big)\big(U^\dagger_{z_1} - U^\dagger_{z_2}\big)\big\}
\bigg)
\nonumber\\
=\!\!\!&& 0
\label{FiF2F2primeSquqre-1tran-b}
\end{eqnarray}
where we used $q^+=\sqrt{s/2}$. Therefore, the contribution of the operators 
$\calf'(z_\perp)$, $\calf_2(z_\perp)$, $\calf_{2'}(z_\perp)$, and $\calf_i(z_\perp)$ to the $F_T$ structure function is zero.

\subsection{Asymmetry polarization with the $\calf'(z_\perp)$, $\calf_2(z_\perp)$, $\calf_{2'}(z_\perp)$, and $\calf_i(z_\perp)$ operators}

The last case we have to consider with $\calf'(z_\perp)$, $\calf_2(z_\perp)$, $\calf_{2'}(z_\perp)$, and $\calf_i(z_\perp)$ operators
is the asymmetric combination of photon helicity. From eq.~\eqref{FiF2F2primeSquqre-1tran}
we have
\begin{eqnarray}
&&{1\over 2\pi\delta(0)}\Big|\langle q(p)\barq(k)|\gamma^*_A(q)\rangle_{\rm eik} + \langle q(p)\barq(k)|\gamma^*_A(q)\rangle_{\rm{F}_2}\Big|^2
\nonumber\\
= \!\!\!&& {2e^2\over \pi s^2}\sum e^2_f\int_0^1 dz \int d^2z_1 d^2z_2 \,
{\dhd^2 q_1\dhd^2 q_2\,e^{i(q_1-q_2,z_1-z_2)}\over [q^2_{1\perp} + Q^2 z\barz][q^2_{2\perp} + Q^2 z\barz]}
\sum_{\lambda={\rm asym.}}
\nonumber\\
&& \times 
\Big[(z-\barz)^2(\varepsilon_\lambda,q_1)(\varepsilon^*_\lambda,q_2) +(\vec{\varepsilon}_\lambda\times\vecq_1)(\vec{\varepsilon^*}_\lambda\times\vecq_2)\Big]
\Bigg[s^2N_c\Big(1 - {1\over N_c}\Tr\{U_{z_1}U^\dagger_{z_2}\}\Big)
\nonumber\\
&& + {q^+\over \barz}\bigg(2q_1^i\Tr\{U_{z_1}\calf^\dagger_{iz_2}\} + \Tr\{U_{z_1} \calf'^\dagger_{z_2}\} 
- \Tr\{[U_{z_1}-U_{z_2}]\calf^\dagger_{2'z_2}\}\bigg)
\nonumber\\
&& - {q^+\over z}\bigg(2q_1^i\Tr\{\calf_{iz_1}U^\dagger_{z_2}\} + \Tr\{\calf'_{z_1}U^\dagger_{z_2}\} 
- \Tr\{\calf_{2'z_1}[U^\dagger_{z_2}-U^\dagger_{z_1}]\}\bigg)
\nonumber\\
&&+ {q^+\over \barz}\bigg(2q_2^i\Tr\{U^\dagger_{z_1}\calf_{iz_2}\} + \Tr\{U^\dagger_{z_1} \calf'_{z_2}\} 
- \Tr\{[U^\dagger_{z_1}-U^\dagger_{z_2}]\calf_{2'z_2}\}\bigg)
\nonumber\\
&& - {q^+\over z}\Big(2q_2^i\Tr\{\calf^\dagger_{iz_1}U_{z_2}\} + \Tr\{\calf'^\dagger_{z_1}U_{z_2}\} 
- \Tr\{\calf^\dagger_{2'z_1}[U_{z_2}-U_{z_1}]\}\bigg)\Bigg]
\label{FiF2F2primeSquqre-1Asym}
\end{eqnarray}
where we defined the asymmetric sum over photon polarization as
\begin{eqnarray}
\sum_{\lambda={\rm asym.}}\varepsilon^i_\lambda\varepsilon^{*j}_\lambda
\equiv \varepsilon^i_+\varepsilon^{*j}_+-\varepsilon^i_-\varepsilon^{*j}_- = i\epsilon^{ji}
\label{asymsum}
\end{eqnarray}
Thus,
\begin{eqnarray}
&&\hspace{-1cm}\sum_{\lambda={\rm asym.}}
\Big[(z-\barz)^2(\varepsilon_\lambda,q_1)(\varepsilon^*_\lambda,q_2) +(\vec{\varepsilon}_\lambda\times\vecq_1)(\vec{\varepsilon^*}_\lambda\times\vecq_2)\Big]
= 2i(z^2+\barz^2)\vecq_2\times\vecq_1
\label{AsymSumFi}
\end{eqnarray}
Using the asymmetric sum \eqref{AsymSumFi}, \eqref{FiF2F2primeSquqre-1Asym} becomes
\begin{eqnarray}
&&\hspace{-0.5cm}{1\over 2\pi\delta(0)}\Big|\langle q(p)\barq(k)|\gamma^*_A(q)\rangle_{\rm eik} + \langle q(p)\barq(k)|\gamma^*_A(q)\rangle_{\rm{F}_2}\Big|^2
\nonumber\\
&&\hspace{-1.0cm} =  i{8q^+e^2\over \pi s^2}\sum e^2_f\int_0^1 dz \,(z^2+\barz^2)\int d^2z_1 d^2z_2 \,
{\dhd^2 q_1\dhd^2 q_2\,e^{i(q_1-q_2,z_1-z_2)}\over [q^2_{1\perp} + Q^2 z\barz][q^2_{2\perp} + Q^2 z\barz]}
\vecq_2\times\vecq_1
\nonumber\\
&&\hspace{-0.7cm}\times\Bigg[
q_1^i\Big({1\over \barz}\Tr\{U_{z_1}\calf^\dagger_{iz_2}\} 
- {1\over z}\Tr\{\calf_{iz_1}U^\dagger_{z_2}\} \Big)
+ q_2^i\Big({1\over \barz}\Tr\{U^\dagger_{z_1}\calf_{iz_2}\} 
 - {1\over z}\Tr\{\calf^\dagger_{iz_1}U_{z_2}\}\Big)\Bigg]
\label{FiF2F2primeSquqre-1Asym-1}
\end{eqnarray}
where we dropped all the terms that are zero under integration over $q_1$ and $q_2$.
Using the symmetry in $z\leftrightarrow \barz$ and the asymmetry in $q_{1\perp}\leftrightarrow q_{2\perp}$ we arrive at
\begin{eqnarray}
&&\hspace{-0.5cm}{1\over 2\pi\delta(0)}\Big|\langle q(p)\barq(k)|\gamma^*_A(q)\rangle_{\rm eik} + \langle q(p)\barq(k)|\gamma^*_A(q)\rangle_{\rm{F}_2}\Big|^2
\nonumber\\
&&\hspace{-1.0cm} =  i\,32\alpha_{\rm em}{q^+\over s^2}\sum e^2_f\int_0^1 dz \,{z^2+\barz^2\over z\barz}\int d^2z_1 d^2z_2 \,
{\dhd^2 q_1\dhd^2 q_2\,e^{i(q_1-q_2,z_1-z_2)}\over [q^2_{1\perp} + Q^2 z\barz][q^2_{2\perp} + Q^2 z\barz]}
\vecq_2\times\vecq_1
\nonumber\\
&&\hspace{-0.7cm}\times
q_1^i\Big(\Tr\{U_{z_1}\calf^\dagger_{iz_2}\} - \Tr\{\calf_{iz_1}U^\dagger_{z_2}\} \Big)
\nonumber\\
&&\hspace{-1.0cm}\equiv \calm^A_{\rm F_i}
\label{FiF2F2primeSquqre-1Asym-2}
\end{eqnarray}
Using the standard modified-Bessel derivative identities, we can rewrite \eqref{FiF2F2primeSquqre-1Asym-2} as
\begin{eqnarray}
\hspace{-1.0cm}\calm^A_{\rm F_i}
=\!\!\!&& 8\,\alpha_{\rm em}Q^2{q^+\over  s^2\pi^2}\sum_f e^2_f \int_0^1dz(z^2 + \barz^2)
\nonumber\\
&&\times\!\int d^2z_1 d^2z_2\Big| K_1(\bar{Q}|z_{12}|)\Big|^2{z_{12}^j\epsilon^{ji} \over  z^2_{12}}
\Big(\Tr\{\calf_{iz_2}^\dagger U_{z_1}\} - \Tr\{\calf_{iz_1}U^\dagger_{z_2}\} \Big)\,.
\end{eqnarray}
We conclude that the only operator considered in this subsection that contributes to the $g_1$ structure function
is the $\calf_i(z_1,z_2)$ operator.

\subsection{The gluon sub-eikonal correction: the $\calf(z_\perp)$ operator}

We now turn to the gluonic sub-eikonal correction associated with the operator
$\calf(z_\perp)$ defined in eq. \eqref{calfoperator_a}, which, unlike the previous case, contributes directly to the helicity-sensitive
part of the dipole cross-section:
\begin{eqnarray}
&&\hspace{-0.7cm}\langle q(p)\barq(k)|\gamma^*(q)\rangle_{\rm Gluon}
\nonumber\\
\hspace{-2cm}
\hspace{-1cm}=\!\!\!&&  - {ee_f \over s}\,\theta(p^+)\theta(k^+)\dbar\!\left({p^+\over q^+} + {k^+\over q^+} -1\right)\!\!\!
\int d^2z_1 d^2z_2 {\dhd^2 q_1\over q^2_{1\perp} + {2\over s}Q^2 p^+k^+ }
e^{i(q_1-p,z_1)_\perp - i (k + q_1,z_2)}
\nonumber\\
&&\hspace{-0.5cm}\times\baru(p,\sigma)\ssn_2\gamma^5[p^+\ssn_1 + \ssq_{1\perp}]\sslash{\varepsilon}(q)[k^+\ssn_1-\ssq_{1\perp}]\ssn_2
\Big({1\over 2sp^+}\calf_{z_1}U^\dagger_{z_2} - {1\over 2 s k^+}U_{z_1}\calf^\dagger_{z_2}\Big)v(k,\sigma')
\label{GluonME-1}
\end{eqnarray}
In the next two subsections we square the longitudinal and transverse photon polarization contributions.

\subsubsection{Longitudinal polarization with $\calf(z_\perp)$ operator}
 \label{sec:Longy-calf}
Let us consider the longitudinal photon polarization. 
The scattering amplitude,  eq. (\ref{GluonME-1}), with longitudinal photon polarization is proportional to the following Dirac matrix element
\begin{eqnarray}
&&\baru(p,\sigma)[\ssn_2\gamma^5(p^+\ssn_1+\ssq_{1\perp})]\sslash{\varepsilon}^L[(k^+\ssn_1-\ssq_{1\perp})]\ssn_2v(k,\sigma')
\nonumber\\
=\!\!\!&&2q^+\baru^i(p)\Big({2\over s}p^+k^+ \,Q - {q^2_{1\perp}\over Q}\Big)\ssn_2\gamma^5v(k,\sigma')
\label{sub-diracCalF-longy}
\end{eqnarray}
and using eq. (\ref{sub-diracCalF-longy}) into the gluon contribution to the scattering amplitude, eq. (\ref{GluonME-1}), we obtain
\begin{eqnarray}
&&\langle q(p)\barq(k)|\gamma_L^*(q)\rangle_{\rm Gluon}
\nonumber\\
=\!\!\!&& -  {ee_f\over s}\, \theta(p^+)\theta(k^+)\dbar\!\left({p^+\over q^+} + {k^+\over q^+} -1\right)\!\!\!
\int d^2z_1 d^2z_2 {\dhd^2 q_1\over q^2_{1\perp} + {2\over s}Q^2 p^+k^+ }
\,e^{ i(q_1-p,z_1)_\perp - i (k + q_1,z_2)}
\nonumber\\
\hspace{-2cm}&&\times
2q^+\Big({2\over s}p^+k^+Q - {q^2_{1\perp}\over Q}\Big)\baru(p,\sigma)\ssn_2\gamma^5
\Big({1\over 2s p^+}\calf_{z_1}U^\dagger_{z_2} - {1\over 2s k^+}U_{z_1}\calf^\dagger_{z_2}\Big)v(k,\sigma')
\label{GluonME-2}
\end{eqnarray}
Squaring the sum of the matrix element of the eikonal, eq. (\ref{dipoleikonallongy}), and the Gluon-sub-eikonal, eq. (\ref{GluonME-2})
with longitudinal photon polarization, we have
\begin{eqnarray}
&&\hspace{-2cm} {1\over 2\pi\delta(0)}\int\!\dhd^4k\dhd^4p\,\dbar(k^2)\dbar(p^2) \theta(p^+)\theta(k^+)
\Big|\langle q(p)\barq(k)|\gamma_L^*(q)\rangle_{\rm eik+Gluon}\Big|^2
\nonumber\\
&&\hspace{-2cm}
= \calm^L_{\rm eikonal} + \calm^L_{\rm Gluon} + O(\lambda^{-2})
\label{squaringGluon}
\end{eqnarray}
where we divided by the usual infinite volume $2\pi\delta(0)$ which will cancel one of the two delta-functions times $(2\pi)$ coming from squaring the amplitude.
In eq. (\ref{squaringGluon}), $\calm^L_{\rm eikonal}$ is the dipole cross-section with longitudinal polarization, eq. (\ref{LOdipoleLongy}), 
and $\calm^L_{\rm Gluon}$ is
\begin{eqnarray}
&&\calm^L_{\rm Gluon}
\nonumber\\
\equiv\!\!\!&& \int\!\dhd^4k\dhd^4p\,\dbar(k^2)\dbar(p^2) \theta(p^+)\theta(k^+)\dbar\!\left({p^+\over q^+} + {k^+\over q^+} -1\right)
\nonumber\\
&&\times \sum_{f,\sigma,\sigma'}
e^2e^2_f {2\over s^2} \int d^2z_1 d^2z_2 \dhd^2 q_1
{{2\over s}Q p^+k^+ - {q_{1\perp}^2\over Q}\over q^2_{1\perp} + {2\over s}Q^2 p^+k^+ }
\,e^{ i(q_1-p,z_1)_\perp - i (k + q_1,z_2)}
\nonumber\\
&&\times\!\!
\int d^2z_3 d^2z_4 \dhd^2 q_2
{{2\over s}Q p^+k^+ - {q_{2\perp}^2\over Q}\over q^2_{2\perp} + {2\over s}Q^2 p^+k^+ }
\,e^{ - i(q_2-p,z_3)_\perp + i (k + q_2,z_4)}
\nonumber\\
&&\times \Bigg\{
\baru(p, \sigma)\Big(U_{z_1}U^\dagger_{z_2} -1\Big)\ssn_2 v(k, \sigma') 
\bigg[\baru^i(p, \sigma)\ssn_2\gamma^5\Big({1\over 2p^+ }\calf_{z_3}U^\dagger_{z_4}
- {1\over 2k^+ }U_{z_3}\calf^\dagger_{z_4} \Big)v(k, \sigma')\bigg]^\dagger
\nonumber\\
&& + \baru(p, \sigma)\ssn_2\gamma^5\Big({1\over 2p^+}\calf_{z_1}U^\dagger_{z_2} 
- {1\over 2k^+ }U_{z_1}\calf^\dagger_{z_2}\Big)v(k, \sigma')
\bigg[\baru(p, \sigma)\Big(U_{z_3}U^\dagger_{z_4} -1\Big)\ssn_2v(k,\sigma')\bigg]^\dagger
\Bigg\}
\nonumber\\
&& + O(\lambda^{-2})
\end{eqnarray}
So, summing over the helicity $\sigma,\sigma'$, we have that square matrix element with longitudinal polarization for the operators
$U_{z_1}\calf_{z_2}^\dagger$ and $\calf_{z_1}U^\dagger_{z_2}$ is proportional to $\tr\{\ssp\ssn_2\gamma^5\ssk\ssn_2\} = 0$, 
so it does not contribute.

\subsubsection{Transverse polarization with $\calf(z_\perp)$ operator}

Let us consider the transverse photon polarization $\varepsilon^k_\lambda = -{1\over \sqrt{2}}(\lambda,i)$ with transverse index $k=1,2$.
The scattering amplitude, eq. (\ref{GluonME-1}), with transverse polarization is proportional to the following Dirac matrix element
\begin{eqnarray}
&&\baru^i(p,\sigma)[\ssn_2\gamma^5(p^+\ssn_1+\ssq_{1\perp})]\sslash{\varepsilon}_\perp[(k^+\ssn_1-\ssq_{1\perp})\ssn_2]v^l(k,\sigma')
\nonumber\\
=\!\!\!&&2\baru^i(p,\sigma)\Big[(p^+ - k^+)\ssn_2\gamma^5(\varepsilon,q_1)_\perp 
+ i (p^+ + k^+)\ssn_2\vec{\varepsilon}_\perp\times \vecq_1\Big]v^l(k,\sigma')\,.
\label{sub-diracCalF-transy}
\end{eqnarray}
With result (\ref{sub-diracCalF-transy}), the Gluon contribution with transverse polarization is
\begin{eqnarray}
&&\langle q(p)\barq(k)|\gamma_T^*(q)\rangle_{\rm Gluon}
\nonumber\\
=\!\!\!&& - ee_f {2\over s^2}\,\theta(p^+)\theta(k^+)\dbar\!\left({p^+\over q^+} + {k^+\over q^+} -1\right)\!\!\!
\int d^2z_1 d^2z_2 {\dhd^2 q_1\over q^2_{1\perp} + {2\over s}Q^2 p^+k^+}
\,e^{ i(q_1-p,z_1)_\perp - i (k + q_1,z_2)}
\nonumber\\
&&\times
\baru(p,\sigma)\Big[\ssn_2\gamma^5(\varepsilon,q_1)_\perp (p^+-k^+)+ i q^+\ssn_2\vec{\varepsilon}_\perp\times \vecq_1\Big]
\nonumber\\
&&\times
\Big({1\over 2p^+ }\calf_{z_1}U^\dagger_{z_2} - {1\over 2k^+ }U_{z_1}\calf^\dagger_{z_2} \Big)v(k,\sigma')\,.
\label{GluonME-T}
\end{eqnarray}

Let us take the square of the sum of the eikonal dipole amplitude with transverse polarization, eq. (\ref{dipoleikonaltransver}), 
and the Gluon-sub-eikonal term, eq. (\ref{GluonME-T}) 
\begin{eqnarray}
&& {1\over 2\pi\delta(0)}\int\!\dhd^4k\dhd^4p\,\dbar(k^2)\dbar(p^2) \theta(p^+)\theta(k^+)
\Big|\langle q(p)\barq(k)|\gamma_T^*(q)\rangle_{\rm eik+Gluon}\Big|^2
\nonumber\\
=\!\!\!&&\int\!\dhd^4k\dhd^4p\,\dbar(k^2)\dbar(p^2) \theta(p^+)\theta(k^+)\dbar\!\left({p^+\over q^+} + {k^+\over q^+} -1\right)
\half\sum_{\lambda=\pm 1}\sum_{f,\sigma,\sigma'}
\nonumber\\
&&\times\Bigg| -  ee_f{2\over s^2} \int d^2z_1 d^2z_2 {\dhd^2 q_1\over q^2_{1\perp} + {2\over s}Q^2 p^+k^+ }
\,e^{ i(q_1-p,z_1)_\perp - i (k + q_1,z_2)}
\nonumber\\
&&\times
\Bigg[
s\,\baru(p,\sigma)\Big(U_{z_1}U^\dagger_{z_2}-1\Big) 
\Big((p^+ - k^+)(\varepsilon, q_1)_\perp\ssn_2 v(k,\sigma') - i q^+(\vec{\varepsilon}_\perp\times \vec{q}_1)\gamma^5\ssn_2 v(k,\sigma')\Big)
\nonumber\\
&&
+\baru(p,\sigma)\Big[\ssn_2\gamma^5(\varepsilon,q_1)_\perp(p^+ - k^+) + i q^+\ssn_2\vec{\varepsilon}_\perp\times \vecq_1\Big]
\Big({1\over 2p^+ }\calf_{z_1}U^\dagger_{z_2} - {1\over 2k^+ }U_{z_1}\calf^\dagger_{z_2}\Big)v(k,\sigma')\Bigg]
\Bigg|^2 
\nonumber\\
&&+O(\lambda^{-2})
\nonumber\\
=\!\!\!&& \calm^T_{\rm eikonal} + \calm^T_{\rm Gluon} + O(\lambda^{-2})
\end{eqnarray}
where $\calm^T_{\rm eikonal}$ is the eikonal contribution given in eq. (\ref{LOdipoleTransver}), and $\calm^T_{\rm Gluon}$ is
\begin{eqnarray}
&&\hspace{-2cm}\calm^T_{\rm Gluon} \equiv  - {2\,q^+\,i\over \pi s^2}\sum_fe^2e^2_f\int_0^1 dz(z-\barz) \int d^2z_1 d^2z_2 \,
{\dhd^2 q_1\dhd^2 q_2\,e^{i(q_1-q_2,z_2-z_1)}\over [q^2_{1\perp} + Q^2 z\barz][q^2_{2\perp} + Q^2 z\barz]}
\nonumber\\
&&\hspace{-0.5cm}
~~\times \half\!\!\sum_{\lambda=\pm 1}\Big[\Big((q_2,\varepsilon^*_\lambda)\vec{\varepsilon}_\lambda\times \vecq_1 - (q_1,\varepsilon_\lambda)\vec{\varepsilon^*_\lambda}\times \vecq_2\Big)
- \Big((q_2,\varepsilon_\lambda)\vec{\varepsilon}^*_\lambda\times \vecq_1 - (q_1,\varepsilon^*_\lambda)\vec{\varepsilon_\lambda}\times \vecq_2\Big)\Big]
\nonumber\\
&&\hspace{-0.5cm}
~~\times\Big({1\over 2z}\Tr\{\calf_{z_1}U^\dagger_{z_2}\} - {1\over 2\barz}\Tr\{U_{z_1}\calf^\dagger_{z_2}\}\Big)
\label{dipolesubGluon-1}
\end{eqnarray}
where, to get eq. (\ref{dipolesubGluon-1}), we summed over the helicity $\sigma,\sigma'$ and used the traces of Dirac matrices
$\tr\{\ssp\ssn_2\gamma^5\ssk\ssn_2\gamma^5\}=$ $\tr\{\ssp\ssn_2\ssk\ssn_2\} = 8p^+k^+$
and $\tr\{\ssp\ssn_2\gamma^5\ssk\ssn_2\}=0$.
Let us sum over $\lambda =\pm 1$, and use
\begin{eqnarray}
&&\sum_{\lambda=\pm1}\Big[(q_2,\varepsilon^*_\lambda)\vec{\varepsilon}_\lambda\times \vecq_1 
- (q_1,\varepsilon_\lambda)\vec{\varepsilon^*_\lambda}\times \vecq_2\Big]
= 2\vecq_2\times \vecq_1
\end{eqnarray}
and arrive at
\begin{eqnarray}
\calm^T_{\rm Gluon} 
=\!\!\!&& - {4\,q^+\,i\over \pi s^2}\sum_fe^2e^2_f\int_0^1 dz(z-\barz) \int d^2z_1 d^2z_2 \,
{\dhd^2 q_1\dhd^2 q_2\,e^{i(q_1-q_2,z_2-z_1)}\over [q^2_{1\perp} + Q^2 z\barz][q^2_{2\perp} + Q^2 z\barz]}
\vecq_2\times \vecq_1
\nonumber\\
&&\hspace{-0.3cm}
~~\times\!
\Big({1\over 2z}\Tr\{\calf_{z_1}U^\dagger_{z_2}\} - {1\over 2\barz}\Tr\{U_{z_1}\calf^\dagger_{z_2}\}\Big)
\nonumber\\
&& \hspace{-0.5cm} = 0
\label{dipolesubGluon-2}
\end{eqnarray}
So, the sub-eikonal operator $\calf(z_\perp)$, does not contribute to the unpolarized DIS structure functions.
This is consistent with the fact that operators with different parity do not mix.
Next, we consider the asymmetry.

\subsubsection{Asymmetry polarization with $\calf(z_\perp)$ operator}

In this subsection, we consider the asymmetry contribution due to the sub-eikonal term $\calf(z_\perp)$.
To calculate the asymmetry contribution, instead of summing over the helicity $\lambda$, we have to consider the following difference
\begin{eqnarray}
&&\sum_{\lambda = {\rm asym.}}
\Big[(q_2,\varepsilon^*_\lambda)\vec{\varepsilon}_\lambda\times \vecq_1 - (q_1,\varepsilon_\lambda)\vec{\varepsilon^*_\lambda}\times \vecq_2\Big]
= - 2i (q_1,q_2)
\label{asymmetrysum2}
\end{eqnarray}
Thus, the contribution of $\calf(z_\perp)$ to the asymmetry, 
which is obtained using (\ref{asymmetrysum2}) in (\ref{dipolesubGluon-1}) instead of the averaged sum over $\lambda$, is
\begin{eqnarray}
&&\hspace{-2cm}\calm^A_{\rm Gluon} \equiv - 8\,q^+\sum_f{e^2e^2_f\over \pi s^2}\int_0^1 dz\,(z-\barz) \int d^2z_1 d^2z_2 \,
{\dhd^2 q_1\dhd^2 q_2\,e^{i(q_1-q_2,z_2-z_1)}\over [q^2_{1\perp} + Q^2 z\barz][q^2_{2\perp} + Q^2 z\barz]}
\nonumber\\
&&
\times(q_1,q_2)
\bigg({1\over 2z}\Tr\{\calf_{z_1}U^\dagger_{z_2}\} - {1\over 2\barz}\Tr\{U_{z_1}\calf^\dagger_{z_2}\} \bigg)
\label{asymmGluonsub-1}
\end{eqnarray}
Symmetrizing the result with respect to $z\leftrightarrow\barz$, which means quark and anti-quark symmetry,
result (\ref{asymmGluonsub-1}) can be put in the form
\begin{eqnarray}
\calm^A_{\rm Gluon}=\!\!\!&& {2\, Q^2\alpha_{\rm em}\over \pi^2}\sum_fe^2_f\int_0^1 dz \, (z-\barz)^2\int d^2z_1 d^2z_2 
[K_1(\bar{Q}|z_{12}|)]^2
\nonumber\\
&&\times{\sqrt{s/2}\over s^2}\Big(\Tr\{\calf_{z_1}U_{z_2}^\dagger\} + \Tr\{\calf_{z_1}^\dagger U_{z_2}\}\Big)
\label{asymmGluonsub}
\end{eqnarray}
Result (\ref{asymmGluonsub}) is one of the results of this paper and it represents the contribution of the $\calf(z_\perp)$
operator to the structure function $g_1$.
In the next section we are going to consider the contribution of quark field in the background.

\subsection{Quark sub-eikonal corrections}
Finally, we consider the sub-eikonal correction in which the background field contains
quark fields, eq. (\ref{diolesubeikoQ}):
\begin{eqnarray}
&&\langle q(p)\barq(k)|\gamma^*(q)\rangle_{\rm Quark}
\nonumber\\
=\!\!\!&& - ee_f {2\over s^2}\, \theta(p^+)\theta(k^+)\dbar\!\left({p^+\over q^+} + {k^+\over q^+} - 1\right) \!\!
\int d^2z_1 d^2z_2 {\dhd^2 q_1\over q^2_{1\perp} + {2\over s}Q^2 p^+k^+ }
e^{i(q_1-p,z_1)_\perp - i (k + q_1,z_2)_\perp}
\nonumber\\
&&\hspace{0.3cm}\times
\Bigg\{
{1\over 2p^+}\baru(p,\sigma)\gamma_\perp^\mu Q(z_{1\perp})U^\dagger_{z_2}\gamma^\perp_\mu 
[p^+\ssn_1 + \ssq_{1\perp}]\sslash{\varepsilon}(q)[k^+\ssn_1-\ssq_{1\perp}]\ssn_2v(k,\sigma')
\nonumber\\
&&\hspace{0.7cm}
-{1\over 2k^+}\baru(p,\sigma)\ssn_2[p^+\ssn_1 + \ssq_{1\perp}]\sslash{\varepsilon}(q)[k^+\ssn_1-\ssq_{1\perp}]
\gamma_\perp^\mu U_{z_1}\tilde{Q}(z_{2\perp})\gamma^\perp_\mu v(k,\sigma')
\Bigg\}
\label{diolesubeikoQb}
\end{eqnarray}
This gives the fermionic contribution to the dipole cross-section at
sub-eikonal level. 
In the next sections, like we did for the gluon contribution,  we will calculate the longitudinal and transverse polarization of eq. (\ref{diolesubeikoQb}).
\subsubsection{Longitudinal polarization for Quark contribution}

The longitudinal polarization component of (\ref{diolesubeikoQb}) 
is proportional to the following two Dirac matrix elements\footnote{Note that, to obtain eqs.~(\ref{quarkmatrix1}) and~(\ref{quarkmatrix2}), we have neglected
	sub-sub-eikonal terms using~$\ssn_2\ssn_1\psi~=~\bigl(2-\ssn_1\ssn_2\bigr)\psi \simeq 2\psi. $
	If this approximation is not performed at this stage, more terms have to be calculated
	when multiplying the amplitude by the complex-conjugate of the eikonal amplitude. 
However, after retaining only the sub-eikonal contribution, one recovers the same result
as that obtained by performing the approximation already used in deriving
eqs.~(\ref{quarkmatrix1}) and~(\ref{quarkmatrix2}).}
\begin{eqnarray}
&&\baru^i(p)\ssn_2(p^+\ssn_1+\ssq_{1\perp})\sslash{\varepsilon}^L(q)(k^+\ssn_1-\ssq_{1\perp})\gamma^\mu_\perp 
[U_{z_1}\tilde{Q}_{z_2}]^{il}\gamma^\perp_\mu v^l(k)
\nonumber\\
=\!\!\!&&  2q^+\Big({2\over s}Qp^+k^+ - {q_{1\perp}^2\over Q}  \Big)\baru^i(p)\gamma_\perp^\mu 
[U_{z_1}\tilde{Q}_{z_2}]^{il}\gamma_\mu^\perp v^l(k) + O(1/\lambda^2)
\label{quarkmatrix1}
\end{eqnarray}
and
\begin{eqnarray}
&&\baru^i(p)[\gamma^\mu_\perp [Q_{z_1}U^\dagger_{z_2}]^{il}\gamma^\perp_\mu
(p^+\ssn_1+\ssq_{1\perp})]\sslash{\varepsilon}^L(q)[(k^+\ssn_1-\ssq_{1\perp})\ssn_2]v^l(k) 
\nonumber\\
=\!\!\!&& 2q^+\Big({2\over s}Qp^+k^+
- { q^2_{1\perp}\over Q} \Big)\baru^i(p)\gamma^\mu_\perp [ Q_{z_1}U^\dagger_{z_2}]^{il}\gamma^\perp_\mu v^l(k)
+ O(1/\lambda^2)
\label{quarkmatrix2}
\end{eqnarray}
Using (\ref{quarkmatrix1}) and (\ref{quarkmatrix2}), the longitudinal contribution to the dipole amplitude becomes
\begin{eqnarray}
&&\langle q(p)\barq(k)|\gamma_L^*(q)\rangle_{\rm Quark}
\nonumber\\
=\!\!\!&& - ee_f {2\over s^2}\theta(p^+)\theta(k^+) \dbar\!\left({p^+\over q^+} + {k^+\over q^+} - 1\right)
\nonumber\\
&&\times\!\int d^2z_1 d^2z_2 
\dhd^2 q_1{{2\over s}Q p^+k^+ - {q_{1\perp}^2\over Q}\over q^2_{1\perp} + {2\over s}Q^2 p^+k^+}
\,e^{ i(q_1-p,z_1)_\perp - i (k + q_1,z_2)}
\nonumber\\
&&\times
\Bigg[
\baru(p,\sigma)\gamma^\perp_\mu \Big({q^+\over p^+} [Q_{z_1}U^\dagger_{z_2}] 
- {q^+\over k^+}[U_{z_1}\tilde{Q}_{z_2}]\Big)\gamma_\perp^\mu v(k,\sigma')
\Bigg]
\label{dipolesubeikoQlongy}
\end{eqnarray}
We now have to square the sum of the eikonal longitudinal dipole amplitude, eq. (\ref{dipoleikonallongy}),
and the sub-eikonal quark contribution, eq. (\ref{dipolesubeikoQlongy}), and obtain
\begin{eqnarray}
&& {1\over 2\pi\delta(0)}\int\!\dhd^4k\dhd^4p\,\dbar(k^2)\dbar(p^2) \theta(p^+)\theta(k^+)
\Big|\langle q(p)\barq(k)|\gamma_L^*(q)\rangle_{\rm eikonal+Quark}\Big|^2
\nonumber\\
\hspace{-2cm}
=\!\!&&\int\!\dhd^4k\dhd^4p\,\dbar(k^2)\dbar(p^2) \theta(p^+)\theta(k^+)\dbar\!\left({p^+\over q^+} + {k^+\over q^+} - 1\right)
\nonumber\\
&&\times
\Bigg|- ee_f {2\over s^2}\,
\int d^2z_1 d^2z_2 
\dhd^2 q_1{{2\over s}Q p^+k^+ - {q_{1\perp}^2\over Q}\over q^2_{1\perp}+ {2\over s}Q^2 p^+k^+ }
\,e^{ i(q_1-p,z_1)_\perp - i (k + q_1,z_2)}
\nonumber\\
&&\times
\bigg[
s\,q^+\baru(p,\sigma)\Big(U_{z_1}U^\dagger_{z_2} -1\Big)\ssn_2 v(k,\sigma') 
\nonumber\\
\hspace{-2cm}&&
+q^+\baru^i(p,\sigma)\gamma^\mu_\perp \Big[{1\over p^+} [Q_{z_1}U^\dagger_{z_2}]^{il} 
- {1\over k^+}[U_{z_1}\tilde{Q}_{z_2}]^{il}\Big]\gamma^\perp_\mu v^l(k,\sigma')\bigg]
\Bigg|^2 + O(\lambda^{-2})
\label{sqareLquark}
\end{eqnarray}
In the product we need terms only up to $\lambda^{-1}$, so from (\ref{sqareLquark}) we have
\begin{eqnarray}
&&{1\over 2\pi\delta(0)}\int\!\dhd^4k\dhd^4p\,\dbar(k^2)\dbar(p^2) \theta(p^+)\theta(k^+)
\Big|\langle q(p)\barq(k)|\gamma_L^*(q)\rangle_{\rm eikonal+Quark}\Big|^2
\nonumber\\
=\!\!\!&&\calm^L_{\rm eikonal} + \calm^{q\barq\,,L}_{\rm Quark}
\end{eqnarray}
where $\calm^L_{\rm eikonal}$ is given in eq. (\ref{LOdipoleLongy}), and the dipole-quark contribution to the polarized amplitude square, 
$\calm^{q\barq\,,L}_{\rm Quark}$, is
\begin{eqnarray}
 \calm^{q\barq\,,L}_{\rm Quark} \equiv\!\!\!&& 
 {1\over 2\pi s^3\pi}\sum_{f,\sigma,\sigma'} e^2e^2_f \int_0^1{dz\over z\barz}
 \int d^2z_1 d^2z_2 \dhd^2 q_1\dhd^2 q_2\,e^{ i(q_1-q_2,z_1-z_2)_\perp }
\nonumber\\
&&\times 
{{2\over s}Q p^+k^+ - {q_{1\perp}^2\over Q}\over q^2_{1\perp} + {2\over s}Q^2 p^+k^+ }
{{2\over s}Q p^+k^+ - {q_{2\perp}^2\over Q}\over q^2_{2\perp}+ {2\over s}Q^2 p^+k^+ }
 \Bigg\{
\baru(p,\sigma)\Big(U_{z_1}U^\dagger_{z_2} -1\Big)\ssn_2 v(k,\sigma') 
\nonumber\\
&&~~~\times\bigg[\baru(p,\sigma)\gamma_\mu^\perp \Big[{q^+\over p^+} [Q_{z_1}U^\dagger_{z_2}] 
- {q^+\over k^+}[U_{z_1}\tilde{Q}_{z_2}]\Big]\gamma_\perp^\mu v(k,\sigma')\bigg]^\dagger
\nonumber\\
&&~+ \bigg[\baru(p,\sigma)\gamma_\mu^\perp \Big[{q^+\over p^+} [Q_{z_1}U^\dagger_{z_2}] 
- {q^+\over k^+}[U_{z_1}\tilde{Q}_{z_2}]\Big]\gamma_\perp^\mu v(k,\sigma')\bigg]
\nonumber\\
&&~~~\times\bigg[\baru(p,\sigma)\Big(U_{z_1}U^\dagger_{z_2} -1\Big)\ssn_2v(k,\sigma')\bigg]^\dagger
\Bigg\}
\label{QdipoleLongy}
\end{eqnarray}
Summing over helicity $\sigma,\sigma'$, we get two Dirac matrices from which we keep again the leading contribution in 
large boost parameter $\lambda$
\begin{eqnarray}
&&\hspace{-1cm}\sum_{\sigma,\sigma'}\,\baru(p,\sigma)\Big(U_{z_1}U^\dagger_{z_2} -1\Big)\ssn_2 v(k,\sigma') 
\bigg[\baru(p,\sigma)\gamma^\mu_\perp 
\Big[ {1\over p^+} [Q_{z_1}U^\dagger_{z_2}] - {1\over k^+}[U_{z_1}\tildeQ_{z_2}]  \Big]\gamma^\perp_\mu v(k,\sigma')\Bigg]^\dagger
\nonumber\\
&&\hspace{-1cm}
= {1\over p^+}\Tr\{\Big(U_{z_1}U^\dagger_{z_2} -1\Big)U_{z_2}\tr\{\ssp\ssn_2\ssk\gamma^\mu_\perp \tilde{Q}_{z_1}\gamma^\perp_\mu\}\}
- {1\over k^+}\Tr\{\Big(U_{z_1}U^\dagger_{z_2} -1\Big)\tr\{\ssp\ssn_2\ssk\gamma^\mu_\perp Q_{z_2}\gamma^\perp_\mu\}U^\dagger_{z_1}\}
\nonumber\\
&&\hspace{-1cm}
=4\,p^+k^+\Bigg[
{1\over p^+}\Big(\Tr\{U_{z_2}\calq^\dagger_{1z_1}\} + C_FQ^\dagger_{1z_1}\Big)
- {1\over k^+}\Big(\Tr\{\calq_{1z_2}U^\dagger_{z_1}\} + C_F Q_{1z_2}\Big)
\Bigg]
+ O(\lambda^{-2})
\label{Diracalgebra1}
\end{eqnarray}
where, to get (\ref{Diracalgebra1}), we used
\begin{eqnarray}
&&\tr\{\ssp\ssn_2\ssk\gamma_\perp^\mu Q_{z_2}\gamma_\mu^\perp\}
\nonumber\\
&& = - 4p^+k^+\tr\{\ssn_1 Q_{z_2}\} - 2(k,p)_\perp\tr\{\ssn_2 Q_{z_2}\} + 2i(\vecp\times\veck)\tr\{\ssn_2\gamma^5Q_{z_2}\}
\nonumber\\
&&= - 4p^+k^+\tr\{\ssn_1 Q_{z_2}\} + \calo(1/\lambda^2)
\label{DiractrQ1a}
\end{eqnarray}
and defined (recall we use $\tr$ for trace over spinor index and $\Tr$ for trace over color index in the fundamental representation)
\begin{eqnarray}
&&\hspace{-0.4cm}\calq_{1\,ij}(x_\perp) \equiv \tr\{\ssn_1 Q_{ij}(x_\perp)\} 
\nonumber\\
=\!\!\!&& g^2{s\over 2} \!\!\int_{-\infty}^{+\infty}\!\!\! dz^+\! \int_{-\infty}^{z^+}\!\!dz'^+
\nonumber\\
&&\times\Big([\infty n_1,z^+]_x t^a\,\tr\big\{\ssn_1\,  \psi(z^+,x_\perp)[z^+,z'^+]^{ab}_x\bar{\psi}(z'^+,x_\perp)\big\} 
t^b[z'^+,-\infty n_1]_x\Big)_{\!\!ij}~~~~~
\end{eqnarray}
where the operator $Q_{ij}$ is defined in eq. (\ref{Q}).
Similarly,
\begin{eqnarray}
&&\hspace{-1cm}\sum_{\sigma,\sigma'}\baru(p,\sigma)\gamma^\mu_\perp \Big[{1\over p^+} [Q_{z_1}U^\dagger_{z_2}]
- {1\over k^+}[U_{z_1}\tildeQ_{z_2}]\Big]\gamma^\perp_\mu v(k,\sigma')
\Bigg[\baru(p,\sigma)\Big(U_{z_1}U^\dagger_{z_2} -1\Big)\ssn_2 v(k,\sigma') \Bigg]^\dagger
\nonumber\\
&&\hspace{-1cm} = 
{1\over p^+}\Tr\{U^\dagger_{z_2}\Big(U_{z_2}U^\dagger_{z_1}-1\Big)\tr\{Q_{z_1}\gamma^\perp_\mu\ssk\ssn_2\ssp\gamma_\perp^\mu\}\}
- {1\over k^+}\Tr\{(U_{z_2}U^\dagger_{z_1}-1)U_{z_1}\tr\{\tildeQ_{z_2}\gamma^\perp_\mu\ssk\ssn_2\ssp\gamma^\mu_\perp\}\}
\nonumber\\
&&\hspace{-1cm} = 4p^+k^+ \Bigg[{1\over p^+}\Big(\Tr\{U^\dagger_{z_2}\calq_{1z_1}\} + C_FQ_{1z_1}\Big)
- {1\over k^+}\Big(\Tr\{U_{z_1}\calq^\dagger_{1z_2}\} + C_F Q^\dagger_{1z_2}\Big)
\Bigg]
+ O(\lambda^{-2})
\label{Diracalgebr2}
\end{eqnarray}
where, to get (\ref{Diracalgebr2}), we used
\begin{eqnarray}
&&\tr\{\ssk\ssn_2\ssp\gamma_\perp^\mu \tilde{Q}_{z_2}\gamma_\mu^\perp\}
\nonumber\\
&& = - 4p^+k^+\tr\{\ssn_1 \tilde{Q}_{z_2}\} - 2(k,p)_\perp\tr\{\ssn_2 \tilde{Q}_{z_2}\} + 2i(\veck\times\vecp)\tr\{\ssn_2\gamma^5\tilde{Q}_{z_2}\}
\nonumber\\
&&= - 4p^+k^+\tr\{\ssn_1 \tilde{Q}_{z_2}\} + \calo(1/\lambda^2)
\label{DiractrQ1b}
\end{eqnarray}
and defined
\begin{eqnarray}
&&\calq^\dagger_{1\,ij}(x_\perp) \equiv \tr\{\ssn_1 \tildeQ_{ij}(x_\perp)\} 
\nonumber\\
=\!\!\!&& g^2{s\over 2}\!\!\int_{-\infty}^{+\infty}\!\!\! dz^+\! \int^{+\infty}_{z^+}\!\!dz'^+
\nonumber\\
&&~~\times\Big([-\infty n_1,z^+]_x \, t^a \tr\{\ssn_1\,\psi(z^+,x_\perp)
[z^+,z'^+]^{ab}_x\bar{\psi}(z'^+,x_\perp)\} t^b[z'^+,+\infty n_1]_x\Big)_{\!\! ij}~~~~~
\end{eqnarray}
where, again, we indicate explicitly the color indexes in the fundamental representation with $i,j$, and the operator $\tildeQ_{ij}$ is defined 
in eq. (\ref{tildeQ}).

When the transverse coordinates coincide, we also have
\begin{eqnarray}
&&\Tr\{U^\dagger_{z_1}\calq_{1z_1}\}
\nonumber\\
=\!\!\!&& - g^2C_F{s\over 2}\!\!\int_{-\infty}^{+\infty}\!\!dz^+\!\!\int_{-\infty}^{z^+}\!\! dz'^+
\barpsi(z'^+,z_{1\perp})\ssn_1[z'^+,z^+]_{z_{1\perp}}\psi(z^+,z_{1\perp}) = - C_F Q_1(z_{1\perp})~~~~~
\label{trCfQ1}
\end{eqnarray}
and
\begin{eqnarray}
&&\Tr\{U_{z_1}\calq^\dagger_{1z_1}\}
\nonumber\\
=\!\!\!&& - g^2C_F{s\over 2}\!\!\int_{-\infty}^{+\infty}\!\!dz^+\!\!\int_{-\infty}^{z^+}\!\! dz'^+
\barpsi(z^+,z_{1\perp})\ssn_1[z^+,z'^+]_{z_{1\perp}}\psi(z'^+,z_{1\perp}) = - C_F Q^\dagger_1(z_{1\perp})~~~~~
\label{trCfQ1dag}
\end{eqnarray}
where
$Q_1(z_\perp) = Q_1(z_\perp,x_B=0)$, and $Q^\dagger_1(z_\perp) = \bar{Q}_1(z_\perp,x_B=0)$,
and
\begin{eqnarray}
&&\hspace{-1.3cm}Q_{1,f}(x_\perp,x_B) \equiv 
g^2{s\over 2}\!\!\int_{-\infty}^{+\infty}\!\! dx^+ \!\!\int_{-\infty}^{x^+}\!\! dy^+ \,e^{ix_B P^-\Delta^+}
\barpsi_f(y^+,x_\perp)[y^+,x^+]_x\ssn_1\psi_f(x^+,x_\perp)\,,
\label{Q1}
\end{eqnarray}
and 
\begin{eqnarray}
&&\hspace{-1.3cm}\bar{Q}_{1,f}(x_\perp,x_B) \equiv 
g^2{s\over 2}\!\!\int_{-\infty}^{+\infty}\!\! dy^+ \!\!\int_{-\infty}^{y^+}\!\! dx^+ \,e^{ix_B P^-\Delta^+}
\,\barpsi_f(y^+,x_\perp)[y^+,x^+]_x\ssn_1\psi_f(x^+,x_\perp)\,.
\label{Q1dagger}
\end{eqnarray}
with $\Delta^+=x^+-y^+$. Operators \eqref{Q1} and \eqref{Q1dagger} were obtained in
ref.~\cite{Chirilli:2026pkv}, and at $x_B=0$ in ref.~\cite{Chirilli:2021lif}. They are the operators
that appear in the first sub-eikonal correction to the DIS cross section,
given by the one-particle final states $\gamma^*\to q$ and
$\gamma^*\to\bar q$~\cite{Chirilli:2026pkv}. There the bilocal quark operator enters with
the factor $2/(g^2 s)$ implied by the normalization of eqs.~\eqref{Q1} and
\eqref{Q1dagger}, while in the $q\bar q$ channel considered here it enters
through the quark-background propagator \eqref{qprpinquark-1}, i.e., with an
additional factor $g^2$.

From eq. (\ref{QdipoleLongy}), using (\ref{Diracalgebra1}) and (\ref{Diracalgebr2}), we arrive at
\begin{eqnarray}
\calm^{q\barq\,,L}_{\rm Quark} =\!\!\!&&
{q^+\over \pi}\int_0^1 dz\,
\sum_f {e^2e^2_f\over s^2} \int d^2z_1 d^2z_2 \dhd^2 q_1\dhd^2 q_2\,e^{ i(q_1 - q_2, z_2-z_1)_\perp }
\nonumber\\
&&\times{Q z\barz - {q_{1\perp}^2\over Q}\over q^2_{1\perp}+Q^2 z\barz }
{Q z\barz - {q_{2\perp}^2\over Q}\over q^2_{2\perp}+Q^2 z\barz }
\Bigg[{1\over z}\Big(\Tr\{U_{z_2}\calq^\dagger_{1z_1}\} + C_FQ^\dagger_1(z_{1\perp})\Big)
\nonumber\\
&&- {1\over \barz}\Big(\Tr\{\calq_{1z_2}U^\dagger_{z_1}\} +  C_FQ_1(z_{2\perp}) \Big) 
+ {1\over z}\Big(\Tr\{\calq_{1z_1}U^\dagger_{z_2}\} +  C_FQ_1(z_{1\perp}) \Big)
\nonumber\\
&& 
- {1\over \barz}\Big(\Tr\{U_{z_1}\calq^\dagger_{1z_2}\} + C_FQ^\dagger_1(z_{2\perp})\Big)\Bigg]
\label{QdipoleLongy-1}
\end{eqnarray}
We observe that performing a change of dummy variable in the last two terms
of the eq. (\ref{QdipoleLongy-1}) we obtain
\begin{eqnarray}
\calm^{q\barq\,,L}_{\rm Quark} = 0
\label{QdipoleLongy-2}
\end{eqnarray}
thus, it does not contribute to the $F_L$ structure function.
In the next section we consider the transverse polarization.

\subsubsection{Transverse polarization for Quark contribution}

The quark contribution to the dipole scattering amplitude with transverse photon polarization is obtained from (\ref{diolesubeikoQb})
\begin{eqnarray}
&&\langle q(p)\barq(k)|\gamma_T^*(q)\rangle_{\rm Quark}
\nonumber\\
=\!\!&& - ee_f {2\over s^2}\, \theta(p^+)\theta(k^+)\dbar\!\left({p^+\over q^+} + {k^+\over q^+} - 1\right) \!\!
\int d^2z_1 d^2z_2 {\dhd^2 q_1\over q^2_{1\perp} + {2\over s}Q^2 p^+k^+ }
\,e^{ i(q_1-p,z_1)_\perp - i (k + q_1,z_2)}
\nonumber\\
&&\times
\Bigg\{
\Big[\epsilon^\mu_{~~\alpha}\, \vecq_1\times\vec{\varepsilon}_\lambda
+ g_{\alpha\perp}^\mu {p^+ - k^+\over q^+}(q_1,\varepsilon_\lambda)_\perp \Big]
\nonumber\\
&&~~~~\times\baru^i(p,\sigma)\gamma^\perp_\mu\Big(
{q^+\over p^+}[Q_{z_1}U^\dagger_{z_2}]^{il}
- {q^+\over k^+}[U_{z_1}\tilde{Q}_{z_2}]
\Big)\gamma^\alpha_\perp v^l(k,\sigma')
\Bigg\}
\label{Quark-trans-aplitude}
\end{eqnarray}
The square of the sum of the eikonal dipole scattering amplitude with transverse polarization eq. (\ref{dipoleikonaltransver}),
and the sub-eikonal (\ref{Quark-trans-aplitude}), is
\begin{eqnarray}
&&{1\over 2\pi\delta(0)} \int\!\dhd^4k\dhd^4p\,\dbar(k^2)\dbar(p^2)
\Big|\langle q(p)\barq(k)|\gamma^*_T(q)\rangle_{\rm eikonal+Quark}\Big|^2
\nonumber\\
=\!\!\!&& \int\!\dhd^4k\dhd^4p\,\dbar(k^2)\dbar(p^2) \theta(p^+)\theta(k^+) \dbar\!\left({p^+\over q^+} + {k^+\over q^+} - 1\right)
\half\!\!\sum_{\lambda=\pm 1}\sum_{f,\sigma,\sigma'}
\nonumber\\
&&\times\Bigg| - ee_f {2\over s^2} \int d^2z_1 d^2z_2 {\dhd^2 q_1\over q^2_{1\perp} + {2\over s}Q^2 p^+k^+}
\,e^{ i(q_1-p,z_1)_\perp - i (k + q_1,z_2)}
\nonumber\\
&&\times
\Bigg[
s\,\baru(p,\sigma)\Big(U_{z_1}U^\dagger_{z_2}-1\Big)
\Big((p^+-k^+)(\varepsilon, q_1)_\perp\ssn_2 v(k,\sigma') - iq^+(\vec{\varepsilon}_\perp\times \vec{q}_1)\gamma^5\ssn_2 v(k,\sigma')\Big)
\nonumber\\
&&+ \Big[\epsilon^\mu_{~~\alpha}\, \vecq_1\times\vec{\varepsilon}_\lambda
+ g_{\alpha\perp}^\mu {p^+ - k^+\over q^+}(q_1,\varepsilon_\lambda)_\perp\Big]
\baru(p)\gamma^\perp_\mu\Big({q^+\over p^+}[Q_{z_1}U^\dagger_{z_2}]
- {q^+\over k^+}[U_{z_1}\tilde{Q}_{z_2}]\Big)\gamma^\alpha_\perp v(k,\sigma')\Bigg]
\Bigg|^2
\nonumber\\
=\!\!\!&&\calm^T_{\rm eikonal} + \calm^{q\barq\,,T}_{\rm Quark} + O(\lambda^{-2})
\label{Quark-trans-aplitude-1}
\end{eqnarray}
where $\calm^T_{\rm eikonal}$ is eq.  (\ref{LOdipoleTransver}), and $\calm^{q\barq\,,T}_{\rm Quark}$ is the product of eikonal amplitude, eq. (\ref{dipoleikonaltransver}),
times the quark contribution, eq. (\ref{Quark-trans-aplitude}). 
Thus, after performing the integration over $\dhd^4 k$ and $\dhd^4p$, and summing over the helicity $\sigma,\sigma'$, we obtain
\begin{eqnarray}
\calm^{q\barq\,,T}_{\rm Quark}
\equiv && 
{1\over 2s^3}\sum_f e^2e^2_f\int \dhd p^+\dhd k^+ \theta(p^+)\theta(k^+)\dbar\Big({p^+\over q^+}+ {k^+\over q^+} - 1\Big)
{q^+\over p^+k^+} 
\nonumber\\
&&\times\!\int {d^2z_1 d^2z_2 \dhd^2 q_1\dhd^2 q_2 \, e^{i(q_1-q_2,z_1-z_2)_\perp}\over 
\Big(q_{1\perp}^2 + {2\over s}Q^2 p^+k^+\Big)\Big(q_{2\perp}^2 + {2\over s}Q^2 p^+k^+\Big)}
\nonumber\\
&&\times\Bigg[
\tr\{\ssp\ssn_2\ssk\gamma^\beta_\perp\Big({q^+\over p^+}\Tr\{(U_{z_1}U^\dagger_{z_2}-1)[U_{z_2}\tildeQ_{z_1}]\}
- {q^+\over k^+}\Tr\{(U_{z_1}U^\dagger_{z_2}-1)[Q_{z_2}U^\dagger_{z_1}]\}\Big)\gamma^\perp_\nu\}
\nonumber\\
&&\times{p^+-k^+\over q^+}\Big(\vecq_1\times\vecq_2\,\epsilon^{~~\nu}_\beta + {p^+-k^+\over q^+}(q_1,q_2)_\perp g^\nu_{\perp\beta}\Big)
\nonumber\\
&& + i\tr\{\ssp\ssn_2\gamma^5\ssk\gamma^\beta_\perp\Big({q^+\over p^+}\Tr\{(U_{z_1}U^\dagger_{z_2}-1)[U_{z_2}\tildeQ_{z_1}]\}
- {q^+\over k^+}\Tr\{(U_{z_1}U^\dagger_{z_2}-1)[Q_{z_2}U^\dagger_{z_1}]\}\Big)\gamma^\perp_\nu\}
\nonumber\\
&&\times\Big((q_1,q_2)\epsilon^{~~\nu}_\beta - {p^+-k^+\over q^+}\vecq_1\times\vecq_2\,g^\nu_{\perp\beta}\Big)
\nonumber\\
&& + \tr\{\ssp\gamma^\perp_\mu\ \Big({q^+\over p^+}\Tr\{[Q_{z_1}U^\dagger_{z_2}](U_{z_2}U^\dagger_{z_1}-1)\}
- {q^+\over k^+}\Tr\{[U_{z_1}\tildeQ_{z_2}](U_{z_2}U^\dagger_{z_1}-1)\}\Big)\gamma^\alpha_\perp\ssk\ssn_2\}
\nonumber\\
&&\times {p^+-k^+\over q^+}\Big(\vecq_1\times\vecq_2\, \epsilon^\mu_{~~\alpha} + {p^+-k^+\over q^+}(q_1,q_2)_\perp g^\mu_{\perp\alpha}\Big)
\nonumber\\
&& +i\tr\{\ssp\gamma^\perp_\mu\ \Big({q^+\over p^+}\Tr\{[Q_{z_1}U^\dagger_{z_2}](U_{z_2}U^\dagger_{z_1}-1)\}
- {q^+\over k^+}\Tr\{[U_{z_1}\tildeQ_{z_2}](U_{z_2}U^\dagger_{z_1}-1)\}\Big)\gamma^\alpha_\perp\ssk\ssn_2\gamma^5\}
\nonumber\\
&&\times\Big((q_1,q_2)_\perp \epsilon^\mu_{~~\alpha} - {p^+-k^+\over q^+}\vecq_1\times\vecq_2 \, g^\mu_{\perp\alpha}\Big)
\Bigg]
\label{calmQt-a}
\end{eqnarray}

Using the Dirac matrix elements
\begin{eqnarray}
&&\hspace{-2cm}\tr\{\ssp\ssn_2\ssk\gamma_\perp^\beta\Big({q^+\over p^+}\tildeQ_{z_1}
- {q^+\over k^+}Q_{z_2}\Big)\gamma_\nu^\perp\}
= -2p^+k^+\tr\{\ssn_1\Big(g^\beta_{\perp\nu} + i\gamma^5\epsilon_\nu^{~~\beta}\Big)
\Big({q^+\over p^+}\tildeQ_{z_1} - {q^+\over k^+}Q_{z_2}\Big)\}
\label{Qtransv-tr1}
\end{eqnarray}

\begin{eqnarray}
&&\hspace{-2.5cm}\tr\{\ssp\ssn_2\gamma^5\ssk\gamma^\beta_\perp\Big({q^+\over p^+}\tildeQ_{z_1}
- {q^+\over k^+}Q_{z_2}\Big)\gamma^\perp_\nu\}
 = 2p^+k^+\tr\{\Big(\gamma^5\ssn_1 g^\beta_{\perp\nu} - i\epsilon_\nu^{~~\beta}\ssn_1\Big)
\Big({q^+\over p^+}\tildeQ_{z_1} - {q^+\over k^+}Q_{z_2}\Big)\}
\label{Qtransv-tr2}
\end{eqnarray}
and the sum over the photon helicity 
\begin{eqnarray}
&&\sum_{\lambda=\pm}\Big((p^+-k^+)(\varepsilon,q_1)_\perp\ssn_2 - iq^+\vec{\varepsilon}\times\vecq_1\gamma^5\ssn_2\Big)
\Big(\epsilon^\nu_{~~\beta}\,\vecq_2\times\vec{\varepsilon}^* + (\varepsilon^*,q_2)_\perp g^\nu_{\perp\beta}\,{p^+-k^+\over q^+}\Big)
\nonumber\\
&&= (p^+-k^+)\ssn_2\Big(\vecq_2\times\vecq_1 \epsilon^\nu_{~~\beta} + {p^+-k^+\over q^+}(q_1,q_2)_\perp g^\nu_{\perp\beta}\Big)
\nonumber\\
&&~~+ i\ssn_2\gamma^5q^+\Big({p^+-k^+\over q^+}\vecq_2\times\vecq_1\,g^\nu_{\perp\beta} - (q_1,q_2)_\perp \epsilon^\nu_{~~\beta}\Big)
\label{sumT-helicity1}
\end{eqnarray}
and
\begin{eqnarray}
&&\sum_{\lambda=\pm}\Big((p^+-k^+)(\varepsilon^*,q_2)_\perp\ssn_2 + iq^+\vec{\varepsilon}^*\times\vecq_2\gamma^5\ssn_2\Big)
\Big(\epsilon^\mu_{~~\alpha}\vecq_1\times\vec{\varepsilon} + {p^+-k^+\over q^+}(\varepsilon,q_1)_\perp g^\mu_{\perp\alpha}\Big)
\nonumber\\
&& = (p^+-k^+)\ssn_2\Big(\vecq_1\times\vecq_2 \epsilon^\mu_{~~\alpha} + {p^+-k^+\over q^+}(q_1,q_2)_\perp g^\mu_{\perp\alpha}\Big)
\nonumber\\
&&~~+ i\ssn_2\gamma^5 q^+\Big((q_1,q_2)_\perp \epsilon^\mu_{~~\alpha} - {p^+-k^+\over q^+} \vecq_1\times\vecq_2 g^\mu_{\perp\alpha}\Big) 
\label{sumT-helicity2}
\end{eqnarray}
from \eqref{calmQt-a} we arrive at
\begin{eqnarray}
\hspace{-2cm}\calm^{q\barq\,,T}_{\rm Quark}
=\!\!\!&&
{q^+\over \pi s^2}\sum_f e^2 e^2_f\int_0^1 dz\int {d^2z_1 d^2z_2 \dhd^2 q_1\dhd^2 q_2 \,
 e^{i(q_1-q_2,z_1-z_2)_\perp}\over 
\Big(q_{1\perp}^2 + {2\over s}Q^2 p^+k^+\Big)\Big(q_{2\perp}^2 + {2\over s}Q^2 p^+k^+\Big)}
\nonumber\\
&&\times \Bigg[
(z^2+\barz^2)(q_1,q_2)_\perp\Big({1\over z}\Tr\{U_{z_2}\calq^\dagger_{1z_1}\} + {1\over z}C_FQ^\dagger_{1z_1}
- {1\over \barz}\Tr\{\calq_{1z_2}U^\dagger_{z_1}\} - {1\over \barz}C_FQ_{1z_2}\Big)
\nonumber\\
&& - i(z-\barz)\vecq_2\times \vecq_1\Big({1\over z}\Tr\{U_{z_2}\calq^\dagger_{5z_1}\} + {1\over z}C_FQ^\dagger_{5z_1}
- {1\over \barz}\Tr\{\calq_{5z_2}U^\dagger_{z_1}\} - {1\over \barz}C_FQ_{5z_2}\Big)
\nonumber\\
&&+ (z^2+\barz^2)(q_1,q_2)_\perp\Big({1\over z}\Tr\{\calq_{1z_1}U^\dagger_{z_2}\} + {1\over z}C_FQ_{1z_1}
- {1\over \barz}\Tr\{U_{z_1}\calq^\dagger_{1z_2}\} - {1\over \barz}C_FQ^\dagger_{1z_2}\Big)
\nonumber\\
&&- i (z-\barz)\vecq_2\times\vecq_1\Big({1\over z}\Tr\{\calq_{5z_1}U^\dagger_{z_2}\} + {1\over z}C_FQ_{5z_1}
- {1\over \barz}\Tr\{U_{z_1}\calq^\dagger_{5z_2}\} - {1\over \barz}C_FQ^\dagger_{5z_2}\Big)
\Bigg]
\label{calmQt-b}
\end{eqnarray}
where we used
\begin{eqnarray}
&&\Tr\{\tr\{\ssn_1\Big({1\over z}Q_{z_1}U^\dagger_{z_2} - {1\over \barz}U_{z_1}\tilde{Q}_{z_2}\Big)\}(U_{z_2}U^\dagger_{z_1}-1)\}
\nonumber\\
=\!\!\!&& - {1\over z}\Tr\{{\calq_1}_{z_1} U^\dagger_{z_2}\} + {1\over \barz}\Tr\{U_{z_1}{\calq^\dagger_1}_{z_2}\}
- C_F{1\over z}Q_{1z_1} + C_F{1\over \barz}Q^\dagger_{1z_2}
\label{trQ1}
\end{eqnarray}
and
\begin{eqnarray}
&&\Tr\{\tr\{\gamma^5\ssn_1\Big({1\over z}Q_{z_1}U^\dagger_{z_2} - {1\over \barz}U_{z_1}\tilde{Q}_{z_2}\Big)\}(U_{z_2}U^\dagger_{z_1}-1)\}
\nonumber\\
=\!\!\!&& - {1\over z}\Tr\{{\calq_5}_{z_1}U^\dagger_{z_2}\} + {1\over \barz}\Tr\{{\calq_5}^\dagger_{z_2}U_{z_1}\} 
- C_F{1\over z}Q_{5z_1} + C_F{1\over \barz}Q^\dagger_{5z_2}
\label{trQ5}
\end{eqnarray}
where we defined 
\begin{eqnarray}
\hspace{-0.6cm}\calq_5(z_\perp) \equiv\!\!\!&& g^2{s\over 2}\!\int_{-\infty}^{+\infty}\!\!dz^+\!\int_{-\infty}^{z^+}dz'^+
\nonumber\\
&&\times[\infty n_1,z^+]_zt^a
\tr\{\gamma^5\ssn_1\psi(z^+,z_\perp)[z^+,z'^+]^{ab}_z\barpsi(z'^+,z_\perp)\}t^b[z'^+,-\infty n_1]_z~~~~~~~
\label{calQ5}
\end{eqnarray}
and 
\begin{eqnarray}
&&\hspace{-1.0cm}\Tr\{{\calq_5}_zU^\dagger_z\}
\nonumber\\
&&\hspace{-1.4cm}= - g^2C_F{s\over 2}\int_{-\infty}^{+\infty}\!dz^+\!\int_{-\infty}^{z^+}dz'^+\,
\barpsi(z'^+,z_\perp)\gamma^5\ssn_1[z'^+,z^+]_z\psi(z^+,z_\perp) = - C_FQ_5(z_\perp)
\label{trCfQ5}
\end{eqnarray}
where
$Q_5(z_\perp) = Q_5(z_\perp,x_B=0)$, and $Q^\dagger_5(z_\perp) = \bar{Q}_5(z_\perp,x_B=0)$,
and
\begin{eqnarray}
&&\hspace{-1.5cm}Q_{5,f}(x_\perp,x_B) \equiv 
g^2{s\over 2}\!\!\int_{-\infty}^{+\infty}\!\! dx^+ \!\!\int_{-\infty}^{x^+}\!\! dy^+ \,e^{ix_B P^-\Delta^+}
\barpsi_f(y^+,x_\perp)[y^+,x^+]_x\gamma^5\ssn_1\psi_f(x^+,x_\perp)\,,
\label{Q5}
\end{eqnarray}
and 
\begin{eqnarray}
&&\hspace{-1.3cm}\bar{Q}_{5,f}(x_\perp,x_B) \equiv 
g^2{s\over 2}\!\!\int_{-\infty}^{+\infty}\!\! dy^+ \!\!\int_{-\infty}^{y^+}\!\! dx^+ \,e^{ix_B P^-\Delta^+}
\,\barpsi_f(y^+,x_\perp)[y^+,x^+]_x\gamma^5\ssn_1\psi_f(x^+,x_\perp)\,.
\label{Q5dagger}
\end{eqnarray}
Unlike the vector operators, the axial combination $Q_5+Q_5^\dagger$ is
C-even: in the one-particle final states it survives when the momentum fraction in the operators is set to zero and
gives the leading sub-eikonal contribution to $g_1$~\cite{Chirilli:2026pkv}, to which the
contributions computed below are $\mathcal{O}(\alpha_s)$ corrections.
Now, performing a change of dummy variables in result (\ref{calmQt-b}), we arrive at
\begin{eqnarray}
\calm^{q\barq\,,T}_{\rm Quark} = 0
\label{barqq-transverseQ}
\end{eqnarray}
Therefore, $\calm^{q\barq\,,T}_{\rm Quark}$ does not contribute to the $F_T$ structure function.

\subsubsection{Asymmetry for Quark contribution}

Let us consider the asymmetry for the scattering dipole amplitude with the quark operator, so
instead of the averaged sum over the helicity $\lambda=\pm1$, we consider the 
difference with opposite polarization defined in \eqref{asymsum}.
Thus, from (\ref{Quark-trans-aplitude-1}),  
using 
\begin{eqnarray}
&&\sum_{\lambda={\rm asym.}}
\Big((p^+-k^+)(\varepsilon,q_1)_\perp\ssn_2 - iq^+\vec{\varepsilon}\times\vecq_1\gamma^5\ssn_2\Big)
\Big(\epsilon^\nu_{~~\beta}\,\vecq_2\times\vec{\varepsilon}^* + (\varepsilon^*,q_2)_\perp g^\nu_{\perp\beta}\,{p^+-k^+\over q^+}\Big)
\nonumber\\
&&= i(p^+-k^+)\ssn_2\Big({p^+-k^+\over q^+}\vecq_2\times\vecq_1\,g^\nu_{\perp\beta} - (q_1,q_2)_\perp \epsilon^\nu_{~~\beta}\Big)
\nonumber\\
&&~~+ \ssn_2\gamma^5q^+\Big(\vecq_1\times\vecq_2\, \epsilon^{~~\nu}_\beta + {p^+-k^+\over q^+}(q_1,q_2)_\perp g^\nu_{\perp\beta}\Big)
\label{sumAsym-helicity1}
\end{eqnarray}
and
\begin{eqnarray}
&&\sum_{\lambda={\rm asym.}}
\Big((p^+-k^+)(\varepsilon^*,q_2)_\perp\ssn_2 + iq^+\vec{\varepsilon}^*\times\vecq_2\gamma^5\ssn_2\Big)
\Big(\epsilon^\mu_{~~\alpha}\vecq_1\times\vec{\varepsilon} + {p^+-k^+\over q^+}(\varepsilon,q_1)_\perp g^\mu_{\perp\alpha}\Big)
\nonumber\\
&& = i(p^+-k^+)\ssn_2\Big((q_1,q_2)_\perp \epsilon^\mu_{~~\alpha} - {p^+-k^+\over q^+} \vecq_1\times\vecq_2 g^\mu_{\perp\alpha}\Big)
\nonumber\\
&&~~+ \ssn_2\gamma^5 q^+\Big(\vecq_1\times\vecq_2 \epsilon^\mu_{~~\alpha} + {p^+-k^+\over q^+}(q_1,q_2)_\perp g^\mu_{\perp\alpha}\Big) 
\label{sumAsym-helicity2}
\end{eqnarray}
we have
\begin{eqnarray}
&&\calm^{q\barq\,,A}_{\rm Quark}
\nonumber\\
=\!\!\!\!\!&&
\sum_f e^2e^2_f\int \dhd p^+\dhd k^+ \theta(p^+)\theta(k^+)\dbar\Big({p^+\over q^+}+ {k^+\over q^+} - 1\Big)
\nonumber\\
&&\times\!\int {d^2z_1 d^2z_2 \dhd^2 q_1\dhd^2 q_2 \, e^{i(q_1-q_2,z_1-z_2)_\perp}\over 
\Big(q_{1\perp}^2 + {2\over s}Q^2 p^+k^+\Big)\Big(q_{2\perp}^2 + {2\over s}Q^2 p^+k^+\Big)}{q^+\over s^3p^+k^+} 
\nonumber\\
&&\times\Bigg[
i\,\tr\{\ssp\ssn_2\ssk\gamma^\beta_\perp\Big({q^+\over p^+}\Tr\{(U_{z_1}U^\dagger_{z_2}-1)[U_{z_2}\tildeQ_{z_1}]\}
- {q^+\over k^+}\Tr\{(U_{z_1}U^\dagger_{z_2}-1)[Q_{z_2}U^\dagger_{z_1}]\}\Big)\gamma^\perp_\nu\}
\nonumber\\
&&\times\Big((q_1,q_2)\epsilon^{~~\nu}_\beta - {p^+-k^+\over q^+}\vecq_1\times\vecq_2\,g^\nu_{\perp\beta}\Big){p^+-k^+\over q^+}
\nonumber\\
&& + \tr\{\ssp\ssn_2\gamma^5\ssk\gamma^\beta_\perp\Big({q^+\over p^+}\Tr\{(U_{z_1}U^\dagger_{z_2}-1)[U_{z_2}\tildeQ_{z_1}]\}
- {q^+\over k^+}\Tr\{(U_{z_1}U^\dagger_{z_2}-1)[Q_{z_2}U^\dagger_{z_1}]\}\Big)\gamma^\perp_\nu\}
\nonumber\\
&&\times\Big(\vecq_1\times\vecq_2\epsilon^{~~\nu}_\beta + {p^+-k^+\over q^+}(q_1,q_2)_\perp g^\nu_{\perp\beta}\Big)
\nonumber\\
&& + i\,\tr\{\ssp\gamma^\perp_\mu\ \Big({q^+\over p^+}\Tr\{[Q_{z_1}U^\dagger_{z_2}](U_{z_2}U^\dagger_{z_1}-1)\}
- {q^+\over k^+}\Tr\{[U_{z_1}\tildeQ_{z_2}](U_{z_2}U^\dagger_{z_1}-1)\}\Big)\gamma^\alpha_\perp\ssk\ssn_2\}
\nonumber\\
&&\times\Big((q_1,q_2)_\perp \epsilon^\mu_{~~\alpha} - {p^+-k^+\over q^+}\vecq_1\times\vecq_2 \, g^\mu_{\perp\alpha}\Big)
{p^+-k^+\over q^+}
\nonumber\\
&& + \tr\{\ssp\gamma^\perp_\mu\ \Big({q^+\over p^+}\Tr\{[Q_{z_1}U^\dagger_{z_2}](U_{z_2}U^\dagger_{z_1}-1)\}
- {q^+\over k^+}\Tr\{[U_{z_1}\tildeQ_{z_2}](U_{z_2}U^\dagger_{z_1}-1)\}\Big)\gamma^\alpha_\perp\ssk\ssn_2\gamma^5\}
\nonumber\\
&&\times\Big(\vecq_1\times\vecq_2\, \epsilon^\mu_{~~\alpha} + {p^+-k^+\over q^+}(q_1,q_2)_\perp g^\mu_{\perp\alpha}\Big)
\Bigg]
\end{eqnarray}
Using eqs. \eqref{Qtransv-tr1}, \eqref{Qtransv-tr2}, and performing the contraction of Lorentz indices,  we arrive at
\begin{eqnarray}
&&\hspace{-1.4cm}\calm^{q\barq\,,A}_{\rm Quark} = 
{2q^+\over \pi s^2 }\sum_f e^2 e_f^2\int_0^1 dz 
\int {d^2z_1 d^2z_2 \dhd^2 q_1\dhd^2 q_2 \, e^{i(q_1-q_2,z_1-z_2)_\perp}\over 
\Big(q_{1\perp}^2 + Q^2 z\barz\Big)\Big(q_{2\perp}^2 + Q^2 z\barz\Big)}
\nonumber\\
&&\times\Bigg[
i(z^2+\barz^2)\vecq_2\times\vecq_1\Big({1\over z}\Tr\{U_{z_2}\calq^\dagger_{1z_1}\} + {1\over z}C_FQ^\dagger_{1z_1}
- {1\over \barz}\Tr\{\calq_{1z_2}U^\dagger_{z_1}\} - {1\over \barz}C_FQ_{1z_2}\Big)
\nonumber\\
&&- (z-\barz)(q_1,q_2)_\perp\Big({1\over z}\Tr\{U_{z_2}\calq^\dagger_{5z_1}\} + {1\over z}C_FQ^\dagger_{5z_1}
- {1\over \barz}\Tr\{\calq_{5z_2}U^\dagger_{z_1}\} - {1\over \barz}C_FQ_{5z_2}\Big)
\nonumber\\
&& + i(z^2+\barz^2)\vecq_2\times \vecq_1\Big({1\over z}\Tr\{\calq_{1z_1}U^\dagger_{z_2}\} + {1\over z}C_FQ_{1z_1}
- {1\over \barz}\Tr\{U_{z_1}\calq^\dagger_{1z_2}\} - {1\over \barz}C_FQ^\dagger_{1z_2}\Big)
\nonumber\\
&&- (z-\barz)(q_1,q_2)_\perp\Big({1\over z}\Tr\{\calq_{5z_1}U^\dagger_{z_2}\} + {1\over z}C_FQ_{5z_1}
- {1\over \barz}\Tr\{U_{z_1}\calq^\dagger_{5z_2}\} - {1\over \barz}C_FQ^\dagger_{5z_2}\Big)
\Bigg]
\end{eqnarray}
we can now observe that the two terms proportional to $\vecq_2\times\vecq_1$ are zero under integration, while
the two terms proportional to $(q_1,q_2)_\perp$, performing a change of integration variable, sum up giving
\begin{eqnarray}
&&\hspace{-1.4cm}\calm^{q\barq\,,A}_{\rm Quark} = 
{2q^+\over \pi s^2 }\sum_f e^2 e_f^2\int_0^1 dz {(z-\barz)^2\over z\barz}
\int {d^2z_1 d^2z_2 \dhd^2 q_1\dhd^2 q_2 \, e^{i(q_1-q_2,z_1-z_2)_\perp}\over 
\Big(q_{1\perp}^2 + Q^2 z\barz\Big)\Big(q_{2\perp}^2 + Q^2 z\barz\Big)}
\nonumber\\
&&~~\times
(q_1,q_2)_\perp\Big(\Tr\{U_{z_2}\calq^\dagger_{5z_1}\} + C_FQ^\dagger_{5z_1}
+ \Tr\{\calq_{5z_2}U^\dagger_{z_1}\} + C_FQ_{5z_2}\Big)
\nonumber\\
&& = {2\alpha_{\rm em}Q^2q^+\over \pi^2 s^2 }\sum_f e_f^2\int_0^1 dz (z-\barz)^2
\int d^2z_1 d^2z_2 \left|K_1(\bar{Q}|z_{12}|)\right|^2
\nonumber\\
&&~~\times\!\!
\Big(\Tr\{U_{z_2}\calq^\dagger_{5z_1}\} + C_FQ^\dagger_{5z_1}
+ \Tr\{\calq_{5z_2}U^\dagger_{z_1}\} + C_FQ_{5z_2}\Big)
\label{qqbar-QuarkAmpsquare}
\end{eqnarray}
with $z_{12}=z_1-z_2$.
Equation~\eqref{qqbar-QuarkAmpsquare} is the quark-background contribution of the $q\bar q$
final state to the helicity-dependent asymmetry. In contrast to the
longitudinal and transverse projections, it does not vanish and contains
both the color-trace terms and the local contributions proportional to
$C_F({\cal Q}_5+{\cal Q}_5^\dagger)$. The latter will play an important
role when this result is combined with the $qg$ and $\bar q g$ channels
in Sec.~\ref{sec:crossection-qg}.

\section{Dipole cross-section with quark + gluon in the final state}
\label{sec:crossection-qg}

\begin{figure}[h]
	\centering
	\includegraphics[width=0.9\linewidth]{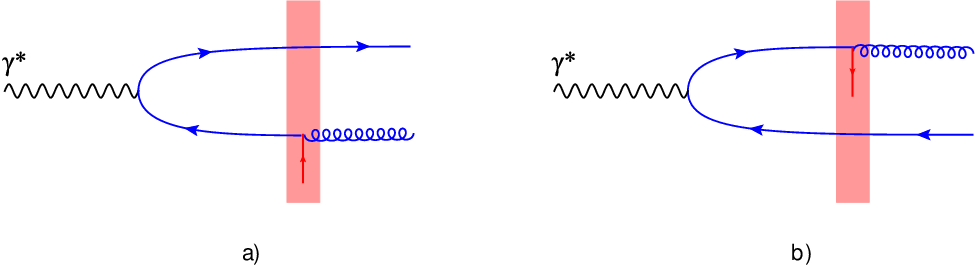}
\caption{Diagrams with a quark or an antiquark and a gluon in the final
	state. Left panel, a): $\gamma^*(q)\to q(p)\,g(k)$. The antiquark line of the
	photon fluctuation couples, at a vertex inside the shock wave, to the
	background quark field $\psi_{\rm cl}$ and is converted into the outgoing
	gluon, while the quark crosses the shock wave. Right panel, b): the
	charge-conjugate process $\gamma^*(q)\to\bar q(p)\,g(k)$, in which the quark
	line couples to $\bar\psi_{\rm cl}$ and is converted into the gluon, while the
	antiquark crosses the shock wave. Blue lines denote quantum fields, the red
	line the background quark field, and the shaded band the shock wave.}
	\label{fig:dipole-q2g}
\end{figure}

Let us consider the diagrams with quark and gluon or anti-quark and gluon in the final state represented in Fig.~\ref{fig:dipole-q2g}. 
To this end we have to consider the quark and gluon propagators with one end-point in the shock-wave.

\subsection{Propagators with one end-point in the shock-wave}

\subsubsection{Quark propagator with one end-point in the shock-wave}

We start with the quark propagator with one end-point in the external field and the other outside~\cite{Chirilli:2018kkw}
\begin{eqnarray}
\hspace{-1cm}\bra{x^{\rm ef}}{i\over \Sp+i\epsilon}\kety =
\!\!\!&& \Big[\int_0^{+\infty} {\dhd p^+\over 4(p^+)^2}\theta(x^+-y^+) - \int_{-\infty}^0{\dhd p^+\over 4(p^+)^2}\theta(y^+-x^+)\Big]
e^{-ip^+(x^--y^-)}
\label{qprop-xinef-a}\\
&&\times\braxp\Big([x^+,y^+]\hat{\ssp} - g\int_{y^+}^{x^+}\!\! dw^+\gamma^i[x^+,w^+]F_i^{~~-}[w^+,y^+]\Big)
\ssn_2\hat{\ssp}\,e^{-i{\hatp^2_\perp\over 2p^+}(x^+-y^+)}\ketyp
\nonumber
\end{eqnarray}
where the superscript $\rm ef$ reminds us that the point $x$ is in the external field. 
We can equivalently write the propagator \eqref{qprop-xinef-a} as\footnote{Note that we use interchangeably the 
equivalent notations $\brax{i\over \sslash{\,P}+i\epsilon}\kety \equiv \langle {\rm T}\{\psi(x)\barpsi(y)\}\rangle_A$.}
\begin{eqnarray}
\bra{x^{\rm ef}}{i\over \Sp+i\epsilon}\kety =\!\!\!&&
\Big([x^+,y^+]_x\ssk\ssn_2 - g\int_{y^+}^{x^+}dw^+\gamma^i[x^+,w^+]_x F_i^{~~-}[w^+,y^+]_x\ssn_2\Big)
\nonumber\\
&&\times
\int \dhd^4k\,{i\,\ssk\over k^2+i\epsilon}{e^{-ik\cdot(x-y)}\over 2k^+}
\label{qprop-xinef-b}
\end{eqnarray}
where \eqref{qprop-xinef-a} can be obtained from \eqref{qprop-xinef-b} by calculating the residue over the $k^-$ variable.

We can now take the shock-wave limit: the gauge field outside the external field is a pure gauge, so we can extend to infinity the limit of integration and
obtain
\begin{eqnarray}
\bra{x^{\rm sw}}{i\over \Sp+i\epsilon}\kety =\!\!\!&&
\Bigg[
\theta(x^+-y^+)\Big([x^+,-\infty n_1]_x\ssk\ssn_2 - g\int_{-\infty}^{x^+}dw^+\gamma^i[x^+,w^+]_x F_i^{~~-}[w^+,-\infty n_1]_x\ssn_2\Big)
\nonumber\\
\nonumber\\
&&+ \theta(y^+-x^+)\Big([x^+,+\infty n_1]_x\ssk\ssn_2+ g\int^{+\infty}_{x^+}dw^+\gamma^i[x^+,w^+]_x F_i^{~~-}[w^+, +\infty n_1]_x\ssn_2\Big)
\Bigg]
\nonumber\\
&&\times\!\int \dhd^4k\,{i\,\ssk\over k^2+i\epsilon}{e^{-ik\cdot(x-y)}\over 2k^+}
\label{qprop-xinsw}
\end{eqnarray}
The quark propagator with $y$ in the external field is
\begin{eqnarray}
\hspace{-1cm}\brax{i\over \Sp+i\epsilon}\ket{y^{\rm ef}} =
\!\!\!&& \Big[\int_0^{+\infty} {\dhd p^+\over 4(p^+)^2}\theta(x^+-y^+) - \int_{-\infty}^0{\dhd p^+\over 4(p^+)^2}\theta(y^+-x^+)\Big]
e^{-ip^+(x^--y^-)}
\label{qprop-yinef-a}\\
&&\times\braxp e^{-i{\hatp^2_\perp\over 2p^+}(x^+-y^+)}
\hat{\ssp}\ssn_2\Big(\ssp[x^+,y^+] + g\int_{y^+}^{x^+}\!\! dw^+\gamma^i[x^+,w^+]F_i^{~~-}[w^+,y^+]\Big)
\ketyp
\nonumber
\end{eqnarray}
Performing similar steps as we did for the case of $x$ in the external field, we can obtain the quark propagator with $y$ end-point in the shock-wave
\begin{eqnarray}
\brax{i\over \Sp+i\epsilon}\ket{y^{\rm sw}} =\!\!\!&&
\int\dhd^4 k{i\,\ssk \over k^2+i\epsilon}{e^{-ik\cdot (x-y)}\over 2k^+}
\label{qprop-yinsw}\\
&&\times\Bigg[
\theta(x^+-y^+)\Big(\ssn_2\ssk[+\infty n_1, y^+]_y + g\int_{y^+}^{+\infty}dw^+\ssn_2\gamma^i[+\infty n_1, w^+]_y F_i^{~~-}[w^+,y^+]_y\Big)
\nonumber\\
&&+ \theta(y^+-x^+)\Big(\ssn_2\ssk[-\infty n_1, y^+]_y - g\int^{y^+}_{-\infty}dw^+\ssn_2\gamma^i[-\infty n_1, w^+]_y F_i^{~~-}[w^+,y^+]_y\Big)
\Bigg]
\nonumber
\end{eqnarray}
\subsubsection{Gluon propagator with one end-point in the shock-wave}

To get the gluon propagator with one end-point in the shock-wave, we proceed in the same way as we did for the quark propagator.
The gluon propagator with one end-point in the external field is~\cite{Chirilli:2018kkw} (see also \cite{Balitsky:2015qba})
\begin{eqnarray}
\hspace{-1.3cm}\langle {\rm T}\{A_\mu^a(x^{\rm ef})A_\nu^b(y)\}\rangle_A =\!\!\! &&\left[-\int_0^{+\infty}
{\dhd p^+\over 2p^+}\theta(x^+-y^+) + \int^0_{-\infty}{\dhd p^+\over 2p^+}\theta(y^+-x^+)\right]e^{-ip^+(x^--y^-)}
\nonumber\\
&&\!\!\!\times\braxp \left(g_{\perp\mu}^\xi - {n_{2\mu}\hatp^\xi\over p^+}\right)[x^+,y^+]\left(g^\perp_{\nu\xi} - {n_{2\nu}\hatp_\xi\over p^+}\right)
e^{-i{\hatp^2_\perp\over 2p^+}(x^+-y^+)}\ketyp^{ab}
\label{gluonprop-xef}
\end{eqnarray}
Similarly to what we did for the quark propagator, in the shock-wave limit we obtain
\begin{eqnarray}
&&\hspace{-1cm}\langle {\rm T}\{A_\mu^a(x^{\rm sw})A_\nu^b(y)\}\rangle_A
\nonumber\\
&&\hspace{-1.3cm}= - i\int\dhd^4k{d^\xi_\mu(k)\over k^2+i\epsilon}\,e^{-ik\cdot(x-y)}
\nonumber\\
&&\hspace{-1.3cm}\times\bigg[\theta(x^+-y^+)\Big(d_{\xi\nu}(k)[x^+,-\infty n_1]_x
+ g{n_{2\nu}g_{\perp\xi}^i\over k^+}\int_{-\infty}^{x^+}dz^+[x^+,z^+]_x{F_i}^{\;-}[z^+,-\infty n_1]_x\Big)
\nonumber\\
&&\hspace{-1.3cm}+ \theta(y^+-x^+)\Big(d_{\xi\nu}(k)[x^+,+\infty n_1]_x
- g{n_{2\nu}g_{\perp\xi}^i\over k^+}\int_{x^+}^{+\infty}dz^+[x^+,z^+]_x{F_i}^{\;-}[z^+,+\infty n_1]_x\Big)
\bigg]^{ab}
\label{gluonprop-xsw}
\end{eqnarray}
For the case of $y$ in the external field we have
\begin{eqnarray}
\hspace{-1.5cm}\langle {\rm T}\{A_\mu^a(x)A_\nu^b(y^{\rm ef})\}\rangle =\!\!\! &&\left[-\int_0^{+\infty}
{\dhd p^+\over 2p^+}\theta(x^+-y^+) + \int^0_{-\infty}{\dhd p^+\over 2p^+}\theta(y^+-x^+)\right]e^{-ip^+(x^- - y^-)}
\nonumber\\
&&\hspace{-0.3cm}\times\braxp e^{-i{\hatp^2_\perp\over 2p^+}(x^+-y^+)}
\left(g_{\perp\mu}^\xi - {n_{2\mu}\hatp^\xi\over p^+}\right)[x^+,y^+]\left(g^\perp_{\nu\xi} - {n_{2\nu}\hatp_\xi\over p^+}\right)\ketyp^{ab}
\label{gluonprop-yef}
\end{eqnarray}
and in the shock-wave limit we obtain
\begin{eqnarray}
&&\hspace{-1.2cm}\langle {\rm T}\{ A_\mu^a(x)A_\nu^b(y^{\rm sw})\}\rangle
\nonumber\\
&&\hspace{-1.5cm}= -i\int\dhd^4k{d^\xi_\mu(k)\over k^2+i\epsilon}\,e^{-ik\cdot(x-y)}
\nonumber\\
&&\hspace{-1.5cm}\times\bigg[\theta(x^+-y^+)\Big(d_{\xi\nu}(k)[+\infty n_1,y^+]_y 
- g{n_{2\nu}g_{\perp\xi}^i\over k^+}\int_{y^+}^{+\infty}dz^+[+\infty n_1,z^+]_y{F_i}^{\;-}[z^+,y^+]_y\Big)
\nonumber\\
&&\hspace{-1.5cm}+ \theta(y^+-x^+)\Big(d_{\xi\nu}(k)[-\infty n_1,y^+]_y 
+ g{n_{2\nu}g_{\perp\xi}^i\over k^+}\int^{y^+}_{-\infty}dz^+[-\infty n_1,z^+]_y{F_i}^{\;-}[z^+,y^+]_y\Big)
\bigg]^{ab}
\label{gluonprop-ysw}
\end{eqnarray}
with
\begin{eqnarray}
d_{\mu\nu}(k) = g_{\mu\nu} - {k_\mu n_{2\nu} + k_\nu n_{2\mu}\over k^+}
\label{dmunu}
\end{eqnarray}
the typical tensor appearing in the axial gauge with a spurious divergence in the light-cone coordinate $k^+$ whose choice of 
prescription is dictated by imposing sub-gauge conditions~\cite{Chirilli:2015fza}.

\subsection{$\gamma^*\to qg$}

We now consider the contribution with a quark of momentum $p$ and a gluon of momentum $k$ in the final state, shown in Fig.~\ref{fig:dipole-q2g}.
We work with massless quarks, $p^+,k^+>0$, and retain the leading contribution of this channel to the sub-eikonal cross section. The
background fields are independent of $x^-$ at the accuracy considered and have a narrow support in $x^+$ around the shock-wave position.
The outgoing gluon is described in the light-cone gauge $n_2\cdot A=0$, with physical polarization satisfying
$n_2\cdot\varepsilon(k)=k\cdot\varepsilon(k)=0$. The amplitude is obtained by applying the LSZ reduction formula to the
outgoing quark and gluon and amputating the incoming photon leg. Since the photon is virtual, $q^2=-Q^2$, no photon mass-shell limit is taken.
The inverse propagators $[-i\tilde D]$ and the operator $i\slashd^y$ in the following expression implement these amputations.

We start with the diagram with quark and gluon in the final state. The matrix element defining the process in Fig.~\ref{fig:dipole-q2g}a is
\begin{eqnarray}
&&\langle q(p) g(k)|\gamma^*(q)\rangle\theta(p^+)\theta(k^+) 
\nonumber\\
=&& \int d^4xd^4yd^4z e^{-iq\cdot x + ip\cdot y+ ik\cdot z}\theta(p^+)\theta(k^+) 
\label{correlator-qg}\\
&&\times\varepsilon^{*a}_\rho(k)[-i\tildeD_{ac}]^{\rho\nu}[-i\tildeD]^\mu_{~\sigma}\varepsilon_\mu(q)
\baru(p)(i\slashd^y) \bra{0}{\rm T}A_\nu^c(z)\psi(y)A^\sigma(x)\ket{0}
\nonumber
\end{eqnarray}

To evaluate the correlator \eqref{correlator-qg}, we expand to first order in the electromagnetic interaction and first order in the quark--gluon
interaction containing one background quark field $\psi_{\rm cl}$. The electromagnetic vertex is denoted by $w_1$ and the vertex containing
the background quark by $w_2$. 

The photon amputation yields $\varepsilon_\mu(q)e^{-iq\cdot w_1}$. 
For the outgoing quark, which
crosses the shock wave, we use the Feynman rule given in eq.~\eqref{FeynRule1a}. This introduces
the fundamental Wilson line $U_{z_1}$, the transverse momentum $p'$
before crossing the shock-wave, and the factor $\theta(-w_1^+)$.
 
The other quark has its endpoint $w_2$ inside the shock wave, so we need the propagator
\eqref{qprop-yinsw} which selects the interval $w_1^+<w_2^+$. 
For the outgoing gluon, instead, we use the propagator~\eqref{gluonprop-ysw}, with $z^+>w_2^+$.
Thus, we have
\begin{eqnarray}
&&\hspace{-0.6cm}\langle q(p) g(k)|\gamma^*(q)\rangle\theta(p^+)\theta(k^+) 
\nonumber\\
&&\hspace{-0.6cm} = ee_fg \int d^4w_1\,e^{-iq\cdot w_1}\varepsilon_\mu(q)
\int d^4z\, e^{i k\cdot z}d^4w_2 \varepsilon^{*a}_\rho(k)
[-i\tildeD_{ac}]^{\rho\nu}\int \dhd^4k'{(-i)d_{\nu\xi}(k')\over k'^2+i\epsilon}e^{-ik'\cdot (z-w_2)}
\nonumber\\
&&\hspace{-0.6cm}\times \theta(z^+-w^+_2)\Big[d^\xi_\lambda(k')[+\infty n_1,w_2^+]^{cd}_{w_2}
- g{n_{2\lambda}g^{i\xi}_\perp\over k'^+}\int_{w_2^+}^{+\infty} dw^+[+\infty n_1,w^+]_{w_2}{F_i}^{\;-}[w^+,w_2^+]_{w_2}\Big]
\nonumber\\
&&\hspace{-0.6cm}\times i\theta(p^+)\theta(-w^+_1)\int d^2z_1\dhd^2p'\baru(p)\ssn_2U_{z_1}{p^+\ssn_1+\ssp'_\perp\over 2p^+}\gamma^\mu
\,e^{ip^+w^-_1 - i(p-p',z_1)_\perp - i(p',w_1)_\perp + i{p'^2_\perp\over 2p^+}w_1^+}
\nonumber\\
&&\hspace{-0.6cm}\times\!\int\dhd^4k_1{i\ssk_1\over 2k^+_1(k^2_1+i\epsilon)}e^{-ik_1\cdot(w_1-w_2)}\theta(w_2^+-w_1^+)
\nonumber\\
&&\hspace{-0.6cm}\times\Big(\ssn_2\ssk_1[-\infty n_1,w^+_2]_{w_2} - g\int_{-\infty}^{w_2^+} dw'^+
\ssn_2\gamma^i[-\infty n_1,w'^+]_{w_2}{F_i}^{\;-}[w'^+,w_2^+]_{w_2}\Big)
\nonumber\\
&&\hspace{-0.6cm}\times\gamma^\lambda t^d\psi_{\rm cl}(w_2^+,w_{2\perp})
\end{eqnarray}
After contraction with the physical gluon polarization, the longitudinal part of the gluon numerator and its explicit
$F_i{}^{-}$ insertion give terms proportional to $\ssn_2\psi_{\rm cl}$. The field-strength insertion in the endpoint
quark propagator likewise contains $\ssn_2\gamma^i\gamma_\perp^\lambda\psi_{\rm cl}
=\gamma^i\gamma_\perp^\lambda\ssn_2\psi_{\rm cl}$. These terms are beyond the accuracy retained in this channel.
The surviving gluon coupling is therefore transverse and gives $\varepsilon_\xi^{*c}(k)g^\xi_{\perp\lambda}\gamma_\perp^\lambda$.

We can also integrate over $w_1^-$ and $w_2^-$. These give, respectively, the constraints $p^+-q^+-k_1^+=0$ and $k^++k_1^+=0$.
Consequently, $k_1^+=-k^+<0$, and the remaining longitudinal delta function is $\dbar(p^++k^+-q^+)$. 
Next, we calculate the residue integrating over $k_1^-$ taking into consideration the constraint $w_1^+<w_2^+$, and that 
the $k_1^-$ dependence cancels from the numerator. We thus obtain

\begin{eqnarray}
&&\langle q(p) g(k)|\gamma^*(q)\rangle\theta(p^+)\theta(k^+) 
\nonumber\\
&& = - i{ee_fg\over 2}\int d^2w_1d^2w_2\int_{-\infty}^{+\infty}dw_2^+\int_{-\infty}^{w_2^+}dw_1^+
\nonumber\\
&&\times\int d^2z_1 \dhd^2p'\dhd^2 k_1\theta(-w^+_1)
\theta(k^+)\theta(p^+)\dbar(p^+-q^++k^+)
e^{iw^+_2\left(k^--{k^2_{1\perp}\over 2k^+}\right)}e^{i\left({k^2_{1\perp}\over 2k^+}+{p'^2_\perp\over 2p^+}-q^-\right)w_1^+}
\nonumber\\
&&\times 
{e^{i(w_2,-k-k_1)_\perp}\over 2(k^+)^2}e^{i(w_1,q+k_1-p')_\perp - i(p-p',z_1)_\perp}
\varepsilon^{*c}_\xi(k) g^\xi_{\perp\lambda}[+\infty n_1,w_2^+]_{w_2}^{cd}
\baru(p) U_{z_1}\ssn_2{p^+\ssn_1+\ssp'_\perp\over 2p^+}
\nonumber\\
&&\times\sslash{\varepsilon}(q)(-k^+\ssn_1+\ssk_{1\perp})\ssn_2(-k^+\ssn_1+\ssk_{1\perp})
\gamma^\lambda_\perp [-\infty n_1,w_2^+]_{w_2}t^d\psi_{\rm cl}(w_2^+,w_{2\perp})
\label{qg-production}
\end{eqnarray}
We can further simplify result \eqref{qg-production}.

The integration over $w_{1\perp}$ gives $\dbar^{(2)}(q_\perp+k_{1\perp}-p'_\perp)$;
choosing $q_\perp=0$ this fixes $p'_\perp=k_{1\perp}$, and together with $p^++k^+=q^+$ and $q^-=-Q^2/(2q^+)$, we obtain 
the usual light-cone energy denominator.

Now, since the sub-eikonal correction is represented by the presence of the quark field in the external field, to avoid going beyond the
approximation we are interested in, we can make the following two approximations.
First, since $e^{-i{\hatp^2_\perp\over 2p^+}z^+}\psi(z^+,z_\perp)e^{-i{\hatp^2_\perp\over 2p^+}z^+} = \psi(z^+,z_\perp)+\calo (\lambda^{-1})$, then
we can set $e^{iw^+_2\left(k^--{k^2_{1\perp}\over 2k^+}\right)} \sim 1$.

Second, in the shock-wave approximation, recalling that the quark field depends on $w_2^+$, and that the thickness, $L^+$, of the shock-wave
goes to zero, we have ($q^-=-{Q^2 \over2q^+}$)
\begin{eqnarray}
&&\int_{-\infty}^{+\infty}dw_2^+\int_{-\infty}^{w_2^+}dw_1^+\theta(-w^+_1)e^{i\left({k^2_{1\perp}q^+\over 2k^+p^+}+ {Q^2\over 2q^+}\right)w_1^+}
\nonumber\\
=&& - i\int_{-\infty}^0dw^+_2{ e^{i\left({k^2_{1\perp}q^+\over 2k^+p^+}+ {Q^2\over 2q^+}\right)w_2^+} -1
	\over {k^2_{1\perp}q^+\over 2k^+p^+}+ {Q^2\over 2q^+}} - i \int_{-\infty}^{+\infty}dw_2^+{1\over {k^2_{1\perp}q^+\over 2k^+p^+}+ {Q^2\over 2q^+}}
\nonumber\\
\simeq && - i {2k^+p^+\over q^+}\int_{-\infty}^{+\infty}dw_2^+\,{1\over k^2_{1\perp} + {2\over s}Q^2k^+p^+}
\label{phase-approxy}
\end{eqnarray}

With the above approximation, from \eqref{qg-production} we arrive at
\begin{eqnarray}
&&\hspace{-0.4cm}\langle q(p) g(k)|\gamma^*(q)\rangle\theta(p^+)\theta(k^+) 
\label{qg-production-2}\\
&&\hspace{-0.4cm} = - {ee_fg\over 4q^+}\int d^2w_2\int_{-\infty}^{+\infty}dw_2^+
\int d^2z_1 \theta(k^+)\theta(p^+){\dhd^2 k_1\over k^+}
{\dbar(p^++k^+-q^+) \over k^2_{1\perp} + {2\over s}Q^2k^+p^+}
\nonumber\\
&&\hspace{-0.4cm}\times 
e^{i(w_2,-k-k_1)_\perp - i(p-k_1,z_1)_\perp}
\varepsilon^{*c}_\xi(k) g^\xi_{\perp\lambda}[+\infty n_1,w_2^+]_{w_2}^{cd}\baru(p) U_{z_1}
\nonumber\\
&&\hspace{-0.4cm}\times\ssn_2(p^+\ssn_1+\ssk_{1\perp})\sslash{\varepsilon}(q)
(-k^+\ssn_1+\ssk_{1\perp})\ssn_2(-k^+\ssn_1+\ssk_{1\perp})\gamma^\lambda_\perp [-\infty n_1,w_2^+]_{w_2}t^d\psi_{\rm cl}(w_2^+,w_{2\perp})
\nonumber
\end{eqnarray}
At this point we can further simplify the 
background-quark vertex by retaining the leading spinor component
according to the power counting used in eq.~\eqref{goodbadcomp}
\begin{eqnarray}
&&(-k^+\ssn_1+\ssk_{1\perp})\ssn_2(-k^+\ssn_1+\ssk_{1\perp})\gamma^\lambda_\perp t^d\psi_{\rm cl}(w_2^+,w_{2\perp})
\nonumber\\
=\!\!\!&& \Big(2(k^+)^2\ssn_1-2k^+\ssk_{1\perp}+k^2_{1\perp}\ssn_2\Big)\gamma^\lambda_\perp t^d\psi_{\rm cl}(w_2^+,w_{2\perp})
\nonumber\\
=\!\!\!&& 2k^+\Big(k^+\ssn_1-\ssk_{1\perp}\Big)\gamma^\lambda_\perp t^d\psi_{\rm cl}(w_2^+,w_{2\perp})
+ \calo(\lambda^{-1})
\end{eqnarray}
and, from \eqref{qg-production-2} we finally arrive at
\begin{eqnarray}
&&\hspace{-0.4cm}\langle q(p) g(k)|\gamma^*(q)\rangle\theta(p^+)\theta(k^+) 
\label{qg-production-final}\\
=\!\!\!&& - {ee_fg\over s}\int d^2w_2d^2z_1 \dhd^2 k_1
\theta(k^+)\theta(p^+)\dbar\left({p^+\over q^+}+{k^+\over q^+} - 1\right)
\nonumber\\
&&\times {e^{-i(w_2,k+k_1)_\perp - i(p-k_1,z_1)_\perp}\over k^2_{1\perp} + {2\over s}Q^2k^+p^+}
\int_{-\infty}^{+\infty}dw_2^+[+\infty n_1,w_2^+]_{w_2}^{cd}
\nonumber\\
&&\times\!\baru(p,\sigma) U_{z_1}\ssn_2(p^+\ssn_1+\ssk_{1\perp})\sslash{\varepsilon}(q)
\Big(k^+\ssn_1-\ssk_{1\perp}\Big) \sslash{\varepsilon}^{*c}_\perp(k) [-\infty n_1,w_2^+]_{w_2}t^d\psi_{\rm cl}(w_2^+,w_{2\perp})
\nonumber
\end{eqnarray}
The above amplitude is also part of the NLO differential cross-section for the production of quark, as well as LO cross-section for the quark plus gluon production.

Here, however, we are interested in the inclusive DIS cross-section, so the inclusive amplitude square is
\begin{eqnarray}
&&\hspace{-1cm}{1\over (2\pi)\delta(0)}\int \dhd^4p\dhd^4k\dbar(p^2)\dbar(k^2)\Big|\langle q(p) g(k)|\gamma^*(q)\rangle\Big|^2\theta(p^+)\theta(k^+)
\nonumber\\
&&
\hspace{-1cm}=\int \dhd^4p\dhd^4k\dbar(p^2)\dbar(k^2)\theta(k^+)\theta(p^+)\dbar\left({p^+\over q^+}+{k^+\over q^+} - 1\right)
\nonumber\\
&&\hspace{-1cm}\times\Bigg|{ee_fg\over s}\int d^2w_2 d^2z_1 \dhd^2 k_1
{e^{-i(w_2,k+k_1)_\perp- i(p-k_1,z_1)_\perp}\over k^2_{1\perp} + {2\over s}Q^2k^+p^+}
\int_{-\infty}^{+\infty}dw_2^+[+\infty n_1,w_2^+]_{w_2}^{cd}
\nonumber\\
&&\hspace{-1cm}\times\baru(p,\sigma) U_{z_1}\ssn_2(p^+\ssn_1+\ssk_{1\perp})\sslash{\varepsilon}(q)
\Big(k^+\ssn_1-\ssk_{1\perp}\Big) \sslash{\varepsilon}^{*c}_\perp(k) [-\infty n_1,w_2^+]_{w_2}t^d\psi_{\rm cl}(w_2^+,w_{2\perp})\Bigg|^2
\label{qg-production-final-sqr}
\end{eqnarray}

In the next sections we will consider the longitudinal, transverse and anti-symmetric photon polarization.

\subsubsection{Longitudinal polarization for $\gamma^*\to qg$}

Let us consider the longitudinal polarization. We have to evaluate the Dirac matrix element squared. 
Using $\sum_\delta \varepsilon_{\xi,\delta}^{*c}\varepsilon_{\xi',\delta}^{c'} = - g_{\xi\xi'}^\perp\delta^{cc'}$,
and the longitudinal polarization vector \eqref{epsL-final}, we have
\begin{eqnarray}
&&\hspace{-0.7cm}\sum_{\delta,\sigma} 
\Big|\varepsilon_{\xi,\delta}^{*c}g^\xi_{\perp\lambda}\baru(p,\sigma) U_{z_1}\ssn_2\Big(p^+\ssn_1+\ssk_{1\perp}\Big)\sslash{\varepsilon}_L(q)
\Big(k^+\ssn_1-\ssk_{1\perp}\Big)\gamma^\lambda_\perp [-\infty n_1,w_2^+]_{w_2}t^d\psi_{\rm cl}(w_2^+,w_{2\perp})\Big|^2
\nonumber\\
&&\hspace{-0.7cm} = 8p^+(q^+)^2\Big({2\over s}Qp^+k^+ - {k^2_{1\perp}\over Q}\Big)\Big({2\over s}Qp^+k^+ - {k^2_{2\perp}\over Q}\Big)
\nonumber\\
&&\hspace{-0.4cm}
\times\barpsi_{\rm cl}(w'^+_2,w'_{2\perp})t^{d'}[w'^+_2,-\infty n_1]_{w_2}U^\dagger_{z_2}\ssn_1U_{z_1}[-\infty n_1,w_2^+]_{w_2}t^d\psi_{\rm cl}(w_2^+,w_{2\perp})
+O(\lambda^{-2})
\label{qg-LongyDirMat}
\end{eqnarray}
Inserting result \eqref{qg-LongyDirMat} in the squared amplitude \eqref{qg-production-final-sqr}, we obtain
\begin{eqnarray}
\hspace{-0.5cm}\calm^{qg,L}_{\rm Quark}\equiv\!\!\!&&{1\over (2\pi)\delta(0)}\int \dhd^4p \dhd^4k\dbar(p^2)\dbar(k^2)\Big|\langle q(p) g(k)|\gamma^*_L(q)\rangle\Big|^2\theta(p^+)\theta(k^+)
\nonumber\\
&& = - \sum_f {e^2e_f^2 g^2\over s^2}\int \dhd^4p \dhd^4k\dbar(p^2)\dbar(k^2)\theta(k^+)\theta(p^+)\dbar\!\left({p^+\over q^+}+{k^+\over q^+} - 1\right)
\nonumber\\
&&\times\int d^2w_2d^2w_1\int_{-\infty}^{+\infty}dw_2^+dw_1^+
\int d^2z_1 \dhd^2 k_1 d^2z_2 \dhd^2 k_2
\nonumber\\
&&\times {e^{-i(w_2,k+k_1)_\perp}\over k^2_{1\perp} + {2\over s}Q^2k^+p^+}
e^{- i(p-k_1,z_1)_\perp}[+\infty n_1,w_2^+]_{w_2}^{cd}
{e^{i(w_1,k+k_2)_\perp}\over k^2_{2\perp} + {2\over s}Q^2k^+p^+}
e^{i(p-k_2,z_2)_\perp}
[w_1^+,+\infty n_1]_{w_1}^{d'c}
\nonumber\\
&&\times  8p^+(q^+)^2\Big({2\over s}Qp^+k^+ - {k^2_{1\perp}\over Q}\Big)\Big({2\over s}Qp^+k^+ - {k^2_{2\perp}\over Q}\Big)
\nonumber\\
&&\times\tr\Tr\{U^\dagger_{z_2}U_{z_1}
[-\infty n_1,w_2^+]_{w_2}t^d\ssn_1\psi_{\rm cl}(w_2^+,w_{2\perp})\barpsi_{\rm cl}(w^+_1,w_{1\perp})t^{d'}[w^+_1,-\infty n_1]_{w_1}\}
\label{qg-production-squared}
\end{eqnarray}
We perform the integration over $\dhd^4k\dhd^2p\dhd p^-$ and from \eqref{qg-production-squared} we arrive at
\begin{eqnarray}
&&\hspace{-1cm}\calm^{qg,L}_{\rm Quark}
\label{qg-production-squared-2}\\
&&\hspace{-1cm} = - {q^+\over 2\pi s}\sum_f e^2e_f^2 g^2\int_0^1{dz\over z}
\int d^2w_1d^2z_1\!\int_{-\infty}^{+\infty}dw_2^+dw_1^+
\int \dhd^2 k_1 \dhd^2 k_2\, e^{-i(w_1-z_1,k_1-k_2)_\perp}
\nonumber\\
&&\hspace{-1cm}\times {\Big(Qz\barz - {k^2_{1\perp}\over Q}\Big)\over k^2_{1\perp} + Q^2z\barz}
{\Big(Qz\barz - {k^2_{2\perp}\over Q}\Big)\over k^2_{2\perp} + Q^2z\barz}
\tr\Tr\{t^d\ssn_1\psi_{\rm cl}(w_2^+,w_{1\perp})[w_2^+,w_1^+]_{w_1}^{dd'}
\barpsi_{\rm cl}(w^+_1,w_{1\perp})t^{d'}[w^+_1,w_2^+]_{w_1}\}
\nonumber
\end{eqnarray}
Using
\begin{eqnarray}
&&\tr\Tr\{t^d\ssn_1\psi_{\rm cl}(w_2^+,w_{1\perp})[w_2^+,w_1^+]_{w_1}^{dd'}
\barpsi_{\rm cl}(w^+_1,w_{1\perp})t^{d'}[w^+_1,w_2^+]_{w_1}\}
\nonumber\\
=\!\!\! && - C_F\barpsi_{\rm cl}(w^+_1,w_{1\perp})[w^+_1,w_2^+]_{w_1}\ssn_1\psi_{\rm cl}(w_2^+,w_{1\perp})
\end{eqnarray}
and the definition of $Q_1(x_\perp)$ and $Q_1^\dagger(x_\perp)$ given in eqs. \eqref{trCfQ1}, 
and \eqref{trCfQ1dag}, respectively, the squared amplitude with longitudinal photon polarization, \eqref{qg-production-squared-2}, becomes
\begin{eqnarray}
\hspace{-1cm}\calm^{qg,L}_{\rm Quark}
 =\!\!&& {q^+\over \pi s^2}\sum_f e^2e_f^2 \int_0^1{dz\over z}
\int d^2w_1d^2z_1
\int \dhd^2 k_1 \dhd^2 k_2\, e^{-i(w_1-z_1,k_1-k_2)_\perp}
\nonumber\\
&&~~\times {\Big(Qz\barz - {k^2_{1\perp}\over Q}\Big)\over k^2_{1\perp} + Q^2z\barz}
{\Big(Qz\barz - {k^2_{2\perp}\over Q}\Big)\over k^2_{2\perp} + Q^2z\barz}
C_F\Big(Q_1(w_{1\perp}) + Q_1^\dagger(w_{1\perp})\Big)\,.
\label{Mquark-qqL}
\end{eqnarray}

\subsubsection{Transverse polarization for $\gamma^*\to qg$}

Let us consider the amplitude square with transverse photon polarization
\begin{eqnarray}
&&\hspace{-1.7cm}\calm^{qg,T}_{\rm Quark}\equiv {1\over (2\pi)\delta(0)}\int \dhd^4 p \dhd^4k\dbar(p^2)\dbar(k^2)\Big|\langle q(p) g(k)|\gamma^*_T(q)\rangle\Big|^2\theta(p^+)\theta(k^+)
\nonumber\\
&&\hspace{-0.3cm}
=\int \dhd^4p \dhd^4k\dbar(p^2)\dbar(k^2)\theta(k^+)\theta(p^+)\dbar\left({p^+\over q^+}+{k^+\over q^+} - 1\right)
\nonumber\\
&&\hspace{-0.3cm}\times\half\sum_{\lambda=\pm}\sum_{f,\sigma,\delta}\Bigg|- {ee_fg\over s}\int d^2w_2 d^2z_1 \dhd^2 k_1
{e^{-i(w_2,k+k_1)_\perp- i(p-k_1,z_1)_\perp}\over k^2_{1\perp} + {2\over s}Q^2k^+p^+}
\varepsilon^{*c}_{\xi,\delta}(k) g^\xi_{\perp\alpha}
\nonumber\\
&&\hspace{-0.3cm}\times\int_{-\infty}^{+\infty}dw_2^+\baru(p,\sigma) U_{z_1}\ssn_2(p^+\ssn_1+\ssk_{1\perp})\sslash{\varepsilon}_T(q)
\nonumber\\
&&\hspace{-0.3cm}\times
\Big(k^+\ssn_1-\ssk_{1\perp}\Big) \gamma^\alpha_\perp [-\infty n_1,w_2^+]_{w_2}[+\infty n_1,w_2^+]_{w_2}^{cd}
t^d\psi_{\rm cl}(w_2^+,w_{2\perp})\Bigg|^2
\label{qg-ampl-sqr-T}
\end{eqnarray} 
We have to simplify the Dirac matrix element with the transverse photon polarization,
and keep only the leading contribution. After simple Dirac and color algebra, we obtain
\begin{eqnarray}
&&\hspace{-0.8cm}\sum_{\delta,\sigma} 
\Big|\varepsilon_{\alpha,\delta}^{*c}\baru(p,\sigma) U_{z_1}\ssn_2\Big(p^+\ssn_1+\ssk_{1\perp}\Big)\sslash{\varepsilon}_T(q)
\Big(k^+\ssn_1-\ssk_{1\perp}\Big)\gamma^\alpha_\perp [-\infty n_1,w_2^+]_{w_2}t^d\psi_{\rm cl}(w_2^+,w_{2\perp})\Big|^2
\nonumber\\
&&\hspace{-1.3cm}=  8\delta^{cc'}p^+\barpsi_{\rm cl}t^{d'}[w_1^+,-\infty n_1]_{w_1}U^\dagger_{z_2}
\nonumber\\
&&\hspace{-0.8cm}
\times\bigg\{\bigg[(p^++k^+)^2 \veck_2\!\times\!\varepsilon^*_\lambda\, \veck_1\!\times\!\vec{\varepsilon}_\lambda
+ (p^+-k^+)^2(\varepsilon^*_\lambda,k_2)_\perp(\varepsilon_\lambda,k_1)_\perp
\bigg]\ssn_1
\nonumber\\
&&\hspace{-0.6cm}
- i\bigg[[(p^+)^2-(k^+)^2]\Big(\veck_2\!\times\! \varepsilon^*_\lambda (\varepsilon_\lambda,k_1)_\perp 
- (\varepsilon^*_\lambda,k_2)_\perp \veck_1\!\times \!\varepsilon_\lambda \Big)\bigg]\gamma^5\ssn_1\bigg\}
\nonumber\\
&&\hspace{-0.8cm}
\times U_{z_1}[-\infty n_1,w_2^+]_{w_2}t^d\psi_{\rm cl}  + O(\lambda^{-2})
\label{qgDiracMatr-squ-T}
\end{eqnarray}
So, using \eqref{qgDiracMatr-squ-T}, and integrating over $\dhd^4k\dhd^2p\dhd p^-$, from \eqref{qg-ampl-sqr-T} we obtain
\begin{eqnarray}
\hspace{-1cm}\calm^{qg,T}_{\rm Quark} =\!\!\!&&  {q^+\over 4\pi s}\sum_{\lambda=\pm}\sum_f e^2e^2_f g^2
\int_0^1{dz\over z}
\int d^2w_1d^2z_1
{\dhd^2k_1\dhd^2k_2\over k^2_{1\perp} + Q^2z\barz}
{e^{-i(w_1-z_1,k_1-k_2)_\perp}\over k^2_{2\perp} + Q^2z\barz}
\nonumber\\
&&\times\int_{-\infty}^{+\infty}\!dw_1^+dw_2^+\barpsi_{\rm cl}(w_{1\perp},w_1^+)t^{d'}[w_1^+,w_2^+]_{w_1}^{d'd}
[w_1^+,-\infty n_1]_{w_1}
\nonumber\\
&&\times\bigg[\bigg((\barz-z)^2(\varepsilon^*_\lambda,k_2)_\perp(\varepsilon_\lambda,k_1)_\perp
+ \veck_2\!\times\!\vec{\varepsilon}^{\,*}_\lambda\, \veck_1\!\times\!\vec{\varepsilon}_\lambda \bigg)\ssn_1
\nonumber\\
&&\hspace{0.2cm} - i(\barz^2-z^2)\Big(\veck_2\!\times\! \vec{\varepsilon}^{\,*}_\lambda (\varepsilon_\lambda,k_1)_\perp 
- (\varepsilon^*_\lambda,k_2)_\perp \veck_1\!\times \!\vec{\varepsilon}_\lambda \Big)
\gamma^5\ssn_1\bigg]
\nonumber\\
&&
\times[-\infty n_1,w_2^+]_{w_1}t^d\psi_{\rm cl}(w_{1\perp},w_2^+)+ O(\lambda^{-2})
\label{qgsquare-transverse-2}
\end{eqnarray}
Let us sum over photon helicities using
\begin{eqnarray}
&&\half\sum_{\lambda=\pm} \Big[(\barz-z)^2(\varepsilon^*_\lambda,k_2)_\perp(\varepsilon_\lambda,k_1)_\perp
+ \veck_2\!\times\!\vec{\varepsilon}^{\,*}_\lambda\, \veck_1\!\times\!\vec{\varepsilon}_\lambda \Big] = (z^2+\barz^2)(k_1,k_2)_\perp
\nonumber\\
&&\half\sum_{\lambda=\pm} \Big(\veck_2\!\times\! \vec{\varepsilon}^{\,*}_\lambda (\varepsilon_\lambda,k_1)_\perp 
- (\varepsilon^*_\lambda,k_2)_\perp \veck_1\!\times \!\varepsilon_\lambda \Big) = \veck_2\times\veck_1
\end{eqnarray}
Thus, from \eqref{qgsquare-transverse-2}, we have
\begin{eqnarray}
&&\hspace{-0.6cm}\calm^{qg,T}_{\rm Quark}
\nonumber\\
&&\hspace{-1cm}=  {e^2q^+\over 2\pi s}\sum_f e^2_f g^2\int_0^1{dz\over z}
\int d^2w_1d^2z_1
{\dhd^2k_1\dhd^2k_2\over k^2_{1\perp} + Q^2z\barz}
{e^{-i(w_1-z_1,k_1-k_2)_\perp}\over k^2_{2\perp} + Q^2z\barz}
\nonumber\\
&&\hspace{-1cm} \times
\int_{-\infty}^{+\infty}\!dw_1^+dw_2^+\,[w_1^+,w_2^+]_{w_1}^{d'd}
\Bigg[(z^2+\barz^2)(k_1,k_2)_\perp\barpsi_{\rm cl}(w_{1\perp},w_1^+)t^{d'}[w_1^+,w^+_2]_{w_1}
\ssn_1
t^d\psi_{\rm cl}(w_{1\perp},w_2^+)
\nonumber\\
&&\hspace{-0.8cm} - i(\barz-z)\veck_2\times\veck_1\barpsi_{\rm cl}(w_{1\perp},w_1^+)t^{d'}[w_1^+,w^+_2]_{w_1}
\gamma^5\ssn_1t^d\psi_{\rm cl}(w_{1\perp},w_2^+)\Bigg]  + O(\lambda^{-2})
\label{qgsquare-transverse-3}
\end{eqnarray}
The term proportional to $\veck_2\times\veck_1$ is zero under integration, so we finally obtain
\begin{eqnarray}
&&~~~\calm^{qg,T}_{\rm Quark}
\nonumber\\
&&
= 2C_F{\alpha_{\rm em}q^+\over s^2}\sum_fe^2_f \int_0^1dz\,{(z^2+\barz^2)\over z\barz}
\int d^2z_1d^2z_2\dhd^2k_1\dhd^2k_2
{e^{-i(z_1-z_2,k_1-k_2)_\perp}\over k^2_{1\perp} + Q^2z\barz}
{(k_1,k_2)_\perp\over k^2_{2\perp} + Q^2z\barz}
\nonumber\\
&& ~~~~\times
\!\Big[Q_1(z_{1\perp}) + Q^\dagger_1(z_{1\perp})\Big]  + O(\lambda^{-2})
\nonumber\\
&&
=  q^+C_F{\alpha_{\rm em} \over 2s^2}{Q^2\over \pi^2}\sum_fe^2_f \int_0^1dz\,(z^2+\barz^2)
\int d^2z_1d^2z_2\Big|K_1(\bar{Q}|z_{12}|)\Big|^2
\nonumber\\
&&~~~~\times\!\Big[Q_1(z_{1\perp}) + Q^\dagger_1(z_{1\perp})\Big]  + O(\lambda^{-2})
\label{qgsquare-transverse-4}
\end{eqnarray}
with ($\bar{Q} = \sqrt{Q^2z\barz}$) and where we used the definition of $Q_1(x_\perp)$ and $Q_1^\dagger(x_\perp)$ given in eqs. \eqref{trCfQ1}, and \eqref{trCfQ1dag}, respectively.

\subsubsection{Asymmetry for $\gamma^*\to qg$}

We consider here the asymmetric combination which allows us to isolate the $g_1$ structure function. 
We can start the calculation from \eqref{qgsquare-transverse-2} with sum over anti-symmetric helicity combination rather than average. So, we have
\begin{eqnarray}
&&\hspace{-2.3cm}\calm^{qg,A}_{\rm Quark}\equiv 
{1\over (2\pi)\delta(0)}\int \dhd^4 p \dhd^4k\dbar(p^2)\dbar(k^2)\Big|\langle q(p) g(k)|\gamma^*_A(q)\rangle\Big|^2\theta(p^+)\theta(k^+)
\nonumber\\
&&\hspace{-0.9cm} = \sum_{\lambda={\rm asym.}}\sum_f{e^2e^2_f g^2\over 2 s}
{q^+\over \pi}\int_0^1{dz\over z}
\int d^2w_1d^2z_1
{\dhd^2k_1\dhd^2k_2\over k^2_{1\perp} + Q^2z\barz}
{e^{-i(w_1-z_1,k_1-k_2)_\perp}\over k^2_{2\perp} + Q^2z\barz}
\nonumber\\
&&\hspace{-0.4cm}\times\!\int_{-\infty}^{+\infty}\!dw_1^+dw_2^+\barpsi_{\rm cl}(w_{1\perp},w_1^+)t^{d'}[w_1^+,w_2^+]_{w_1}^{d'd}[w_1^+,-\infty n_1]_{w_1}
\nonumber\\
&&\hspace{-0.4cm}\times\bigg[\Big((\barz-z)^2(\varepsilon_\lambda,k_1)_\perp(\varepsilon^*_\lambda,k_2)_\perp 
+ \veck_1\times \vec{\varepsilon}_\lambda\,\veck_2\times\vec{\varepsilon}^*_\lambda\Big)\ssn_1
\nonumber\\
&&\hspace{-0.4cm}~~ - i[\barz^2-z^2]
\Big((\varepsilon_\lambda,k_1)_\perp\veck_2\times\vec{\varepsilon}^*_\lambda 
- (\varepsilon_\lambda^*,k_2)\veck_1\times\vec{\varepsilon}_\lambda\Big)
\gamma^5\ssn_1\bigg]
\nonumber\\
&&\hspace{-0.4cm}\times[-\infty n_1,w_2^+]_{w_1}t^d\psi_{\rm cl}(w_{1\perp},w_2^+)+ O(\lambda^{-2})\,.
\label{qg-antisymsquare-a}
\end{eqnarray}
The sum over anti-symmetric helicity combination, using definition \eqref{asymsum}, is
\begin{eqnarray}
&&\sum_{\lambda={\rm asym.}}
\Big[(\barz-z)^2(\varepsilon_\lambda,k_1)_\perp(\varepsilon^*_\lambda,k_2)_\perp 
+ \veck_1\times \vec{\varepsilon}_\lambda\,\veck_2\times\vec{\varepsilon}^*_\lambda\Big] = 2i(z^2+\barz^2)\veck_2\!\times\!\veck_1
\nonumber\\
&&\sum_{\lambda={\rm asym.}}
\Big((\varepsilon_\lambda,k_1)_\perp\veck_2\times\vec{\varepsilon}^*_\lambda 
- (\varepsilon_\lambda^*,k_2)\veck_1\times\vec{\varepsilon}_\lambda\Big) = -2i(k_1,k_2)_\perp
\end{eqnarray}
thus, from \eqref{qg-antisymsquare-a} we arrive at
\begin{eqnarray}
&&\hspace{-2cm} \calm^{qg,A}_{\rm Quark} = \sum_f{e^2e^2_f g^2\over s}
{q^+\over \pi}\int_0^1{dz\over z}
\int d^2w_1d^2z_1\,
{\dhd^2k_1\dhd^2k_2\over k^2_{1\perp} + Q^2z\barz}
{e^{-i(w_1-z_1,k_1-k_2)_\perp}\over k^2_{2\perp} + Q^2z\barz}
\nonumber\\
&&\hspace{-0cm} \times
\int_{-\infty}^{+\infty}\!dw_1^+dw_2^+\barpsi_{\rm cl}(w_{1\perp},w_1^+)t^{d'}[w_1^+,w_2^+]_{w_1}^{d'd}[w_1^+,-\infty n_1]_{w_1}
\nonumber\\
&&\hspace{-0cm}\times\Big[i(z^2+\barz^2)\veck_2\!\times\!\veck_1\ssn_1
- (\barz^2-z^2)(k_1,k_2)_\perp\gamma^5\ssn_1\Big]
\nonumber\\
&&\hspace{-0cm}\times[-\infty n_1,w_2^+]_{w_1}t^d\psi_{\rm cl}(w_{1\perp},w_2^+)+ O(\lambda^{-2})
\end{eqnarray}
We observe again that the term proportional to $\veck_1\times\veck_2$ is zero under integration, thus,
performing the color algebra, we have
\begin{eqnarray}
&&\hspace{-2.0cm} \calm^{qg,A}_{\rm Quark} 
= - 8\,q^+\alpha_{\rm em}C_F\sum_f{e^2_f \over s^2}
\int_0^1dz\,(\barz-z)
\int d^2w_1d^2z_1\dhd^2k_1\dhd^2k_2
\nonumber\\
&&\hspace{-0.4cm}
\times{e^{-i(w_1-z_1,k_1-k_2)_\perp}\over k^2_{1\perp} + Q^2z\barz}
{(k_1,k_2)_\perp\over k^2_{2\perp} + Q^2z\barz}
\Big({1\over z}Q_5(w_{1\perp}) - {1\over \barz}Q^\dagger_5(w_{1\perp})\Big)+ O(\lambda^{-2})
\nonumber\\
&&\hspace{-0.4cm} = - C_Fq^+{\alpha_{\rm em} \over s^2}{Q^2\over \pi^2}\sum_f e^2_f
\int_0^1dz\,(z-\barz)^2
\int d^2z_1d^2z_2\Big|K_1(\bar{Q}|z_{12}|)\Big|^2
\nonumber\\
&&~~\times\Big(Q_5(z_{1\perp}) + Q^\dagger_5(z_{1\perp})\Big)+ O(\lambda^{-2})\,.
\label{qg-antisymsquare-final}
\end{eqnarray}
In the last step we used that $\bar Q$ and the transverse integral are symmetric under $z\leftrightarrow\barz$.

\subsection{$\gamma^*\to \barq g$}

Here we calculate the process with $\barq g$ in the final state (see Fig.~\ref{fig:dipole-q2g}b).
The matrix element that describes the process is
\begin{eqnarray}
\hspace{-0.4cm}\theta(q^+)\langle\barq(p)g(k)|\gamma^*(q)\rangle
=\!\!\!&&\theta(q^+)\int d^4x\,e^{-iq\cdot x}\varepsilon_\mu(q)[-i\tilde{D}]^\mu_{~\sigma}
\int d^4z\,e^{ik\cdot z}\varepsilon^{*a}_\rho(k)[-i\tildeD_{ac}]^\rho_{~\nu}
\nonumber\\
&&\times\!
\int d^4y\bra{0} A^\nu_c(z)\barpsi^m(y)A^\sigma(x)\ket{0}(-i\overleftarrow{\sslash{\partial}}_y)v(p) e^{ip\cdot y}
\end{eqnarray}
We now proceed in the same way as we did for the $qg$ channel: we need the propagator
\eqref{qprop-xinsw} with interval $w_1^+<w_2^+$, and 
the outgoing gluon propagator~\eqref{gluonprop-ysw}, with $z^+>w_2^+$. Thus, we have
\begin{eqnarray}
&&\theta(q^+)\langle\barq(p)g(k)|\gamma^*(q)\rangle
\label{antiq-g}\\
=\!\!\!&&ee_fg\int d^4w_1\,e^{-iq\cdot w_1}\varepsilon_\mu(q)\int d^4w_2 d^4z\,e^{ik\cdot z}\varepsilon^{*a}_\rho(k)[-i\tildeD_{ac}]^\rho_{~\nu}
\int\dhd^4k'{(-i)d^{\nu\xi}(k')\over k'^2+i\epsilon}\,e^{-ik'\cdot (z-w_2)}
\nonumber\\
&&
\times\Big(d_{\xi\lambda}(k')[+\infty n_1,w_2^+]^{cd}_{w_2} - g{n_{2\lambda}g^i_{\perp\xi}\over k'^+}\int_{w_2^+}^{+\infty}dw^+[+\infty n_1,w^+]_{w_2}
{F_i}^{\;-}[w^+,w_2^+]_{w_2}\Big)
\nonumber\\
&&\times \barpsi_{\rm cl}(w_2^+,w_{2\perp})\gamma^\lambda t^d
\theta(w_2^+-w_1^+)\bigg[[w_2^+, -\infty n_1]_{w_2}\ssk_1\ssn_2
\nonumber\\
&& - g\!\int_{-\infty}^{w_2^+}dz^+\gamma^i\ssn_2[w_2^+,z^+]_{w_2}{F_i}^{\;-}(z^+,w_{2\perp})[z^+,-\infty n_1]_{w_2}
\bigg]\int \dhd^4 k_1\, {i\ssk_1\over 2k_1^+(k^2_1+i\epsilon)}\,e^{-ik_1\cdot (w_2-w_1)}
\nonumber\\
&&\times \gamma^\mu\bigg[i\theta(p^+)\theta(-w_1^+)U^\dagger_{z_1}{p^+ \ssn_1-\ssk_{2\perp}\over2p^+}\ssn_2
d^2z_1\dhd^2k_2\,e^{ip^+w_1^- + {i{k^2_{2\perp}\over 2p^+}w_1^+} - i(p+k_2,z_1)_\perp + i(k_2,w_1)_\perp}v(p)
\bigg]\,.
\nonumber
\end{eqnarray}
Using
\begin{eqnarray}
\barpsi_{\rm cl}\gamma_\perp^\lambda(k^+\ssn_1+\ssk_{1\perp})\ssn_2(k^+\ssn_1 + \ssk_{1\perp})
= 2k^+\barpsi_{\rm cl}\gamma_\perp^\lambda(k^+\ssn_1+\ssk_{1\perp}) + O(\lambda^{-2})
\end{eqnarray}
we obtain
\begin{eqnarray}
&&\hspace{-1.2cm}\theta(q^+)\langle\barq(p)g(k)|\gamma^*(q)\rangle
\nonumber\\
&&\hspace{-1.7cm}= - {ee_fg\over s}\int d^2w_2 d^2z_1 \dhd^2k_1\theta(p^+)\theta(k^+)\dbar\left({p^+\over q^+}+{k^+\over q^+} - 1\right)
\nonumber\\
&&\hspace{-1.5cm}\times
{e^{-i(w_2,k+k_1)_\perp}\over k^2_{1\perp}+{2\over s}Q^2k^+p^+}\,e^{- i(p-k_1,z_1)_\perp}\varepsilon^{*c}_\xi g^\xi_{\perp\lambda}[+\infty n_1,w_2^+]^{cd}_{w_2}
\nonumber\\
&&\hspace{-1.5cm}\times\!
\int_{-\infty}^{+\infty}dw^+_2\barpsi_{\rm cl}(w_2^+,w_{2\perp})\gamma_\perp^\lambda t^d[w_2^+,-\infty n_1]_{w_2}
(k^+\ssn_1-\ssk_{1\perp})\sslash{\varepsilon}(q)
(p^+\ssn_1+\ssk_{1\perp})\ssn_2 U^\dagger_{z_1}v(p)
\label{barqg-amplitude}
\end{eqnarray}
Result \eqref{barqg-amplitude} is the amplitude for the $\barq g$ channel.
In the next sections we will consider the longitudinal, transverse, and anti-symmetric photon polarization for the amplitude squared.

\subsubsection{Longitudinal polarization for $\gamma^*\to \barq g$}

Let us start with the longitudinal polarization. The Dirac matrix with the longitudinal photon polarization is
\begin{eqnarray}
&&\barpsi_{cl}\gamma_\perp^\lambda(k^+\ssn_1+\ssk_{1\perp})\Big({q^+\over Q}\ssn_1 + {Q\over 2q^+}\ssn_2\Big)
(p^+\ssn_1-\ssk_{1\perp})\ssn_2 v(p)
\nonumber\\
=\!\!\!&& \barpsi_{cl}\gamma_\perp^\lambda\Big(-{q^+\over Q}k^2_{1\perp} + {p^+k^+Q\over q^+}\Big)2v(p) + O(\lambda^{-2})
\end{eqnarray}
So, the amplitude square is
\begin{eqnarray}
&&\hspace{-2.0cm}\calm^{\barq g,L}_{\rm Quark} \equiv 
{1\over 2\pi\delta(0)} \int \dhd^4p\dhd^4k\dbar(p^2)\dbar(k^2)\Big|\langle\barq(p)g(k)|\gamma^*_L(q)\rangle\Big|^2\theta(p^+)\theta(k^+)
\nonumber\\
&&\hspace{-0.6cm}=\sum_f{e^2e^2_fg^2\over s^2}\int \dhd^4p\dhd^4k\dbar(p^2)\dbar(k^2)
\theta(p^+)\theta(k^+)\dbar\left({p^+\over q^+}+{k^+\over q^+} - 1\right)
\nonumber\\
&&\hspace{-0.2cm}\times
\Bigg| \int d^2w_2 d^2z_1 \dhd^2k_1
{e^{-i(w_2,k+k_1)_\perp}\over k^2_{1\perp}+{2\over s}Q^2k^+p^+}\,e^{- i(p-k_1,z_1)_\perp}\varepsilon^{*c}_\xi g^\xi_{\perp\lambda}[+\infty n_1,w_2^+]^{cd}_{w_2}
\nonumber\\
&&\hspace{-0.2cm}\times\!
\int_{-\infty}^{+\infty}dw^+_2\barpsi_{\rm cl}(w_2^+,w_{2\perp})\gamma_\perp^\lambda t^d[w_2^+,-\infty n_1]_{w_2}
(k^+\ssn_1-\ssk_{1\perp})\sslash{\varepsilon}(q)
(p^+\ssn_1+\ssk_{1\perp})\ssn_2 U^\dagger_{z_1}v(p)\Bigg|^2
\nonumber\\
&&\hspace{-0.6cm}
= - {q^+\over \pi s^2}\sum_f e^2e_f^2 \int_0^1{dz\over z}
\int d^2w_1d^2z_1
\int \dhd^2 k_1 \dhd^2 k_2\, e^{-i(w_1-z_1,k_1-k_2)_\perp}
\nonumber\\
&&\hspace{-0.2cm}~~\times {\Big(Qz\barz - {k^2_{1\perp}\over Q}\Big)\over k^2_{1\perp} + Q^2z\barz}
{\Big(Qz\barz - {k^2_{2\perp}\over Q}\Big)\over k^2_{2\perp} + Q^2z\barz}
C_F\Big(Q_1(w_{1\perp}) + Q_1^\dagger(w_{1\perp})\Big)
\label{barqg-amplsquare}
\end{eqnarray}
So, we obtain that
\begin{eqnarray}
\calm^{q g,L}_{\rm Quark}  = - \calm^{\barq g,L}_{\rm Quark} 
\label{Codd-result}
\end{eqnarray}
Therefore, when $x_B$ is set to zero in the quark operators, the channels $\gamma^*\to qg$ and
$\gamma^*\to\bar qg$ do not contribute to $F_L$. The $\gamma^*\to qg$ channel is the charge conjugate of
$\gamma^*\to\bar qg$, and the vector operator $Q_1+Q_1^\dagger$ is $C$-odd, so result \eqref{Codd-result} was expected.
As shown in Sec.~\ref{sec:xBreal}, this is a property of the operators at zero momentum fraction. With the
$x_B$ dependence kept, the two channels probe the vector light-ray operator at the momentum fractions
$+x(k)$ and $-x(k)$, where $x(k)=x_B(1+k^2/\bar Q^2)\ge x_B$ is fixed by momentum conservation of the minus-component. Their
sum is a $C$-even combination that is odd in the momentum fraction and does not vanish for $x_B>0$.

\subsubsection{Transverse polarization for $\gamma^*\to \barq g$}

In order to calculate the $\gamma^*\to \barq g$ channel with transverse polarization, we need to square the Dirac matrix element
of the amplitude \eqref{barqg-amplitude} and average over the photon helicity. We obtain
\begin{eqnarray}
&&\half\sum_{\lambda=\pm}\sum_{\delta,\sigma}\Big|\varepsilon^{*c}_{\alpha,\delta}
\barpsi_{\rm cl}(w_2^+,w_{2\perp})\gamma_\perp^\alpha t^d[w_2^+,-\infty n_1]_{w_2}
(k^+\ssn_1-\ssk_{1\perp})
\nonumber\\
&&\hspace{1.5cm}\times\sslash{\varepsilon}_T(q)
(p^+\ssn_1+\ssk_{1\perp})\ssn_2 U^\dagger_{z_1}v(p)\Big|^2
\nonumber\\
&& = - 8\delta^{cc'}p^+\Tr\{\barpsi(w_2^+,w_{2\perp})t^d[w_2^+,-\infty n_1]_{w_2}U^\dagger_{z_1}
\nonumber\\
&&~~~\times
\bigg[ \ssn_1\Big((p^+)^2 + (k^+)^2\Big)(k_1,k_2)_\perp+ \,i\,\ssn_1\gamma^5\Big((p^+)^2 - (k^+)^2\Big)\veck_1\times\veck_2\bigg]
\nonumber\\
&&~~~\times U_{z_2}[-\infty n_1, w_1^+]_{w_1}t^{d'}\psi(w_1^+,w_{1\perp})\}
+ \calo(\lambda^{-2})
\label{barqg-matrixsquare-final}
\end{eqnarray}
Thus, using \eqref{barqg-matrixsquare-final}, and repeating the same step that led us to $\calm^{q g, T}_{\rm Quark}$, we obtain
\begin{eqnarray}
&&\hspace{-1.4cm}\calm^{\barq g, T}_{\rm Quark} \equiv
{1\over 2\pi\delta(0)} \int \dhd^4p\dhd^4k\dbar(p^2)\dbar(k^2)\Big|\langle\barq(p)g(k)|\gamma^*_T(q)\rangle\Big|^2\theta(p^+)\theta(k^+)
\nonumber\\
&& = - q^+C_F{\alpha_{\rm em} \over 2s^2}{Q^2\over \pi^2}\sum_fe^2_f \int_0^1dz\,(z^2+\barz^2)
\int d^2z_1d^2z_2\Big|K_1(\bar{Q}|z_{12}|)\Big|^2
\nonumber\\
&&~~~~\times\!\Big[Q_1(z_{1\perp}) + Q^\dagger_1(z_{1\perp})\Big]  + O(\lambda^{-2})
\label{barqgsquare-transverse}
\end{eqnarray}
Comparing \eqref{barqgsquare-transverse} with \eqref{qgsquare-transverse-4} we have
\begin{eqnarray}
\calm^{\barq g, T}_{\rm Quark} = - \calm^{q g, T}_{\rm Quark}
\label{Codd-transverse}
\end{eqnarray}
Therefore their sum cancels when the quark operators are evaluated at $x_B=0$, for the same reason as in
eq.~\eqref{Codd-result}.

\subsubsection{Asymmetry for $\gamma^*\to \barq g$}

Finally, let us consider the asymmetric sum of the photon helicity.  The square matrix element with asymmetric sum is
\begin{eqnarray}
&&\sum_{\lambda = {\rm asym.}}\sum_{\delta,\sigma}\Big|\varepsilon^{*c}_{\alpha,\delta}
\barpsi_{\rm cl}(w_2^+,w_{2\perp})\gamma_\perp^\alpha t^d[w_2^+,-\infty n_1]_{w_2}
(k^+\ssn_1-\ssk_{1\perp})
\nonumber\\
&&\hspace{1.8cm}\times\sslash{\varepsilon}_T(q)
(p^+\ssn_1+\ssk_{1\perp})\ssn_2 U^\dagger_{z_1}v(p)\Big|^2
\nonumber\\
&&\hspace{-0.3cm}
= 16\delta^{cc'}p^+\tr\Tr\{\barpsi(w_2^+,w_{2\perp})t^d[w_2^+,-\infty n_1]_{w_2}U^\dagger_{z_1}
\nonumber\\
&&\hspace{0.1cm}\times\Big[
i\ssn_1\Big((p^+)^2 + (k^+)^2\Big)\veck_1\times\veck_2 - \gamma^5\ssn_1\Big((p^+)^2 - (k^+)^2\Big)(k_1,k_2)_\perp\Big]
\nonumber\\
&&\hspace{0.1cm}\times U_{z_2}[-\infty n_1, w_1^+]_{w_1}t^{d'}\psi(w_1^+,w_{1\perp})\}
+ \calo(\lambda^{-2})
\label{barqg-asymatrixsqua-final}
\end{eqnarray}
Using result \eqref{barqg-asymatrixsqua-final}, and repeating the same steps we performed to obtain \eqref{qg-antisymsquare-final},
we obtain the amplitude square with asymmetric helicity sum for the $\gamma^*\to \barq g$ channel
\begin{eqnarray}
&&\hspace{-1.4cm}\calm^{\barq g,A}_{\rm Quark}\equiv 
{1\over (2\pi)\delta(0)}\int \dhd^4 p \dhd^4k\dbar(p^2)\dbar(k^2)\Big|\langle \barq(p) g(k)|\gamma^*_A(q)\rangle\Big|^2\theta(p^+)\theta(k^+)
\nonumber\\
&&= - C_Fq^+{\alpha_{\rm em} \over s^2}{Q^2\over \pi^2}\sum_f e^2_f
\int_0^1dz\,(z-\barz)^2
\int d^2z_1d^2z_2\Big|K_1(\bar{Q}|z_{12}|)\Big|^2
\nonumber\\
&&~~~\times\!\Big(Q_5(z_{1\perp}) + Q^\dagger_5(z_{1\perp})\Big)+ O(\lambda^{-2})\,.
\label{barqg-antisymsquare-final}
\end{eqnarray}
Thus, comparing \eqref{barqg-antisymsquare-final} with \eqref{qg-antisymsquare-final}, we obtain that
\begin{eqnarray}
&&\calm^{\barq g,A}_{\rm Quark} + \calm^{q g,A}_{\rm Quark}
\nonumber\\
=\!\!\!&& - 2C_Fq^+{\alpha_{\rm em} \over s^2}{Q^2\over \pi^2}\sum_f e^2_f
\int_0^1dz\,(z-\barz)^2
\int d^2z_1d^2z_2\Big|K_1(\bar{Q}|z_{12}|)\Big|^2
\nonumber\\
&&~~~\times\!\Big(Q_5(z_{1\perp}) + Q^\dagger_5(z_{1\perp})\Big)+ O(\lambda^{-2})
\label{Ceven-result}
\end{eqnarray}
An important consequence of eq.~\eqref{Ceven-result} is that the charge-conjugate real-emission channels behave differently in the
unpolarized and helicity-dependent projections. In the longitudinal and transverse cases the contributions proportional to the vector operator $Q_1$
cancel between the $\gamma^*\to qg$ and $\gamma^*\to\bar qg$ channels when $x_B$ is set to zero in the
operators, as shown in eqs.~\eqref{Codd-result} and \eqref{Codd-transverse}. In contrast, the axial combination $Q_5+Q_5^\dagger$ is
$C$-even, and the two contributions add in the helicity-dependent asymmetry. As shown in Sec.~7.4, this
remains true when the $x_B$ dependence is kept: the corresponding $C$-even axial light-ray operator is even
in the momentum fraction of the target quark, whereas the vector one is odd and vanishes when that fraction
is set to zero.

The surviving contribution in eq.~(\ref{Ceven-result}) is singular both in the transverse short-distance region and at the longitudinal
endpoints.  Indeed, for $|z_{12}|\to0$ we have
\begin{eqnarray}
Q^2 (z-\bar z)^2
K_1^2\!\left(\bar Q |z_{12}|\right)
\simeq \frac{1}{z_{12}^2}
{(z-\bar z)^2\over z\barz}
= {1\over z_{12}^2}
\left({1\over z}+{1\over \barz}-4
\right).
\end{eqnarray}
The transverse integration contains a logarithmic
short-distance singularity. The longitudinal endpoint terms appear at both $z=0$ and $z=1$ in this symmetric representation. 
The resulting logarithmic regions are discussed in Sec.~\ref{sec:divergence}.

\subsection{Restoring the $x_B$ dependence: a sketch}
\label{sec:xBreal}

We outline how the $x_B$ dependence enters the real-emission channels; a complete treatment is left for
future work. In eq.~\eqref{phase-approxy} we used the shock-wave approximation and set to one the phase $e^{iEw_2^+}$ of the
photon-vertex integration,\footnote{Once this phase is kept, the constraint $\theta(-w_1^+)$ of eq.~\eqref{qg-production},
which places the photon vertex before a shock wave of zero width, must be replaced by $w_1^+<w_2^+$, which
is already implied by the $w_1^+$ integration. Otherwise the result would depend on the position of the
origin of $w^+$ relative to the target. The added region, $0<w_1^+<w_2^+$, has vanishing measure in the
shock-wave limit.} where
\begin{eqnarray}
E={k^2_\perp+\bar Q^2\over 2z\bar z\,q^+}=x(k)\,P^-,\qquad x(k)\equiv x_B\Big(1+{k^2_\perp\over \bar Q^2}\Big)\ge x_B,
\label{eq:xB-E}
\end{eqnarray}
is the minus component of the momentum that the target quark must supply to produce the on-shell final state with relative
transverse momentum $k_\perp$. Keeping the phase, the quark bilinears of eqs.~\eqref{Mquark-qqL}, \eqref{qgsquare-transverse-4} 
and \eqref{qg-antisymsquare-final} become the full-plane light-ray operators
\begin{equation}
\calo_\Gamma(w_\perp;x)\equiv g^2\frac s2\int\!du^+\!\int\!dv^+\,e^{ixP^-(u^+-v^+)}\,
\bar\psi(v^+,w_\perp)[v^+,u^+]_w\,\Gamma\,\psi(u^+,w_\perp),
\label{caloGamma}
\end{equation}
with $\Gamma=\ssn_1$ ($\calo_V$) or $\gamma^5\ssn_1$ ($\calo_A$). Here $x$ is the
longitudinal momentum fraction of the target quark, conjugate to the light-cone separation $u^+-v^+$, and
$k\equiv|k_\perp|$. In the $qg$ channel $x=x(k)$; in the
$\bar qg$ channel, eqs.~\eqref{barqg-amplsquare}, \eqref{barqgsquare-transverse} and \eqref{barqg-antisymsquare-final}, 
the fields are exchanged between the amplitude and the
conjugate amplitude, and $x=-x(k)$. At zero momentum fraction, $\mathcal O_V(0)=Q_1+Q_1^\dagger$ and
$\mathcal O_A(0)=Q_5+Q_5^\dagger$. Since $C\,\mathcal O_V(x)\,C^{-1}=-\mathcal O_V(-x)$ and
$C\,\mathcal O_A(x)\,C^{-1}=\mathcal O_A(-x)$~\cite{Collins:1981uw, Jaffe:1991ra, Jaffe:1991kp, Diehl:2003ny}, the sum of the two channels
projects onto $\mathcal O_V(x)-\mathcal O_V(-x)$ for $F_L$ and $F_T$, which is odd in $x$ and vanishes at
$x=0$, and onto $\mathcal O_A(x)+\mathcal O_A(-x)$ for $g_1$, which is even in $x$. This is the origin of
eqs.~\eqref{Codd-result}, \eqref{Codd-transverse} and \eqref{Ceven-result}. 
In refs.~\cite{Altinoluk:2025ivn, Altinoluk:2025ang} the $x=0$ operator is promoted to an $x$-dependent one and subsequently evaluated at $x=x_B$, 
a replacement argued there to affect only the next-to-next-to-eikonal accuracy. Here, instead, the momentum fraction 
$x=x(k)$ follows directly from retaining the longitudinal phase of the photon vertex.

The quark and helicity distributions are defined by the light-ray operators of ref.~\cite{Chirilli:2026pkv},
\begin{equation}
	\begin{aligned}
		q_f(x)&=\frac1{4\pi}\!\int\!d\Delta^+e^{ixP^-\Delta^+}\langle N(P)|\bar\psi_f(0)[0,\Delta^+]\ssn_1\psi_f(\Delta^+)|N(P)\rangle,\\
		\Delta q_f(x)&=\frac1{4\pi}\!\int\!d\Delta^+e^{ixP^-\Delta^+}\langle N(P)|\bar\psi_f(0)[0,\Delta^+]\ssn_1\gamma^5\psi_f(\Delta^+)|N(P)\rangle,
	\end{aligned}
	\label{eq:pdfdef}
\end{equation}
for a positive-helicity target, with $-\barq_f(x)$ and $\Delta\barq_f(x)$ the values of the same
integrals at $-x$. For an operator integrated over the transverse plane and over the light-cone positions
of its fields, the forward matrix element is proportional to the longitudinal and transverse volume
factors $\dbar(0)\,\dbar^{(2)}(0)=\int dX^+d^2w_\perp$, as in ref.~\cite{Chirilli:2026pkv}, and we write
\begin{equation}
	\langle N(P)|\!\int\!d^2w_\perp\,\mathcal O(w_\perp)|N(P)\rangle\equiv\dbar(0)\,\dbar^{(2)}(0)\,
	\langle\!\langle\mathcal O\rangle\!\rangle .
	\label{eq:dbracket}
\end{equation}
With eqs.~\eqref{eq:pdfdef}, for $x>0$,
\begin{equation}
	\begin{aligned}
		\langle\!\langle\calo^f_V(x)-\calo^f_V(-x)\rangle\!\rangle&=2\pi g^2s\,\big[q_f+\bar q_f\big](x),\\
		\langle\!\langle\calo^f_A(x)+\calo^f_A(-x)\rangle\!\rangle&=-2\pi g^2s\,\big[\Delta q_f+\Delta\bar q_f\big](x),
	\end{aligned}
	\label{eq:dict}
\end{equation}
where the sign of the axial matrix element follows from $\gamma^5\ssn_1=-\ssn_1\gamma^5$ in
eq.~\eqref{Q5}. In eqs.~\eqref{SumLongy}--\eqref{SumAsym} each operator stands for $\langle\!\langle\mathcal O\rangle\!\rangle/2P^-$, the
factor $2P^-\dbar(0)\dbar^{(2)}(0)=\langle N(P)|N(P)\rangle$ being the normalization of the target
state; with $P^-=q^+$ this is the rule for which eq.~\eqref{SumLongy} is the standard dipole expression of $F_L$. The
two channels are then separately proportional to $q_f(x(k))$ and $\bar q_f(x(k))$ in $F_L$ and $F_T$, as
required by positivity of each squared amplitude.

For the massless, dilute leading-twist matching of $F_L$, we use
the effective longitudinal vector
\begin{eqnarray}
\varepsilon_{L,\mathrm{eff}}^\mu
=\varepsilon_L^\mu- {q^\mu\over Q}
={Q\over q^+}n_2^\mu.
\end{eqnarray}
The electromagnetic Ward identity makes this replacement equivalent
to using eq.~\eqref{epsL-final} in the complete amplitude.%
\footnote{In this matching, the incoming target quark is massless
and collinear, $r^\mu=r^-n_2^\mu$, so that
$\gamma_\mu n_2^\mu u(r)=0$. Consequently, the additional diagram
in which the photon is absorbed by the incoming target quark has
zero contraction with $\varepsilon_{L,\mathrm{eff}}$.
With the polarization of eq.~\eqref{epsL-final}, this diagram instead
cancels the contact term in the longitudinal kernel of Fig.~\ref{fig:dipole-q2g}:
\[
\frac{Qz\bar z-k_\perp^2/Q}{k_\perp^2+Q^2z\bar z}
=
\frac{2Qz\bar z}{k_\perp^2+Q^2z\bar z}-\frac{1}{Q}.
\]
The cancellation takes place at amplitude level, before squaring. This argument uses the collinear leading-twist projection of the background quark.}
Changing variables from $k_\perp^2$ to $x$ using
Eq.~\eqref{eq:xB-E}, we obtain
\begin{eqnarray}
F_L^{qg+\bar qg}&=&{\alpha_sC_F\over \pi}\,x_B\sum_fe_f^2\int_{x_B}^1\frac{dx}{x}\,\frac{x_B}{x}\,
\big[q_f+\bar q_f\big](x),
\label{eqxB-FL}
\\
F_T^{qg+\bar qg}&=&{\alpha_sC_F\over 4\pi}\,x_B\sum_fe_f^2\int_0^1\!dz\,\frac{z^2+\bar z^2}{z\bar z}
\int_{x_B}^1\frac{dx}{x}\Big(1- {x_B\over x}\Big)\big[q_f+\bar q_f\big](x),
\label{eqxB-FT}
\\
g_1^{qg+\bar qg}&=&{\alpha_sC_F\over 8\pi}\sum_fe_f^2\int_0^1\!dz\,\frac{(z-\bar z)^2}{z\bar z}
\int_{x_B}^1\frac{dx}{x}\Big(1-\frac{x_B}{x}\Big)\big[\Delta q_f+\Delta\bar q_f\big](x).
\label{eqxB-g1}
\end{eqnarray}
Notice that the factor $1-x_B/x$ in eqs.~\eqref{eqxB-FT} and \eqref{eqxB-g1} 
vanishes at the lower endpoint $x=x_B$. Using eq.~\eqref{eq:xB-E}, this endpoint corresponds to $k_\perp^2=0$, since
\begin{eqnarray}
1-\frac{x_B}{x} = {k_\perp^2\over k_\perp^2+\bar Q^2}
\end{eqnarray}
Thus, the vanishing occurs only at the boundary of the real-emission phase space and does not imply that the integrated contributions 
to $F_T$ and $g_1$ vanish.

Equation~\eqref{eqxB-FL} reproduces the Altarelli--Martinelli~\cite{Altarelli:1978tq} quark
coefficient in the massless, dilute leading-twist limit.
With the effective longitudinal contraction used above, the
additional photon--background-quark attachment vanishes, and the
diagrams of Fig.~\ref{fig:dipole-q2g} supply the complete $O(\alpha_s)$ quark-initiated
contribution to $F_L$ in this limit.
The virtual correction to the one-particle channel gives no
contribution to $F_L$ at this order because the massless collinear
Born longitudinal amplitude vanishes.
With the polarization of eq.~\eqref{epsL-final}, the additional
real-emission attachment must instead be included to obtain the same result.
Equations \eqref{eqxB-FT} and \eqref{eqxB-g1} are instead only the contribution of the Fig.~\ref{fig:dipole-q2g} diagrams. 
The diagram in which the photon is absorbed by the background quark and the gluon is emitted after the shock wave, and
the virtual corrections to the one-particle channel, contribute to $F_T$ and $g_1$ at the same order.
Without them the coefficient of the endpoint (collinear) logarithm is $C_F(1-x_B/x)$ instead of the full
$P^{(0)}_{qq}=\Delta P^{(0)}_{qq}$, and the result depends on the gauge of the emitted gluon. 
The $z$ integrals in eqs.~\eqref{eqxB-FT} and \eqref{eqxB-g1} diverge logarithmically at the endpoints. This is the
collinear singularity of the incoming target quark; with a cutoff $\delta$ on $z$ and $\bar z$,
\begin{eqnarray}
F_T^{qg+\bar qg}\simeq {\alpha_s\over 2\pi}\ln\frac1\delta\;x_B\sum_fe_f^2\int_{x_B}^1\frac{dx}{x}\,
C_F\Big(1- {x_B\over x}\Big)\big[q_f+\bar q_f\big](x),
\end{eqnarray}
and the same coefficient multiplies $\frac12\sum_fe_f^2[\Delta q_f+\Delta\bar q_f]$ in $g_1^{qg+\bar qg}$.
In the high-energy formalism this logarithm is the one absorbed into the rapidity evolution of the quark
operators, Sec.~\ref{sec:divergence}. For $F_T$ and $g_1$ the
transverse integration is finite: the short-distance logarithm of eqs.~\eqref{qgsquare-transverse-4} and \eqref{Ceven-result} is replaced by the
$x$ integration, which for constant distributions gives $\ln(1/x_B)-1+x_B$. Finally, the vanishing at $x=0$ is not the limit $x_B\to0^+$:
the $C$-even vector combination vanishes at $x=0$, whereas its one-sided
limit as $x\to0^+$ is proportional to $q(0)+\bar q(0)$. For distributions
regular at $x=0$, eq.~\eqref{eqxB-FL} gives
\begin{eqnarray}
F_L^{qg+\bar qg}/x_B\to {\alpha_sC_F\over \pi}\sum_fe_f^2[q_f+\bar q_f](0)\neq 0\,.
\end{eqnarray}
The complete finite-$x_B$
analysis, including the missing diagrams and the $x_B$ dependence of the $q\bar q$ channel, will be
presented elsewhere.

\section{Summary of results for the dipole sub-eikonal corrections}
\label{sec:summaryresults}

In this section we collect the contributions of the $q\barq$,
$qg$, and $\bar q g$ final states, computed with the $x_B$ dependence of the
sub-eikonal operators neglected. These results have to be added to the contribution of
the one-particle final states $\gamma^*\to q$ and $\gamma^*\to\bar q$, in
which one end-point of the quark propagator lies inside the shock wave. 
We computed this channel in ref.~\cite{Chirilli:2026pkv}, obtaining
\begin{eqnarray}
&&F_L^{q,\barq} = O(\lambda^{-2})\,,
\label{one-particle-LO-Fl}\\
&&F_T^{q,\barq}(x_B) = x_B\sum_f e_f^2\,\big[q_f(x_B)+\bar q_f(x_B)\big]\,,
\label{one-particle-LO-Ft}\\
&&g_1^{q,\barq}(x_B) = \half\sum_f e_f^2\,\big[\Delta q_f(x_B)+\Delta\bar q_f(x_B)\big]\,,
\label{one-particle-LO-g1}
\end{eqnarray}
with the quark and helicity distributions defined by the light-ray operators
of ref.~\cite{Chirilli:2026pkv}. In the one-particle channel the momentum fraction of the target quark is $x=x_B$. When $x_B$ is set to
zero in these operators, the quark and antiquark contributions to $F_T^{q,\bar q}$ cancel, because the
$C$-even vector combination vanishes at zero momentum fraction (Sec.~\ref{sec:xBreal}), whereas those to
$g_1^{q,\bar q}$ add, and $g_1^{q,\bar q}$ becomes proportional to
\begin{eqnarray}
{2\over g^2 s}\sum_f e_f^2\int d^2x_\perp\,
\big[Q^f_5(x_\perp)+Q^{f\dagger}_5(x_\perp)\big]\,,
\label{Q5ops}
\end{eqnarray}
with $Q^f_5$ and $Q^{f\dagger}_5$ the operators of eqs.~\eqref{Q5} and
\eqref{Q5dagger} at $x_B=0$. The quark-operator contributions of the
$q\bar q$, $qg$, and $\bar q g$ final states carry instead one more factor
$g^2$: the quark-background propagator \eqref{qprpinquark-1} is of order
$g^2$, and the $qg$ and $\bar q g$ amplitudes are of order $g$. They are
therefore of relative order $\alpha_s$ with respect to
eq.~\eqref{one-particle-LO-g1}, up to the logarithmic enhancements discussed in
Sec.~\ref{sec:divergence}. Moreover, as shown in ref.~\cite{Chirilli:2026pkv}, setting
$x_B=0$ in the operators before the phase-space integration does not commute
with the small-$x_B$ limit of the distributions; the statements below about
$F_L$ and $F_T$ refer to the former.

\subsection{$F_L$ structure function up to sub-eikonal corrections}

The longitudinal and helicity-averaged transverse
projections are related to $F_L$ and $F_T$ by the
normalization established in Sec.~\ref{sec:dipolecrossection}. In the longitudinal
projection, the gluonic contributions vanish, as shown
in Secs.~\ref{sec:Longy-calf} and \ref{sec:LongyFprimeF2Fi}, and the quark-background
$q\bar q$ contribution vanishes according to eq.~\eqref{QdipoleLongy-2}.
The $qg$ and $\bar qg$ contributions cancel according
to eq.~\eqref{Codd-result}. Thus, for the $q\bar q$, $qg$, and $\bar q g$ final states at sub-eikonal
accuracy and setting the longitudinal phases to unity in the operators, $F_L$ is given by its eikonal
dipole expression (the one-particle final states do not contribute to $F_L$
at this order, see eq.~\eqref{one-particle-LO-Fl}):

\begin{eqnarray}
\hspace{-0.8cm}F_L(Q^2) 
=\!\!\!&& {Q^2\over 8\pi^2\alpha_{\rm em}}\calm^L_{\rm Eikonal} + O(\lambda^{-2})
\nonumber\\
=\!\!\!&&{N_cQ^4\over \pi^4}\sum_f e_f^2
\int_0^1\!dz \,z^2\barz^2\int d^2z_1d^2z_2\left|K_0(\bar{Q}|z_{12}|)\right|^2
\calu(z_1,z_2)
 + O(\lambda^{-2})
\label{SumLongy}
\end{eqnarray}
Equation~\eqref{SumLongy} refers to the approximation in which the
longitudinal phases of the sub-eikonal operators are set to unity.
Retaining these phases and performing the massless, dilute
leading-twist matching described in Sec.~\ref{sec:xBreal} gives
the nonzero quark contribution in eq.~\eqref{eqxB-FL}, which
reproduces the Altarelli--Martinelli coefficient.
The cancellation at zero operator momentum fraction therefore
does not establish the small-$x_B$ behaviour of the physical
structure function. A finite-$x_B$ extension in a general background requires the
corresponding kinematics and operator matching for all contributions
retained at the chosen accuracy.

\subsection{$F_T$ structure function up to sub-eikonal corrections}

The same conclusion holds for the helicity-averaged
transverse projection. The gluonic corrections vanish
according to eqs. \eqref{FiF2F2primeSquqre-1tran-b} and \eqref{dipolesubGluon-2}, the quark-background
$q\bar q$ contribution vanishes according to eq.~\eqref{barqq-transverseQ},
and the charge-conjugate real-emission contributions
cancel according to eq.~\eqref{Codd-transverse}. Consequently, when the momentum fraction in the sub-eikonal operators is set to zero, 
the $q\bar q$, $qg$, and $\bar q g$ final states leave $F_T$ with its eikonal dipole expression:
\begin{eqnarray}
\hspace{-1cm}F_T(Q^2) =\!\!\!&& 
{Q^2\over 8\pi^2\alpha_{\rm em}}\calm^T_{\rm Eikonal}  + O(\lambda^{-2})
\nonumber\\
=\!\!\!&& {N_cQ^4\over 4\pi^4}\sum_fe_f^2\int_0^1dz\,z\barz\,(z^2+\barz^2)
\int d^2z_1 d^2z_2\left|K_1(\bar{Q}|z_{12}|)\right|^2\calu(z_1,z_2) 
+ O(\lambda^{-2})
\label{SumTransv}
\end{eqnarray}
At finite $x_B$, eq.~\eqref{SumTransv} must be supplemented by
the one-particle contribution $F_T^{q,\bar q}$ of eq.~\eqref{one-particle-LO-Ft},
and the $qg$ and $\bar qg$ channels no longer cancel (Sec.~\ref{sec:xBreal}).
The cancellation obtained by setting the longitudinal phases
of the operators to unity does not determine the one-sided
small-$x_B$ behaviour of
$F_T^{q,\bar q}=x_B\sum_f e_f^2[q_f+\bar q_f](x_B)$.
This behaviour depends on the parton distributions; in particular,
if they have finite limits as $x_B\to0^+$, then
$F_T^{q,\bar q}=O(x_B)$.

\subsection{$g_1$ structure function: one-particle term and $q\barq$, $qg$, $\barq g$ final states}

The leading sub-eikonal contribution to the helicity-dependent structure
function is the one-particle term $g_1^{q,\barq}$ of eq.~\eqref{one-particle-LO-g1},
which, unlike $F_T^{q,\barq}$, survives setting the longitudinal phases to unity in the operators. The
$q\bar q$, $qg$, and $\bar q g$ final states add the gluonic insertions,
eqs.~\eqref{FiF2F2primeSquqre-1Asym-2} and \eqref{asymmGluonsub}, the
quark-background contribution to the $q\bar q$ final state,
eq.~\eqref{qqbar-QuarkAmpsquare}, and the charge-conjugate real-emission
channels, eq.~\eqref{Ceven-result}. We write

\begin{eqnarray}
g_1(x_B,Q^2) = g_1^{q,\barq}(x_B) + g_1^{q\bar q+qg+\bar q g}(Q^2) + O(\lambda^{-2})\,,
\label{g1-total}
\end{eqnarray}
where the quark sector of the second term is of relative order $\alpha_s$ with
respect to the first. Combining the contributions of the $q\bar q$, $qg$, and
$\bar q g$ final states, and using $q^+=\sqrt{s/2}$, we obtain
\begin{eqnarray}
\hspace{-1cm}g_1^{q\bar q+qg+\bar q g}(Q^2) =\!\!\!&& {Q^2\over 32\pi^2\alpha_{\rm em}x_B}
\Big(\calm^A_{\rm Gluon} + \calm^A_{\rm F_i} + \calm^{q\barq, A}_{\rm Quark}
+ \calm^{qg, A}_{\rm Quark} + \calm^{\barq g,A}_{\rm Quark}\Big) + O(\lambda^{-2})
\nonumber\\
=\!\!\!&& 
{Q^4\over 16\pi^4x_B}{\sqrt{s/2}\over  s^2}
\sum_f e^2_f\int_0^1 \!  dz\int \!d^2z_1 d^2z_2 \,
\left|K_1(\bar{Q}|z_{12}|)\right|^2
\nonumber\\
&&\times \bigg\{(z-\barz)^2\Big(
\Tr\{U_{z_2}\calq^\dagger_{5z_1}\}  + \Tr\{\calq_{5z_2}U^\dagger_{z_1}\} 
+ \Tr\{\calf_{z_1}U^\dagger_{z_2}\} + \Tr\{U_{z_1}\calf_{z_2}^\dagger\}\Big)
\nonumber\\
&&~~ + 4(z^2 + \barz^2){z_{12}^j\epsilon^{ji} \over  z^2_{12}}
\Big(\Tr\{\calf_{iz_2}^\dagger U_{z_1}\} + \Tr\{\calf_{iz_2}U^\dagger_{z_1}\} \Big)\bigg\}
+ O(\lambda^{-2})\,.
\label{SumAsym}
\end{eqnarray}
Notice that eq.~\eqref{SumAsym} is real. Under Hermitian conjugation,
the quark and $\mathcal F$ terms are mapped into their corresponding
partners after the dummy interchange $z_1\leftrightarrow z_2$, while
the $\mathcal F_i$ combination is Hermitian by cyclicity of the color
trace; the integration measure and coefficient functions are invariant
under this relabeling.

Let us define the operators
\begin{eqnarray}
&&\calf_{xy} = \calf(x_\perp,y_\perp) \equiv \Tr\{U^\dagger_x\calf_y\}
\label{calFxy}
\nonumber\\
&&\calf_{i\,xy} = \calf_i(x_\perp,y_\perp) \equiv \Tr\{U^\dagger_x\calf_{iy}\}
\label{calFixy}
\end{eqnarray}
and
\begin{eqnarray}
&&\calq^f_{5\,xy} = \calq^f_5(x_\perp,y_\perp) \equiv Q^f_{5\,x}\calu_{xy}
\label{Q5xy}
\\
&&\Psi^f_{5\,xy} = \Psi^f_5(x_\perp,y_\perp) \equiv \Tr\{\tildeQ^f_{5x}\big(U^\dagger_x-U^\dagger_y\big)\}
\label{Psi5xy}
\end{eqnarray}
with
\begin{eqnarray}
&&\hspace{-2.8cm}\tildeQ^f_{5ij}(x_\perp) \equiv g^2{s\over 2}\!\!\int_{-\infty}^{\infty}dz^+\!\int_{-\infty}^{z^+}\!dz'^+
\big([\infty n_1, z^+]_x\tr\{\psi_f(z^+,x_\perp)\barpsi_f(z'^+,x_\perp)\gamma^5\ssn_1\}[z'^+,-\infty n_1]_x\big)_{ij}
\label{Qt5}
\end{eqnarray}
and observe that
\begin{eqnarray}
\Tr\{\calq^f_{5z_1}U^\dagger_{z_2}\} =\!\!\!&&
-\half\Tr\{U_{z_1}U^\dagger_{z_2}\}Q^f_{5z_1} + {1\over 2N_c}\Psi^f_{5z_1z_2} + {1\over 2N_c}Q^f_{5z_1}
\nonumber\\
=\!\!\!&& {N_c\over 2}\calq^f_{5z_1z_2} + {1\over 2N_c}\Psi^f_{5z_1z_2} - C_FQ^f_{5z_1}\,.
\label{rewritecalQ5}
\end{eqnarray}
Thus, result (\ref{SumAsym}) can be rewritten as
\begin{eqnarray}
\hspace{-0.2cm} g_1^{q\bar q+qg+\bar q g}(Q^2) =
\!\!\!&& {Q^4\over 32\pi^4x_B}{\sqrt{s/2}\over s^2}\sum_f e^2_f\int_0^1 \! dz
\int \!d^2z_1\, d^2z_2\,
\left|K_1(\bar Q|z_{12}|)\right|^2
\nonumber\\
&&\times \bigg\{(z-\bar z)^2\bigg(2\calf_{z_1z_2} 
+ N_c\calq^f_{5z_1z_2} + {1\over N_c}\Psi^f_{5z_1z_2} - 2C_FQ^f_{5z_1}
\nonumber\\
&& \hspace{0.5cm}+ 2\calf^\dagger_{z_1z_2} + N_c{\calq^f}^\dagger_{5z_1z_2} + {1\over N_c}{\Psi^f}^\dagger_{5z_1z_2}
- 2C_F{Q^f}^\dagger_{5z_1}\bigg)
\nonumber\\
&&+ 8(z^2 + \barz^2){z_{12}^j\epsilon^{ji} \over  z^2_{12}}
\Big(\calf_{iz_1z_2} + \calf^\dagger_{iz_1z_2}\Big)
\bigg\}
+ O(\lambda^{-2}) \,.
\label{SumAsym-a}
\end{eqnarray}

The cancellation of explicit local terms between final
states does not remove the local axial quark contribution
from the complete result. The terms proportional to
$C_FQ_5$ and $C_FQ_5^\dagger$ in eq.~\eqref{qqbar-QuarkAmpsquare} cancel the
real-emission contribution in eq.~\eqref{Ceven-result}, after the
transverse integration variables are consistently
relabeled. This leaves the color traces displayed in
eq.~\eqref{SumAsym}. Rewriting those traces with eq.~\eqref{rewritecalQ5} produces
the local terms in eq.~\eqref{SumAsym-a}.

In particular, quark operators
$\calq_{5xy}^f$ and $\Psi_{5xy}^f$ vanish at $x_\perp=y_\perp$,
whereas their combination with the local terms approaches
\begin{eqnarray}
-2C_F\bigl[Q_5^f(x_\perp)+Q_5^{f\dagger}(x_\perp)\bigr]\,.
\label{Q5Cf}
\end{eqnarray}
The quark sector therefore retains a logarithmic
transverse short-distance singularity. The gluonic
$\mathcal F_i$ contribution requires a separate analysis:
its coefficient contains an additional factor
$\epsilon^{ji}r^j/r^2$, so the vanishing of its operator
trace at $r=0$ is not by itself sufficient to establish
ultraviolet finiteness.

Let us consider the flavor non-singlet projection of the quark sector. To this end, we define
\begin{eqnarray}
Q_5^{{\rm NS},(a)}(x_\perp)\equiv \sum_f c_f^{(a)}\,Q_5^f(x_\perp),
\label{NSproj-ops}
\end{eqnarray}
with coefficients $c_f^{(a)}$ satisfying
\begin{eqnarray}
\sum_f c_f^{(a)}=0.
\label{NSproj-coeff}
\end{eqnarray}
Here the label $(a)$ specifies the chosen non-singlet direction in flavor space.

In terms of these projected operators, the non-singlet quark-sector contribution to the
helicity-dependent asymmetry is
\begin{eqnarray}
\hspace{-1cm}g^{{\rm NS},(a)}_1(Q^2)
=\!\!\!&& {Q^4\over 32\pi^4x_B}{\sqrt{s/2}\over s^2}\int_0^1 \! dz\,(z-\bar z)^2
\int d^2z_1\, d^2z_2\,
\left|K_1(\bar Q|z_{12}|)\right|^2
\nonumber\\
&&\hspace{-0.4cm}\times \bigg(N_c\calq^{{\rm NS},(a)}_{5z_1z_2} + {1\over N_c}\Psi^{{\rm NS},(a)}_{5z_1z_2} - 2C_FQ_{5z_1}^{{\rm NS},(a)}
\nonumber\\
&&\hspace{-0.4cm} + N_c{\calq^{{\rm NS},(a)}_{5z_1z_2}}^\dagger
+ {1\over N_c}{\Psi^{{\rm NS},(a)}_{5z_1z_2}}^\dagger - 2C_F{Q_{5z_1}^{{\rm NS},(a)}}^\dagger \bigg)
+ O(\lambda^{-2}) \,.
\label{SumAsym-aNS}
\end{eqnarray}

Equation~(\ref{SumAsym-aNS}) should be understood as the non-singlet projection of the
quark contribution to the asymmetry. Equation (\ref{SumAsym-a}) is weighted by the charges $e_f^2$ and therefore contains both
quark and gluonic contributions, whereas the non-singlet projection isolates the quark
sector and removes the flavor-blind gluonic operator $\calf_{z_1 z_2}$.
Accordingly, eq. (\ref{SumAsym-aNS}) is naturally associated with the non-singlet sector of the
high-energy evolution, see eq.~\eqref{eq:Q5NSevol}; it is not the same object as the physical $g_1$ in eq.~(\ref{SumAsym-a}).

\subsection{Singularities and relation to high-energy evolution}
\label{sec:divergence}

Within the approximation in which the $x_B$ dependence of the
background operators is neglected, the calculated sub-eikonal
contributions cancel in the inclusive longitudinal and
helicity-averaged transverse projections. The surviving
singularities therefore occur in the helicity-dependent
contribution.

In the quark sector of eq.~\eqref{SumAsym-a}, the 
bilocal operators vanish at coincident transverse coordinates,
but the complete operator combination approaches
$-2C_F[Q_5^f(x_\perp)+Q_5^{f\dagger}(x_\perp)]$.
With $r=|z_{12}|$, $\bar z=1-z$, and $\bar Q^2=Q^2z\bar z$,
the corresponding coefficient behaves as
\begin{eqnarray}
Q^2(z-\bar z)^2 K_1^2(\bar Qr)
\underset{r\to0}{\sim}
{1\over r^2}
\left({1\over z}+{1\over\bar z}-4\right).
\label{eq:g1-short-distance}
\end{eqnarray}
Thus, the surviving local axial contribution produces both
a logarithmic transverse short-distance integral and
longitudinal endpoint logarithms. Including the $qg$ and
$\barq g$ channels therefore changes the short-distance
conclusion obtained from the $q\barq$ interference contribution
alone. For the gluonic $\calf_i$ contribution the coefficient is one power more singular,
$\epsilon^{ji}z_{12}^j/z_{12}^4$, while the operator vanishes only linearly at coincidence. The transverse
integration is therefore logarithmic also for this term, with coefficient
$\pi\,\epsilon^{ji}\partial_{z_{12}^j}\big(\calf_{i\,z_1z_2}+\calf^\dagger_{i\,z_1z_2}\big)\big|_{z_{12}=0}$,
which is a $z^+$-weighted $F_i^{\,-}F_j^{\,-}$ correlator contracted with $\epsilon^{ji}$ and does not
vanish.

The logarithmic regions of the quark sector are those generated by the rapidity evolution of the axial
operator that gives the one-particle term \eqref{Q5ops}. The evolution of $Q^f_5$ was derived in
refs.~\cite{Chirilli:2018kkw, Chirilli:2021lif}. Using eqs.~\eqref{calfoperator_a}, \eqref{Q5} and \eqref{calFixy}--\eqref{Qt5} it reads
\begin{eqnarray}
{d\over d\eta}\,Q^f_{5x}
&=& - {\alpha_s\over 2\pi^2}\int\frac{d^2z_\perp}{(x-z)_\perp^2}\,
\Tr\big\{U^\dagger_x\big(\calq^f_{5z}+\calf_z\big)\big\}\nonumber\\
&=&{\alpha_s\over 4\pi^2}\int {d^2z_\perp \over (x-z)_\perp^2}
\Big(2C_F\,Q^f_{5z}-N_c\,\calq^f_{5\,zx} - {1\over N_c}\,\Psi^f_{5\,zx}-2\,\calf_{xz}\Big),
\label{eq:Q5evol}
\end{eqnarray}
where the second line follows
from eq.~\eqref{rewritecalQ5}, and $Q^{f\dagger}_{5x}$ obeys the Hermitian-conjugate equation. The two operators enter
through the same quark--background insertion, $\ssn_2\gamma^5(\calf+\calq_5)$, which is
why they appear with equal weight both in eq.~\eqref{SumAsym} and in eq.~\eqref{eq:Q5evol}. At
$z_\perp=x_\perp$ the bracket reduces to $2C_FQ^f_{5x}$, since $\mathcal U_{xx}=\Psi^f_{5\,xx}=0$ and
$\calf_{xx}=\Tr\{U^\dagger_x\calf_x\}=0$.

Let us observe that the operators in the bracket multiplying
$(z-\bar z)^2$ in eq.~\eqref{SumAsym-a} are equal, term by term, to minus the operators in the bracket of eq.~\eqref{eq:Q5evol}.
At fixed $z_{12}$, $\bar Q|z_{12}|\to0$ for $z\to0$ and for $z\to1$, and by eq.~\eqref{eq:g1-short-distance} the
weight becomes $(1/z+1/\bar z-4)/z_{12}^2$. The $z_{1}$ integration then has the kernel of
eq.~\eqref{eq:Q5evol}. The terms $1/z$ and $1/\bar z$ produce the endpoint logarithms, while the
constant $-4$ is not enhanced at the endpoints and belongs to the coefficient function. With a cutoff
$\sigma$ on $z$ and $\bar z$, the endpoint-enhanced part of eq.~\eqref{SumAsym-a} is therefore proportional to
\begin{eqnarray}
\ln{1\over \sigma}\;\sum_f e_f^2\int d^2x_\perp\,{d\over d\eta}\Big(Q^f_{5x}+Q^{f\dagger}_{5x}\Big),
\label{lnsigmaQ5}
\end{eqnarray}
that is, to one step of the rapidity evolution of the operator \eqref{Q5ops}.
This is the organization described in Sec.~\ref{sec:dipolecrossection} for the eikonal dipole. The logarithm of the
$O(\alpha_s)$ coefficient function is generated by the evolution of the lower-order operator, and it is
absorbed into the rapidity dependence of $\langle Q_5+Q_5^\dagger\rangle$. The absorption covers the
local term \eqref{Q5Cf}, whose transverse short-distance logarithm is the one of the kernel of
eq.~\eqref{eq:Q5evol} at $z_\perp\to x_\perp$. It also covers the gluonic term: the source
$-2\calf_{xz}$ in eq.~\eqref{eq:Q5evol} describes the mixing of $\mathcal F$ into the axial quark
operator, and it accounts for the endpoint logarithm of the $\mathcal F$ term in eq.~\eqref{SumAsym-a}. 

With the normalization of Sec.~\ref{sec:xBreal}, eqs.~\eqref{eq:dbracket}--\eqref{eq:dict}, the one-particle term \eqref{one-particle-LO-g1}
reads
\begin{eqnarray}
g_1^{q,\bar q}= - {1\over 8\pi^2s\,\alpha_s}\sum_fe_f^2\,\langle\!\langle Q^f_5+Q^{f\dagger}_5\rangle\!\rangle ,
\end{eqnarray}
so that eq.~\eqref{lnsigmaQ5} is $\ln(1/\sigma)\,dg_1^{q,\bar q}/d\eta$: the coefficient of the endpoint logarithm is
one step of the evolution of the one-particle term, with unit coefficient when $\eta$ is identified with
the logarithm of the cutoff on the longitudinal momentum fraction. The same identification follows from
Sec.~\ref{sec:xBreal}. In the dilute limit the kernel of eq.~\eqref{eq:Q5evol} reduces to $(\alpha_sC_F/2\pi^2)\int d^2z_\perp/(x-z)^2_\perp$,
and eq.~\eqref{lnsigmaQ5} becomes $(\alpha_sC_F/\pi)\ln(1/\sigma)\int dr/r\;g_1^{q,\bar q}$; 
the endpoint part of eq.~\eqref{eqxB-g1}, for $x\gg x_B$
and slowly varying helicity distributions, gives $(\alpha_s C_F/2\pi)\ln(1/\sigma)\ln(1/x_B)\,
g_1^{q,\bar q}$. The two
agree, the transverse logarithm $\int dr^2/r^2=\ln(1/x_B)$ running from $r^2\sim1/\bar Q^2$ down to
$r^2\sim x_B/\bar Q^2$, which is the bound $x(k)\le1$ of eq.~\eqref{eq:xB-E}.

The $\calf_i$ term of
eq.~\eqref{SumAsym-a} has no counterpart in eq.~\eqref{eq:Q5evol}. Its endpoint logarithm requires either an
additional source in the evolution of $Q_5$ or the evolution of $\mathcal F_i$ itself, and we do not
address it here.

For the flavor non-singlet projection \eqref{NSproj-coeff}--\eqref{SumAsym-aNS} the gluonic term drops out, because $\calf$ is
flavor blind and $\sum_fc^{(a)}_f=0$:
\begin{eqnarray}
{d\over d\eta}\,Q^{{\rm NS},(a)}_{5x}={\alpha_s\over 4\pi^2}\int {d^2z_\perp\over (x-z)_\perp^2}
\Big(2C_F\,Q^{{\rm NS},(a)}_{5z}-N_c\,\calq^{{\rm NS},(a)}_{5\,zx}
- {1\over N_c}\,\Psi^{{\rm NS},(a)}_{5\,zx}\Big).
\label{eq:Q5NSevol}
\end{eqnarray}
With the same identification, the bracket in
eq.~\eqref{eq:Q5NSevol} is minus the bracket in eq.~\eqref{SumAsym-aNS}.
The endpoint logarithm is therefore absorbed into the evolution
of the flavor non-singlet quark operators.
The purely gluonic source cancels in this projection, but the
displayed equation does not close on the local operator
$Q_5^{\mathrm{NS},(a)}$ in a general background:
its right-hand side still contains the bilocal operators
$\calq_5^{\mathrm{NS},(a)}$ and
$\Psi_5^{\mathrm{NS},(a)}$. In the dilute limit $U\to1$ the operators $\calq^{\rm NS}_5$ and $\Psi^{\rm NS}_5$ vanish,
and eq.~\eqref{eq:Q5NSevol} reduces to the ladder kernel
$(\alpha_sC_F/2\pi^2)\int d^2z_\perp/(x-z)^2_\perp$. Its double-logarithmic resummation was analyzed for
$Q_1$ in ref.~\cite{Chirilli:2026pkv}, and we do not repeat it here. For the flavor-singlet combination $\sum_fQ^f_5$ the
gluonic term becomes $-2N_f\calf_{xz}$, and the evolution is coupled to that of $\calf$~\cite{Chirilli:2018kkw, Chirilli:2021lif}.

Equation~\eqref{eq:Q5evol} does not close on $g_1$. In the singlet sector it is coupled to the evolution
of $\mathcal F$, which in turn mixes with quark and gluonic operators~\cite{Chirilli:2018kkw, Chirilli:2021lif}, and the $\mathcal F_i$
sector requires its own evolution equation. The endpoint analysis above fixes the operator content of
the logarithm at this order. The small-$x_B$ behaviour of $g_1$, however, follows only from the solution
of the coupled system, together with matching and subtraction in a common factorization prescription.
This remains beyond the scope of the present calculation.

To determine finite coefficient functions with dimensional regularization,
the calculation must retain the required $D=4-2\epsilon$ dependence. In the helicity-dependent
sector this includes a specified prescription for $\gamma^5$ and the polarized projectors, since
$O(\epsilon)$ numerator terms can multiply poles and contribute to the finite result. 

\section{Conclusions}
\label{sec:conclusions}

In this work, we developed a mixed-space formulation of high-energy DIS in the
shock-wave (Wilson-line) formalism beyond the eikonal approximation and used it to derive
the first sub-eikonal corrections to dipole structure functions. Starting from the quark
propagator in the background field, we obtained the corresponding mixed-space Feynman
rules from the LSZ reduction formula in the presence of a shock wave and then applied the
same formalism to derive the sub-eikonal corrections to $F_L$, $F_T$, and to the
helicity-sensitive asymmetry related to $g_1$.

Our starting point was the quark propagator in the shock-wave background with
sub-eikonal corrections~\cite{Chirilli:2018kkw}, together with the coordinate-space
high-energy OPE at sub-eikonal level derived in ref.~\cite{Chirilli:2021lif}. 
We rewrote the propagator in a form suitable for the direct application of the LSZ reduction formula in
the presence of the shock wave. In this way, we derived the corresponding mixed-space
Feynman rules, including the terms that in light-cone perturbation theory are usually
interpreted as instantaneous interactions. As a first check of the formalism, we also
re-derived the standard eikonal dipole cross sections for longitudinal and transverse photon
polarization. See Appendix~\ref{sec:FeynRules} for the full list of Feynman rules.

Applying this formalism to the inclusive DIS projections of the $q\bar q$,
$qg$, and $\bar q g$ final states, we find that their first sub-eikonal
corrections to $F_L$ and $F_T$ cancel when the $x_B$ dependence of the
operators is neglected. The cancellation involves the gluonic and
quark-background contributions to the $q\bar q$ channel and the sum of the
charge-conjugate $qg$ and $\bar q g$ channels. In this approximation the
longitudinal and transverse structure functions retain their eikonal dipole
expressions, eqs.~\eqref{SumLongy} and \eqref{SumTransv}. 
The one-particle final states $\gamma^*\to q$ and $\gamma^*\to\bar q$~\cite{Chirilli:2026pkv}, which are
of lower order in $\alpha_s$, do not modify this conclusion when the momentum fraction in their quark operators is set to zero;  
at finite $x_B$, the one-particle channel gives
$F_T^{q,\bar q}=x_B\sum_f e_f^2(q_f+\bar q_f)$.

The cancellation has a common origin in charge conjugation: the $C$-even vector-quark combination is odd
in the momentum fraction $x$ of the target quark and vanishes at $x=0$, although its matrix element tends to
$q(0)+\bar q(0)$ as $x\to0^+$. It is therefore a property of the operators at zero momentum fraction, not
the small-$x_B$ limit of the physical structure functions. Keeping the longitudinal phase of the photon vertex, the $qg$ and $\bar qg$ channels reproduce
the Altarelli--Martinelli quark contribution to $F_L$, which checks the formalism and the normalization of
the quark operators.

The helicity-dependent projection starts at sub-eikonal order. Its leading
quark contribution is the one-particle term
$g_1^{q,\barq}=\half\sum_f e_f^2\,(\Delta q_f+\Delta\bar q_f)$, 
which, when the momentum fraction in the operators is set to zero, is proportional to $Q_5+Q_5^\dagger$~\cite{Chirilli:2026pkv}. 
The $q\bar q$, $qg$, and $\bar q g$ final states add contributions from the gluonic operators
$\mathcal{F}$ and $\mathcal{F}_i$ and an $\mathcal{O}(\alpha_s)$ correction
from axial quark operators.
The transverse momentum
integrations can be expressed in terms of modified
Bessel functions. In particular, the $\calf_i$
contribution contains the tensor kernel
$\epsilon^{ji}z_{12}^jK_1^2(\bar Q|z_{12}|)/z_{12}^2$.
The resulting expression for $g_1$ is given in eqs.~\eqref{g1-total},
\eqref{SumAsym}, and \eqref{SumAsym-a}, and the flavor non-singlet projection
of its quark sector in eq.~\eqref{SumAsym-aNS}.

Combining the quark final states cancels the explicit
local terms accompanying the color traces, but it does
not eliminate the local axial contribution contained
in those traces. In the subtracted dipole basis this
contribution is proportional to $Q_5+Q_5^\dagger$ and
produces a logarithmic transverse short-distance
singularity. The helicity-dependent coefficient
functions also contain logarithmic longitudinal
endpoint regions. For the $\calf_i$ contribution, the tensor kernel behaves
as $\epsilon^{ji}r^j/r^4$, where $r=z_1-z_2$.
Its contraction with the linear small-separation term of the
operator can also produce a logarithmic transverse integral,
as discussed in Sec.~\ref{sec:divergence}.
The vanishing of the operator trace at $r=0$ is therefore
insufficient to establish ultraviolet finiteness.

The longitudinal endpoint logarithm of the quark and $\calf$ terms is the one generated by the
rapidity evolution of $Q_5$, eq.~\eqref{eq:Q5evol}. Its coefficient reproduces the right-hand side of that equation
operator by operator, including the gluonic source. The same operator matching holds for the flavor non-singlet
projection and eq.~\eqref{eq:Q5NSevol}. The singularity is therefore absorbed into the evolution of the one-particle
term $g_1^{q,\bar q}$. A complete small-$x_B$ description of $g_1$ still requires the coupled evolution
of $Q_5$, $\mathcal F$ and $\calf_i$ and a common factorization prescription for the finite
remainders.

An additional extension is a systematic treatment of the $x_B$ dependence of the sub-eikonal operators.
The cancellations found here for $F_L$ and $F_T$ hold when this dependence is neglected. In
Sec.~\ref{sec:xBreal} we sketched how it is restored in the real-emission channels, where it follows from
the phase of the photon vertex. For $F_L$ the diagrams of Fig.~\ref{fig:dipole-q2g} give the complete $O(\alpha_s)$ quark-initiated contribution
in the dilute leading-twist limit and
reproduce the Altarelli--Martinelli quark coefficient. For $F_T$ and $g_1$ a complete result also requires
the diagram in which the photon is absorbed by the background quark and the gluon is emitted after the shock
wave, and the virtual corrections to the one-particle channel. In the $q\bar q$ channel the quark and gluon
insertions sit in the same amplitude, and restoring the $x_B$ dependence there requires the longitudinal
momentum transfer to the target, which also affects the eikonal dipole. Keeping the phase in only part of
these contributions does not define a controlled correction, since the remaining ones are of the same
order. A systematic finite-$x_B$ extension, with consistent power counting and the complete set of
contributions at the chosen accuracy, is left for future work.

\section{Acknowledgments}

The author gratefully acknowledges the financial support from the Theoretical Physics Division of the National Centre for Nuclear Research (NCBJ). Part of this 
work was developed during the workshop \emph{Bridging TMD Frameworks: Intersections, Tensions, and Applications} at ECT* in Trento, and the author acknowledges 
the hospitality and stimulating scientific atmosphere of ECT*. The author also thanks ECT* and INFN for support during the workshop. The author 
acknowledges the University of Salento, where part of this work was carried out. The author is grateful to I.~Balitsky, Yu.~Kovchegov, and A.~Vladimirov for 
useful discussions.

\appendix

\section{Notation}
\label{sec:notation-appendix}

In this section we include some of the results we obtained in ref.~\cite{Chirilli:2018kkw} which we will use in this work.

Let us consider the effect of a large longitudinal boost parameter $\lambda$ on the components of the gauge fields. We have
\begin{eqnarray}
&&A^-(x^-, x^+, x_\perp) \to \lambda\, A^-(\lambda^{-1}x^-, \lambda\, x^+, x_\perp)\,,\nonumber\\
&&A^+(x^-, x^+, x_\perp) \to  \lambda^{-1}A^+(\lambda^{-1}x^-, \lambda\, x^+, x_\perp)\,,
\label{boost}\\
&&A_\perp(x^-, x^+, x_\perp)  \to  A_\perp(\lambda^{-1}x^-, \lambda\, x^+, x_\perp)\,.\nonumber
\end{eqnarray}
Consequently, the field strength is rescaled as follows
\begin{eqnarray}
&&{F_i}^{\;-}(x^-, x^+, x_\perp) \to  \lambda\, {F_i}^{\;-}(\lambda^{-1}x^-, \lambda\, x^+, x_\perp)\,,\nonumber\\
&&{F_i}^{\;+}(x^-, x^+, x_\perp) \to  \lambda^{-1}{F_i}^{\;+}(\lambda^{-1}x^-, \lambda\, x^+, x_\perp)\,,
\label{Fboost}\nonumber
\\
&&F^{-+}(x^-, x^+, x_\perp)  \to  F^{-+}(\lambda^{-1}x^-, \lambda\, x^+, x_\perp)\,,
\nonumber\\
&&F_{ij}(x^-, x^+, x_\perp)  \to  F_{ij}(\lambda^{-1}x^-, \lambda\, x^+, x_\perp)\,.
\end{eqnarray}
and the spinor fields as
\begin{eqnarray}
\bar{\psi}t^a\ssn_1\psi \to  \lambda \bar{\psi}t^a\ssn_1\psi\,,~~~~
\bar{\psi}t^a\gamma^\perp_\nu\psi \to  \bar{\psi}t^a\gamma^\perp_\nu\psi\,,~~~~
\bar{\psi}t^a\ssn_2\psi \to  \lambda^{-1} \bar{\psi}t^a\ssn_2\psi\,.
\label{spinorboost}
\end{eqnarray}

In Schwinger representation, which will be frequently used throughout this paper, the free scalar propagator can be written as
\begin{eqnarray}
\brax {i\over p^2 + i\epsilon}\kety = i\!\int\!\dhd^4 k \,{e^{-ik\cdot(x-y)}\over k^2 + i\epsilon}\,,
\label{schwrep}
\end{eqnarray}
with $\langle k\ketx = e^{ix\cdot k}$.

In ref.~\cite{Chirilli:2018kkw}, we introduced the notion of covariant derivative of a Wilson line and
we distinguished it from the standard covariant derivative, using also different symbols.
The derivative of the gauge link with respect to the transverse position is given by
\begin{eqnarray}
{\partial \over \partial z^i} [un_1, vn_1 ]_z
=\!\!\!&& ig A_i(un_1+z_\perp)[un_1,vn_1]_z - ig [un_1,vn_1]_zA_i(vn_1+z_\perp)
\nonumber\\
&&+ ig\!\int^u_v\!\!\! ds\,[un_1,sn_1]_z{F_i}^{\,-}(n_1s+z_\perp)[n_1s,n_1v]_z\,,
\label{deriv-glink}
\end{eqnarray}
with transverse index $i=1,2$.  
From (\ref{deriv-glink}) we may formally define the transverse covariant derivative 
$\mathfrak{D}_i $ that acts on a non-local operator as
\begin{eqnarray}
\hspace{-1cm}i\mathfrak{D}_i\, [un_1, vn_1]_z &\!\equiv\!& 
i{\partial \over \partial z^i} [un_1, vn_1]_z
+g\big[A_i(z_\perp), [un_1,vn_1]_z\big]
\nonumber\\
&\!=\!& g\!\int^u_v\!\!\! ds\,[un_1,sn_1]_z{F^-}_{ i}(n_1s+z_\perp)[n_1s,n_1v]_z\,,
\label{coderiv-glink}
\end{eqnarray}
In eq. (\ref{coderiv-glink}), we have used the implicit notation 
\begin{eqnarray}
&&\big[A_i(z_\perp), [x^+n_1,y^+n_1]_z\big] 
\nonumber\\
=\!\!\!&& A_i(z_\perp+x^+n_1) [x^+n_1,y^+n_1]_z - [x^+n_1,y^+n_1]_z\,A_i(z_\perp + y^+n_1)
\end{eqnarray}

Another notation we will often use is
the Schwinger formalism for a gauge link
\begin{eqnarray}
\braxp [x^+,y^+] \ketyp = [x^+,y^+]_x\,\delta^{(2)}(x-y)\,.
\label{schwi-nota}
\end{eqnarray}
where we used the short-hand notation
\begin{eqnarray}
 [x^+,y^+]_z\equiv[x^+n_1 + z_\perp, y^+n_1 + z_\perp]
\end{eqnarray}

With the notations just introduced, we can then see the
action of the transverse momentum operator $\hat{P}_i = \hat{p}_i + g \hat{A}_i$ on a gauge link
\begin{eqnarray}
&&\braxp \big[\hat{P}_i, [x^+,y^+]\big]\ketyp 
\nonumber\\
=\!\!\!&& \braxp i\mathfrak{D}_i[x^+,y^+] \ketyp
= \braxp g\!\!\int^{x^+}_{y^+}\!\!\! d\omega^+
\,[x^+,\omega^+]{F^-}_i[\omega^+,y^+] \ketyp\,,
\label{defPi}
\end{eqnarray}
where we used the short-hand notation 
$[x^+,\omega^+]{F_i}^{\;-}[\omega^+,y^+] = [x^+,\omega^+]{F_i}^{\;-}(\omega^+)[\omega^+,y^+]$.

It is useful to notice that the covariant derivative $i\mathfrak{D}_i$ given in eq. (\ref{defPi}), 
acts on the gauge link $[x^+,y^+]$ even though the 
transverse coordinate has not been specified because of notation (\ref{schwi-nota}).

\section{The quark propagator up to sub-eikonal corrections in the shock-wave formalism}
\label{sec:quarpropsubeik}

We derived the quark propagator up to sub-eikonal corrections in the shock-wave formalism in ref.~\cite{Chirilli:2018kkw}. 
Let us report the result for the sub-eikonal correction in the background of gluon field. We have
\begin{eqnarray}
\langle {\rm T}\{\psi(x)\barpsi(y)\}\rangle_A
=\!\!\!&& \left[\int_0^{+\infty}\!\!{\dhd p^+\over 4(p^+)^2}\theta(x^+-y^+) - 
\int_{-\infty}^0\!\!{\dhd p^+\over 4(p^+)^2}\theta(y^+-x^+) \right] e^{-ip^+(x^- - y^-)}
\nonumber\\
&&\times \braxp e^{-i{\hatp^2_\perp\over 2p^+}x^+}\Bigg\{
\hat{\ssp}\ssn_2[x^+,y^+]\hat{\ssp}
+ \hat{\ssp}\ssn_2\,\widehat{\mathcal{O}}_1(x^+,y^+;p_\perp)\,\hat{\ssp}
\nonumber\\
&&
+ \hat{\ssp}\ssn_2 \half\widehat{\mathcal{O}}_2(x^+,y^+;p_\perp)
-  \half\hat{\mathcal{O}}_2(x^+,y^+;p_\perp)\ssn_2\hat{\ssp} \Bigg\}e^{i{\hatp^2_\perp\over 2p^+}y^+}\ketyp
\nonumber\\
&&+ O(\lambda^{-2})\,.
\label{quarksubnoedge3}
\end{eqnarray}
where
\begin{eqnarray}
\hspace{-0.5cm} 
\widehat{\mathcal{O}}_1(x^+,y^+;p_\perp) = \!\!\!&&
{ig\over 2p^+}\int^{x^+}_{y^+}\!\!\!d\omega^+\bigg(
[x^+,\omega^+]\half \sigma^{ij}F_{ij}[\omega^+,y^+]
+ \big\{\hat{p}^i,[x^+,\omega^+]\,\,\omega^+\, {F_i}^{\;-}\,[\omega^+,y^+]\big\}
\nonumber\\
&&+ g\!\!\int^{x^+}_{\omega^+}\!\!\!d\omega'^+\big(\omega^+ - \omega'^+\big)
[x^+,\omega'^+]F^{i-}[\omega'^+,\omega^+]\,{F_i}^{\;-}\,[\omega^+,y^+]\bigg)\,,
\label{O1}
\end{eqnarray}
and\footnote{Note that, as we already noted in ref~\cite{Chirilli:2021lif},
the term $\big\{(i\Slash{D}_\perp F_{ij}),\gamma^i\gamma^j\big\}$ is absent from the operator $\widehat{\calo}_2$, because it is identically 0.}
\begin{eqnarray}
\widehat{\mathcal{O}}_2(x^+,y^+;p_\perp) = \!\!\!&&
{ig\over 2p^+}\int^{x^+}_{y^+}\!\!\!d\omega^+
\bigg[\big\{\hat{p}^k,[x^+,\omega^+]i F_{kj}\gamma^j[\omega^+,y^+]\big\}
\nonumber\\
&&\hspace{-0.2cm}
+ [x^+,\omega^+]i F_{kj}\gamma^j(i\,\mathfrak{D}^k[\omega^+,y^+]) 
- (i\,\mathfrak{D}^k[x^+,\omega^+])i F_{kj}\gamma^j[\omega^+,y^+] 
\nonumber\\
&&\hspace{-0.2cm}
- [x^+,\omega^+]\,i\, F^{-+}(i\,\Slash{\mathfrak{D}}_\perp[\omega^+,y^+])
+ (i\,\Slash{\mathfrak{D}}_\perp [x^+,\omega^+])\,i\, F^{-+}[\omega^+,y^+] 
\nonumber\\
&&\hspace{-0.2cm}
+(\hat{p}^+\ssn_1-\hat{\ssp}_\perp)[x^+,\omega^+]\,i\, F^{-+}[\omega^+,y^+]
\bigg]\,,
\label{O2}
\end{eqnarray}
with
\begin{eqnarray}
&&{ig\over 2p^+}\int_{y^+}^{x^+}\!\!\!d\omega^+\Big[[x^+,\omega^+]i F_{kj}\gamma^j(i\,\mathfrak{D}^k[\omega^+,y^+]) 
- (i\,\mathfrak{D}^k[x^+,\omega^+])i F_{kj}\gamma^j[\omega^+,y^+] \Big]
\\
&&\hspace{-0.8cm}= {ig\over 2p^+}\int_{y^+}^{x^+}\!\!\!d\omega^+
\!\int^{x^+}_{\omega^+}\!\!\!d{\omega'^+}
\Big[
[x^+,\omega'^+]gF^{k-}[\omega'^+,\omega^+]iF_{kj}\gamma^j[\omega^+,y^+]
\nonumber\\
&&~~~~~~~~ - [x^+,\omega'^+]iF_{kj}\gamma^j[\omega'^+,\omega^+]gF^{k-}[\omega^+,y^+]
\Big]\nonumber
\end{eqnarray}
and 
\begin{eqnarray}
&&{ig\over 2p^+}\int_{y^+}^{x^+}\!\!\!d\omega^+
\Big[- [x^+,\omega^+]\,i\, F^{-+}(i\,\Slash{\mathfrak{D}}_\perp[\omega^+,y^+])
+ (i\,\Slash{\mathfrak{D}}_\perp [x^+,\omega^+])\,i\, F^{-+}[\omega^+,y^+] \Big]
\\
&&\hspace{-0.8cm} = {ig\over 2p^+}\int_{y^+}^{x^+}\!\!\!d\omega^+
\!\int^{x^+}_{\omega^+}\!\!\!d{\omega'^+}
\Big[
[x^+,\omega'^+]iF^{-+}[\omega'^+,\omega^+]\gamma^kg{F_k}^{\;-}[\omega^+,y^+]
\nonumber\\
&&~~~~~~~~ - [x^+,\omega'^+]\gamma^kg{F_k}^{\;-}[\omega'^+,\omega^+]iF^{-+}[\omega^+,y^+]
\Big]\,,\nonumber
\end{eqnarray}

\subsection{Absence of the $\widehat O_2$ contribution for propagation
	with both endpoints outside the shock wave}
\label{sec:O2}

Let us consider here the sub-eikonal contribution $\widehat{\cal O}_2$ given in \eqref{O2}, and show that such contribution does not
contribute for propagation with both end-points outside the shock-wave.

First, we observe that the propagator is a background-field
Green function, rather than a matrix element between an incoming and
an outgoing quark. However, when both end-points are outside the background field, its free propagation
obeys homogeneous free Dirac equations at
both endpoints.  This imposes a constraint which was not implemented
in the overall symmetrization we used in ref~\cite{Chirilli:2018kkw}.

For definiteness, let us consider $x^+$ after the target and $y^+$ before it, so that
only the $p^+>0$ branch proportional to $\theta(x^+-y^+)$ contributes. 
Thus, assuming that $2\epsilon^+$ is the infinitesimal width of the shock-wave, then
we are considering only the case
\begin{equation}
y^+<-\epsilon^+\,, \qquad x^+>\epsilon^+\,.
\end{equation}
so, the propagator, in the shock-wave limit, is
\begin{eqnarray}
\langle {\rm T}\{\psi(x)\barpsi(y)\}\rangle_A
=\!\!\!&& \int_0^{+\infty}\!\!{\dhd p^+\over 4(p^+)^2}\theta(x^+-y^+)  e^{-ip^+(x^- - y^-)}
\braxp e^{-i{\hatp^2_\perp\over 2p^+}x^+}\Bigg\{
\hat{\ssp}\ssn_2\widehat{U}\hat{\ssp}
\nonumber\\
&&
+ \hat{\ssp}\ssn_2\,\widehat{\mathcal{O}}_1(p_\perp)\,\hat{\ssp}
+ \hat{\ssp}\ssn_2 \half\widehat{\mathcal{O}}_2(p_\perp)
-  \half\widehat{\mathcal{O}}_2(p_\perp)\ssn_2\hat{\ssp} \Bigg\}e^{i{\hatp^2_\perp\over 2p^+}y^+}\ketyp
\nonumber\\
&&+ O(\lambda^{-2})\,.
\label{quarksubnoedge-theta-xy}
\end{eqnarray}
with $\widehat{U} \equiv [\infty n_1,-\infty n_1]$, and $\widehat{\mathcal{O}}_i(p_\perp) = \widehat{\mathcal{O}}_i(\infty,-\infty;p_\perp)$,
$i=1,2$.
Inserting complete sets of transverse-momentum eigenstates on the two
sides of the operator in eq.~\eqref{quarksubnoedge-theta-xy}, we have
\begin{equation}
e^{-ip^+x^-}\langle x_\perp|p'_\perp\rangle
e^{-ip'^{\,2}_\perp x^+/(2p^+)} = e^{-ip'\cdot x},
\end{equation}
with
\begin{equation}
p^{\prime -} = {p'^{\,2}_\perp \over 2p^+},
\qquad
p'^2=0,
\end{equation}
and analogously on the $y$ side. In this way, eq.~\eqref{quarksubnoedge-theta-xy} becomes
\begin{eqnarray}
\langle {\rm T}\{\psi(x)\barpsi(y)\}\rangle_A
=\!\!\!&& \int_0^{+\infty}\!\!{\dhd p^+\over 4(p^+)^2}
\dhd^2p'\dhd^2p e^{-ip'\cdot x + ip\cdot y}\theta(x^+-y^+) 
\nonumber\\
&&\times  \bigg[
\ssp'\ssn_2\bra{p'_\perp}\widehat{U}\ket{p_\perp}\ssp
+ \ssp'\ssn_2\,\bra{p'_\perp}\widehat{\mathcal{O}}_1(p_\perp)\ket{p_\perp}\,\ssp
\nonumber\\
&&
+ \ssp'\ssn_2 \half\bra{p'_\perp}\widehat{\mathcal{O}}_2(p_\perp)\ket{p_\perp}
-  \half\bra{p'_\perp}\widehat{\mathcal{O}}_2(p_\perp)\ket{p_\perp}\ssn_2\ssp\bigg]
\nonumber\\
&&+ O(\lambda^{-2})\,.
\label{quarksubnoedge-theta-xy-1}
\end{eqnarray}
To facilitate the analysis, let us define
\begin{eqnarray}
&&\calk(p', p) \equiv \ssp'\ssn_2\bra{p'_\perp}\Big[\widehat{U} + \widehat{\mathcal{O}}_1(p_\perp)\Big]\ket{p_\perp}\ssp
\label{calk}
\\
&&\calt(p', p) \equiv \ssp'\ssn_2 \half\bra{p'_\perp}\widehat{\mathcal{O}}_2(p_\perp)\ket{p_\perp}
-  \half\bra{p'_\perp}\widehat{\mathcal{O}}_2(p_\perp)\ket{p_\perp}\ssn_2\ssp\,.
\label{calt}
\end{eqnarray}
The kernel $\calk(p', p)$ satisfies the equations
\begin{eqnarray}
\ssp'\calk(p', p) = \calk(p', p)\ssp = 0\,.
\label{eq:ofmotion}
\end{eqnarray}
A property of all kernels of the type \eqref{calk}, that satisfy the equation of motion \eqref{eq:ofmotion}, is
\begin{eqnarray}
\calk(p', p) = {\ssp'\ssn_2\calk(p', p)\ssn_2\ssp \over 4(p^+)^2}
\label{property}
\end{eqnarray}
Since the complete shock-wave kernel $\calk+\calt$ must satisfy
the free Dirac equations at both external endpoints, and $\calk$
already satisfies them separately, $\calt$ must satisfy them as well.
Therefore the identity \eqref{property} must apply also to $\calt$
\begin{eqnarray}
\calt(p', p) = {\ssp'\ssn_2\calt(p', p)\ssn_2\ssp \over 4(p^+)^2}
\label{propertyT}
\end{eqnarray}
but, since $\ssn_2\calt(p',p)\ssn_2 = 0$,
we deduce that 
\begin{eqnarray}
\calt(p',p) =0
\end{eqnarray}
We conclude that,
since the kernel $\calt$ cannot be put in the same form as
kernel $\calk$, then 
the operator $\calo_2$ does not contribute
for the case of a quark propagator in the shock-wave background with free propagator before and after the shock-wave. 
Operators of type $\calo_2$ contribute if the propagation stops or starts in the background field. For this reason, we
can omit operator $\calo_2$ from our present analysis.

\section{Combining the $\calf_i(z_\perp)$, and $\calg_2(z_\perp)$ sub-eikonal terms}
\label{app:pushingP}

Let us consider the following sub-eikonal corrections to the quark propagators and, using derivative 
(\ref{coderiv-glink}), we \textit{push} the $\hat{P}_i$ operator all to the right. 
So, considering only the case $x^+>0>y^+$, the term under consideration is
\begin{eqnarray}
\hspace{-1.4cm}\langle{\rm T}\{\psi(x)\bar{\psi}(y)\}\rangle_{A} &&
\stackrel{x^+>0>y^+}{\ni}
\int_0^{+\infty}\!{\dhd p^+\over 8(p^+)^3}\,e^{-i p^+(x^- -y^-)}\int d^2z
\braxp \ssp \,e^{-i{\hatp^2_\perp\over 2p^+}x^+}\ketzp
\nonumber\\
&&\hspace{-1cm}
\times ig\!\int_{-\infty}^{+\infty}d\omega^+\,\ssn_2\,\Bigg[\{\hat{P}^i\,,[\infty n_1, \omega^+]_z
\omega^+ {F_i}^{\;-}(\omega^+, z_\perp)
[\omega^+, -\infty n_1]_z\}
\nonumber\\
&&\hspace{-1cm}
+ g\int_{\omega^+}^{+\infty}\!\! d\omega'^+(\omega^+-\omega'^+)[\infty n_1,\omega'^+]F^{i-}(\omega'^+)
[\omega'^+,\omega^+]{F_i}^{\;-}(\omega^+)[\omega^+,-\infty n_1] \Bigg]
\nonumber\\
&&\hspace{-1cm}
\times\brazp\ssp\,e^{i{\hatp^2\over 2p^+}y^+}\ketyp
\label{Pi2right-1}
\end{eqnarray}
In eq. (\ref{Pi2right-1}), we have two terms because of $\{\hat{P}^i\,,[x^+,y^+]_z\} = P_i[x^+,y^+]_z + [x^+,y^+]_zP_i$.
Using eq. (\ref{coderiv-glink}), we arrive at
\begin{eqnarray}
\hspace{-0.8cm}(\ref{Pi2right-1})	= \!\!\!&& \int_0^{+\infty}\!{\dhd p^+\over 8(p^+)^3}\,e^{-ip^+(x^- -y^-)}\int d^2z
\braxp \ssp \,e^{-i{\hatp^2_\perp\over 2p^+}x^+}\ketzp
\nonumber\\
&&
\times ig\!\int_{-\infty}^{+\infty}d\omega^+\,\ssn_2\,\Bigg[
[\infty n_1, \omega^+]_z\omega^+ {F_i}^{\;-}(\omega^+, z_\perp)[\omega^+, -\infty n_1]_z\,2\hat{P}^i
\nonumber\\
&&+ [\infty n_1, \omega^+]_z\omega^+ iD^i{F_i}^{\;-}(\omega^+, z_\perp)[\omega^+, -\infty n_1]_z
\nonumber\\
&&- 2g\int_{-\infty}^{\omega^+}\! dz^+[\infty n_1,\omega^+]_z\,\omega^+ {F_i}^{\;-}[\omega^+,z^+]_z F^{i-}[z^+,-\infty n_1]_z
\Bigg]\brazp\ssp\,e^{i{\hatp^2\over 2p^+}y^+}\ketyp
\label{PropP2right-2}
\end{eqnarray}
Using definitions (\ref{Fprime}), (\ref{F2prime}), and (\ref{F2}), from result (\ref{PropP2right-2}) we obtain result (\ref{PropP2right}).

Similarly to the procedure which led us to (\ref{PropP2right-2}), one can consider 
the case in which the operator $\hat{P}_i$ is \textit{pushed} all to the left and get (\ref{PropP2left}).

\section{Feynman rules in the Shock-wave formalism}
\label{sec:FeynRules}
\subsection{Feynman rules for propagation outside the shock-wave}
\label{sec:FeynRulesA}

\begin{eqnarray}
&&\lim_{p^2\to 0}\int \! d^4x \, e^{ip\cdot x}\baru(p)i\slashd_x\Big(\theta(x^+)\theta(y^+)\left\langle{\rm T}\{\psi(x)\barpsi(y)\}\right\rangle\Big)
\nonumber\\
=&& i\, \lim_{p^2\to 0}\baru(p)\theta(p^+)\theta(y^+)\, e^{ip^+y^- + i{p^2_\perp\over 2p^+}y^+ - i(p,y)_\perp}
\end{eqnarray}
\begin{eqnarray}
&&\lim_{p^2\to 0}\int\! d^4x \, e^{ip\cdot x}\baru(p)i\slashd_x\Big(\theta(-x^+)\theta(-y^+)\left\langle{\rm T}\{\psi(x)\barpsi(y)\}\right\rangle\Big)
\nonumber\\
=&& i\, \lim_{p^2\to 0}\baru(p)\theta(-p^+)\theta(-y^+)\, e^{ip^+y^- + i{p^2_\perp\over 2p^+}y^+ - i(p,y)_\perp}
\end{eqnarray}
\begin{eqnarray}
&&\lim_{p^2\to 0}\int\! d^4x \, e^{-ip\cdot x}\barv(p)i\slashd_x\Big(\theta(x^+)\theta(y^+)\left\langle{\rm T}\{\psi(x)\barpsi(y)\}\right\rangle\Big)
\nonumber\\
=&& i\, \lim_{p^2\to 0}\barv(p)\theta(-p^+)\theta(y^+)\, e^{-ip^+y^- - i{p^2_\perp\over 2p^+}y^+ + i(p,y)_\perp}
\end{eqnarray}
\begin{eqnarray}
&&\lim_{p^2\to 0}\int\! d^4x \, e^{-ip\cdot x}\barv(p)i\slashd_x\Big(\theta(-x^+)\theta(-y^+)\left\langle{\rm T}\{\psi(x)\barpsi(y)\}\right\rangle\Big)
\nonumber\\
=&& i \,\lim_{p^2\to 0}\barv(p)\theta(p^+)\theta(-y^+)\, e^{-ip^+y^- - i{p^2_\perp\over 2p^+}y^+ + i(p,y)_\perp}
\end{eqnarray}
\begin{eqnarray}
&&\lim_{p^2\to 0}\int\! d^4y \Big(\theta(x^+)\theta(y^+)\left\langle{\rm T}\{\psi(x)\barpsi(y)\}\right\rangle\Big)(-i\overleftarrow{\slashd}_y)u(p)e^{-ip\cdot y}
\nonumber\\
=&& i\, \lim_{p^2\to 0}u(p) \theta(-p^+)\theta(x^+)\,e^{-ip^+x^- - i{p^2_\perp\over 2p^+}x^+ + i(p,x)_\perp}
\end{eqnarray}
\begin{eqnarray}
&&\lim_{p^2\to 0}\int\!d^4y \Big(\theta(-x^+)\theta(-y^+)\left\langle{\rm T}\{\psi(x)\barpsi(y)\}\right\rangle\Big)(-i\overleftarrow{\slashd}_y)u(p)e^{-ip\cdot y}
\nonumber\\
=&& i \,\lim_{p^2\to 0}u(p)\theta(p^+)\theta(-x^+)\, e^{-ip^+x^- - i{p^2_\perp\over 2p^+}x^+ + i(p,x)_\perp}
\end{eqnarray}
\begin{eqnarray}
&&\lim_{p^2\to 0}\int\! d^4y \Big(\theta(x^+)\theta(y^+)\left\langle{\rm T}\{\psi(x)\barpsi(y)\}\right\rangle\Big)(-i\overleftarrow{\slashd}_y)v(p)e^{ip\cdot y}
\nonumber\\
=&& i \,\lim_{p^2\to 0}v(p) \theta(p^+)\theta(x^+)\,e^{ip^+x^- + i{p^2_\perp\over 2p^+}x^+ - i(p,x)_\perp}
\end{eqnarray}
\begin{eqnarray}
&&\lim_{p^2\to 0}\int\! d^4y \Big(\theta(-x^+)\theta(-y^+)\left\langle{\rm T}\{\psi(x)\barpsi(y)\}\right\rangle\Big)(-i\overleftarrow{\slashd}_y)v(p)e^{ip\cdot y}
\nonumber\\
=&& i \,\lim_{p^2\to 0}v(p) \theta(-p^+)\theta(-x^+)\,e^{ip^+x^- + i{p^2_\perp\over 2p^+}x^+ - i(p,x)_\perp}
\end{eqnarray}

\subsection{Feynman rules for propagation across the shock-wave}
\label{sec:FeynRulesB}

\begin{eqnarray}
&&\hspace{-1.3cm}\lim_{p^2\to 0}\int\! d^4x\,e^{ip\cdot x}\baru(p)i\slashd_x\int d^4z\delta(z^+)\brax{i\hat{\ssp}\over p^2+i\epsilon}
\ketz \ssn_2
\Big(U_z\theta(x^+)\theta(-y^+) - U^\dagger_z\theta(-x^+)\theta(y^+)\Big)\braz{i\hat{\ssp}\over p^2+i\epsilon}\kety
\nonumber\\
&&\hspace{-1.3cm}=i\lim_{p^2\to 0}\int d^4z\,\delta(z^+)\,e^{ip^+ z^- - i(p,z)}
\baru(p)\ssn_2\Big(\theta(p^+)\theta(-y^+)U_z - \theta(-p^+)\theta(y^+)U^\dagger_z\Big)\braz{i\hat{\ssp}\over p^2+i\epsilon}\kety
\\
&&\hspace{-1.3cm} = i\lim_{p^2\to 0}\int d^2z \dhd^2 k\, e^{i p^+ y^- - i(p-k,z)_\perp - i(k,y)_\perp + i{k^2_\perp\over 2p^+}y^+}
\nonumber\\
&&~~~~~~\times\baru(p)\Big(\theta(p^+)\theta(-y^+)U_z + \theta(-p^+)\theta(y^+)U^\dagger_z\Big)
{\ssn_2(p^+\ssn_1+\ssk_\perp)\over 2p^+}
\label{FeynRule1}
\end{eqnarray}

\begin{eqnarray}
&&\hspace{-1.3cm}\lim_{p^2\to 0}\int\! d^4x\,e^{-ip\cdot x}\barv(p)i\slashd_x\int d^4z\delta(z^+)\brax{i\hat{\ssp}\over p^2+i\epsilon}
\ketz \ssn_2
\Big(U_z\theta(x^+)\theta(-y^+) - U^\dagger_z\theta(-x^+)\theta(y^+)\Big)\braz{i\hat{\ssp}\over p^2+i\epsilon}\kety
\nonumber\\
&&\hspace{-1.3cm}=i\lim_{p^2\to 0}\int d^4z\,\delta(z^+)\,e^{-ip^+ z^- + i(p,z)}
\barv(p)\ssn_2\Big(\theta(-p^+)\theta(-y^+)U_z - \theta(p^+)\theta(y^+)U^\dagger_z\Big)\braz{i\hat{\ssp}\over p^2+i\epsilon}\kety
\\
&&\hspace{-1.3cm} = i\lim_{p^2\to 0}\int d^2z \dhd^2 k\, e^{-ip^+ y^- - i{k^2_\perp\over 2p^+}y^+  + i(p+k,z)_\perp - i(k,y)_\perp}
\nonumber\\
&&~~~~~~\times
\barv(p)\Big(\theta(-p^+)\theta(-y^+)U_z + \theta(p^+)\theta(y^+)U^\dagger_z\Big)
{\ssn_2(p^+\ssn_1-\ssk_\perp)\over 2p^+}
\label{FeynRule2}
\end{eqnarray}

\begin{eqnarray}
&&\hspace{-1.3cm}\lim_{p^2\to 0}\int\! d^4y\int d^4z\,\delta(z^+)\brax{i\hat{\ssp}\over p^2+i\epsilon}
\ketz \ssn_2
\Big(U_z\theta(x^+)\theta(-y^+) - U^\dagger_z\theta(-x^+)\theta(y^+)\Big)\braz{i\hat{\ssp}\over p^2+i\epsilon}\kety 
(-i\overleftarrow{\slashd}_y)u(p)\,e^{-ip\cdot y}
\nonumber\\
&&\hspace{-1.3cm}= i\lim_{p^2\to 0}\int\! d^4z\,\delta(z^+)\,e^{-ip^+ z^- + i(p,z)}\brax{i\hat{\ssp}\over p^2+i\epsilon}
\ketz \ssn_2\Big(\theta(p^+)\theta(x^+)U_z - \theta(-p^+)\theta(-x^+)U^\dagger_z\Big)u(p)
\nonumber\\
&&\hspace{-1.3cm}= i\lim_{p^2\to 0}\,\int d^2z \dhd^2 k\, e^{-i p^+ x^- - i{k^2_\perp\over 2p^+}x^++ i(p-k,z)_\perp + i(k,x)_\perp }
\nonumber\\
&&~~~~~~\times
{(p^+\ssn_1+\ssk_\perp)\ssn_2\over 2p^+}
\Big(\theta(p^+)\theta(x^+)U_z + \theta(-p^+)\theta(-x^+)U^\dagger_z\Big)u(p)
\label{FeynRule3}
\end{eqnarray}

\begin{eqnarray}
&&\hspace{-1.3cm}\lim_{p^2\to 0}\int\! d^4y\int d^4z\,\delta(z^+)\brax{i\hat{\ssp}\over p^2+i\epsilon}
\ketz \ssn_2
\Big(U_z\theta(x^+)\theta(-y^+) - U^\dagger_z\theta(-x^+)\theta(y^+)\Big)\braz{i\hat{\ssp}\over p^2+i\epsilon}\kety 
(-i\overleftarrow{\slashd}_y)v(p)\,e^{ip\cdot y}
\nonumber\\
&&\hspace{-1.3cm}= i\lim_{p^2\to 0}\int\! d^4z\,\delta(z^+)\,e^{ip^+ z^- - i(p,z)}\brax{i\hat{\ssp}\over p^2+i\epsilon}
\ketz \ssn_2\Big(\theta(-p^+)\theta(x^+)U_z - \theta(p^+)\theta(-x^+)U^\dagger_z\Big)v(p)
\nonumber\\
&&\hspace{-1.3cm}= i\lim_{p^2\to 0}\,\int d^2z \dhd^2 k\, e^{i p^+ x^- + i{k^2_\perp\over 2p^+}x^+ - i(p+k,z)_\perp + i(k,x)_\perp }
\nonumber\\
&&~~~~~~\times
{(p^+\ssn_1-\ssk_\perp)\ssn_2\over 2p^+}
\Big(\theta(-p^+)\theta(x^+)U_z + \theta(p^+)\theta(-x^+)U^\dagger_z\Big)v(p)
\label{FeynRule4}
\end{eqnarray}


\bibliographystyle{JHEP}
\bibliography{MyReferences}

\end{document}